\documentclass[letterpaper,11pt,fleqn]{article}
\pdfoutput=1
\usepackage{jheppub}

\usepackage{epstopdf}
\usepackage[T1]{fontenc}
\usepackage{extarrows}
\usepackage{bm,amsmath,amssymb}
\usepackage[mathscr]{eucal}
\usepackage{graphicx}
\usepackage{cancel}

\newcommand{\lan}{\langle}
\newcommand{\ran}{\rangle}
\long\def\comment#1{ }

\newcommand{\nlo}{{\rm \scriptscriptstyle NLO}}
\newcommand{\lo}{{\rm \scriptscriptstyle LO}}

\newcommand{\eqn}[1]{Eq.~\eqref{#1}}
\newcommand{\beq}{\begin{equation}}
\newcommand{\eeq}{\end{equation}}

\newcommand{\order}[1]{\mathcal{O}{(#1)}}
\newcommand{\nn}{\nonumber\\}
\newcommand{\rmd}{{\rm d}}
\newcommand{\rme}{{\rm e}}
\newcommand{\bk}{\bm{k}}
\newcommand{\bq}{\bm{q}}
\newcommand{\bp}{\bm{p}}
\newcommand{\bx}{\bm{x}}
\newcommand{\by}{\bm{y}}

\newcommand{\bz}{\bm{z}}
\newcommand{\bw}{\bm{w}}

\title{Forward trijet production in proton-nucleus collisions}
\author{Edmond Iancu, Yair Mulian}
\affiliation{Institut de Physique Theorique, Universite Paris Saclay, CNRS, CEA, F-91191, Gif-sur-Yvette, France}
\emailAdd{edmond.iancu@ipht.fr, yair25m@gmail.com}

\abstract{Using the formalism of the light-cone wave function in perturbative QCD together with the hybrid factorization, we compute the cross-section for three particle production at forward rapidities in proton-nucleus collisions. We focus on the quark channel, in which the three produced partons --- a quark accompanied by a gluon pair, or two quarks plus one antiquark --- are all generated via two successive splittings starting with a quark that was originally collinear with the proton.  The three partons are put on-shell by their scattering off the nuclear target, described as a Lorentz-contracted ``shockwave''.  The three-parton component of the quark light-cone wave function that we compute on this occasion is also an ingredient for other interesting calculations, like the next-to-leading order correction to the cross-section for the production of a pair of jets.}

\begin{document}
\maketitle

\flushbottom

\section{Introduction}
Particle production in proton-nucleus collisions at ``forward rapidities'' (that is, in the proton fragmentation region) represents an important source of information about the small-$x$ part of the nuclear wave function, where gluon occupation numbers are high and non-linear effects like gluon saturation and multiple scattering are expected to be important. Within perturbative QCD, the corresponding cross-sections can be computed using the Color Glass Condensate (CGC) effective theory \cite{Iancu:2003xm,Gelis:2010nm}, which has recently been extended to next-to-leading order (NLO) accuracy (at least for the high-energy evolution and for specific scattering processes), together with the so-called ``hybrid factorization''  \cite{Dumitru:2005gt,Albacete:2013tpa,Albacete:2014fwa}. The physical picture underlying this factorization is that the ``forward'' jets (or hadrons) observed in the final state are generally produced via the fragmentation of a single collinear parton from the incoming proton, which carries a large fraction $x_p\sim\order{1}$ of the longitudinal momentum of the proton and hence is typically a quark. (The gluon channel too becomes important when $x_p\lesssim 0.2$ and this should be kept in mind for the applications to phenomenology. For simplicity, in what follows we shall stick to the quark channel alone. See the Conclusion section for a brief discussion of other channels.)

The collinear quark from the proton scatters off the gluon distribution in the nucleus and in the process it can also radiate other partons --- gluons and quark-antiquark pairs ---, which are put on-shell (up to the effects of hadronisation) by their collision with the target.  As a result of the collision, the  partonic system also acquires a {\it total} transverse momentum of the order of the saturation momentum in the nucleus. In the high-energy regime of interest, the effects of multiple scattering can be resummed to all orders using the eikonal approximation. This amounts to associating a Wilson line built with the color field of the target to each parton from the projectile. The cross-section for parton(s)--nucleus scattering is then obtained by averaging over all the color field configurations in the target with the CGC weight function  \cite{Iancu:2003xm,Gelis:2010nm}. Finally, the cross-section for (multi)-hadron/jet production in proton-nucleus collisions is obtained by convoluting the partonic cross-section with the quark distribution function in the proton and the fragmentation functions for partons fragmenting  into hadrons, or jets.

Using this approach, one has so far computed the cross-section for single inclusive hadron production, first  to leading-order (LO) accuracy \cite{Kovchegov:1998bi,Kovchegov:2001sc,Dumitru:2002qt} and then to NLO \cite{Chirilli:2011km,Chirilli:2012jd,Altinoluk:2014eka,Iancu:2016vyg}, and that for dijet production only at LO \cite{Baier:2005dv,Marquet:2007vb,Dominguez:2011wm,Iancu:2013dta,vanHameren:2016ftb}. The results thus obtained compare quite well with the phenomenology, for both the single inclusive spectra \cite{Albacete:2003iq,Kharzeev:2003wz,Iancu:2004bx,Blaizot:2004wu,Blaizot:2004wv,Dumitru:2005gt,Albacete:2010bs,Tribedy:2011aa,Rezaeian:2012ye,Lappi:2013zma,Stasto:2013cha} and the dijet production
\cite{Albacete:2010pg}. Some issues which appeared in the NLO factorization for single particle production \cite{Stasto:2013cha,Stasto:2014sea,Watanabe:2015tja,Ducloue:2016shw,Ducloue:2017mpb} seem to have been resolved by now \cite{Iancu:2016vyg,Ducloue:2017mpb,Ducloue:2017dit}, but the associated phenomenology is still waiting for an update. At this point, one should also mention the closely related progress with understanding the non-linear evolution in QCD at high energy \cite{Balitsky:1995ub,Kovchegov:1999yj,JalilianMarian:1997jx,JalilianMarian:1997gr,Kovner:2000pt,Iancu:2000hn,Iancu:2001ad,Ferreiro:2001qy} beyond leading order \cite{Balitsky:2008zza,Balitsky:2013fea,Kovner:2013ona,Kovner:2014lca,Beuf:2014uia,Iancu:2015vea,Iancu:2015joa,Lappi:2016fmu,Hatta:2016ujq,Lublinsky:2016meo}, as well as with computing the structure functions for deep inelastic scattering at NLO \cite{Balitsky:2012bs,Beuf:2016wdz,Beuf:2017bpd,Ducloue:2017ftk}.

The multi-particle correlations  in the simultaneous production of several hadrons, or jets, at forward rapidities are particularly sensitive to the high gluon density effects in the nuclear wave function. The example of di-hadron (or dijet) production is instructive in that sense. When produced via a standard $2\to 2$ hard process, the final particles must propagate nearly back-to-back in the transverse plane, by momentum conservation. Accordingly, the di-hadron distribution in  the relative azimuthal angle $\Delta\phi$ shows a pronounced peak at $\Delta\phi=\pi$. However, if the two final particles are produced via a $1\to 2$ splitting in the background of the target gluon field --- the dominant channel for di-hadron production at forward rapidities ---, then one expects a transverse momentum imbalance, due to the momentum transferred by the target. This should lead to a smearing of the peak at $\Delta\phi=\pi$ in $pA$ collisions as compared to $pp$ collisions (for the same kinematical conditions). Besides, the smearing should get stronger when increasing the rapidities of the produced particles (since one is probing larger and larger values of the target saturation momentum) and also when increasing the `centrality' (multiplicity) in the $pA$ collisions. These phenomena have been first predicted by the theory \cite{Marquet:2007vb,Albacete:2010pg} and then confirmed by the RHIC data for d+Au collisions \cite{Braidot:2011zj,Adare:2011sc}.  The agreement between the data and the present, LO, calculation being only qualitative, it becomes urgent to promote the di-hadron  calculation to NLO.  But this is an extremely complex calculation  --- considerably more so than for single particle production. 

The main reason for the additional complication is the fact that, in the final state, two particles are detected instead of just one. For the purposes of the NLO calculation, one must also consider the final states involving three partons, one of which is not measured --- its kinematics is rather ``integrated over'', to yield a ``real'' one-loop correction to the cross-section for two particle production. Besides, there are also ``virtual'' one-loop corrections, which appear already at the level of the amplitude. In this step, we shall achieve a first step towards the full NLO dijet calculation, by computing the amplitude and the cross-section for three parton production at leading order (tree-level) and in the quark channel.

Our new results in this paper are also interesting in themselves, as they refer to a trijet final state that can in principle be measured in the experiments. They extends previous results in the literature where three-particle final state were considered as well, but in somewhat simpler contexts --- namely in the framework of deep inelastic scattering at small $x$ \cite{Ayala:2016lhd,Boussarie:2016ogo}, or in the case where one of the three produced partons is a photon  \cite{Altinoluk:2018uax}. However, in our present analysis we shall not attempt to explore the physical consequences of our results for the three-parton cross-section: albeit fully explicit, they still involve complicated convolutions in transverse coordinate space, built notably with Weisz\"acker-Williams gluon emission kernels and multi-particle correlators of the Wilson lines. Possible simplifications, notably associated with the limit of a large number of colors ($N_c\to\infty$) and/or with specific kinematical configurations for the produced particles, will be briefly mentioned in the Conclusion section and more thoroughly addressed in a subsequent publication.

As already mentioned, we shall only consider the channel initiated by a large-$x_p$ quark collinear with the proton. Then there are two possible final\footnote{By `final' we mean here at partonic level, that is, the outgoing states in (leading-order) perturbation theory.} states with three partons, both produced via two successive parton splittings: in the first state, the original quark is accompanied by two gluons ($qgg$), in the second one, by a quark-antiquark pair ($qq\bar q$). We shall use the light-cone (LC) wave function (WF) formalism together with LC perturbation theory to describe the splittings of the incoming quark and the CGC formalism for the scattering between the ensuing system of partons and the nuclear target. We shall work in a frame in which the target is ultrarelativistic and hence appears to the proton as a Lorentz-contracted shockwave. The effects of the high-energy evolution will be implicitly encoded in the nuclear gluon distribution, as probed by the ``dilute--dense'' scattering via multi-point correlations of Wilson lines.

Light-cone wave-function calculations of a similar complexity have already been presented in the literature
\cite{Altinoluk:2014eka,Beuf:2014uia,Lublinsky:2016meo,Beuf:2016wdz,Beuf:2017bpd,Ayala:2016lhd,Boussarie:2016ogo,Lappi:2016oup,Hanninen:2017ddy,Altinoluk:2018uax}, but what is new about the present calculation is the systematic inclusion of all the (tree-level) Feynman graphs contributing to a final state with three colored partons, including ``initial-state'' (prior to the collisions) and ``final-state'' (after the collision) parton branchings, together with the appropriate contractions of Wilson lines. There are indeed many topologies, notably due to the various time-orderings between the two parton branchings and the collision with the shockwave, which explains why our final results for the cross-section, as summarized in Sect.~\ref{trijetfinal}, look rather cumbersome.

The paper is organized as follows. In Sect.~\ref{sec:prod}, we describe our general strategy for computing particle production in proton-nucleus collisions and notably the construction of the ``outgoing'' state in light-cone perturbation theory.  In Sect.~\ref{locros}, we illustrate our method in the simple context of the dijet production, in which the LCWF includes just one parton branching. For more generality, we consider here both quark and gluon-initiated channels.
Starting with Sect.~\ref{sec:NLOprod}, we present the calculation of the cross-section for three parton production, for the quark channel alone. First, in Sect.~\ref{sec:NLOprod} we discuss the three-parton Fock space components of the quark outgoing LCWF. Then, in Sect.~\ref{trijetfinal} we assembly our results to build the various contributions to the cross-sections for the processes $qA\to qq\bar q+X$ and $qA\to qg\bar q+X$. Via the hybrid factorization, this immediately yields the contribution of the quark channel to the leading-order cross-section for trijet production in $pA$ collisions at forward rapidities. Our conclusions and some perspectives are summarized in Sect.~\ref{Conc}. The seven Appendices play an important role too in the structure of this paper. The first two of them summarize our notations and conventions for the LC perturbation theory. The two subsequent ones collect mathematical ingredients of the calculations: the relevant matrix elements of the interaction Hamiltonian and some formulae for the Fourier transform, respectively. Then follows two appendices devoted to intermediate calculations that we refer to in the main text: for the leading-order (two-parton) components of the gluon LCWF and for the three-parton components of the quark LCWF, respectively. The final Appendix contains the definitions for all the multi-parton $S$-matrices built with Wilson lines which enter our final formulae for the cross-section. We also exhibit the simplifications arising in the multicolor limit.

\section{Particle production in the proton-shockwave scattering}
\label{sec:prod}
As discussed in the introduction, we would like to compute multi-particle production in proton-nucleus collisions at forward rapidities, that is, in the fragmentation region of the proton projectile. For this particular kinematics, the dominant contribution comes from the process where a valence quark from the proton, possibly accompanied by its radiation products, scatters off the gluon distribution in the nucleus and then emerges in the final state.  We shall compute this process within perturbative QCD, so in particular we shall ignore confinement: our `final state' will be built with partons (quarks and gluons), rather than physical hadrons. (For applications to the phenomenology, our final results should be convoluted with parton-to-hadrons fragmentation functions.) Hence, the `produced particles' will be partons from the light-cone wave function of the valence quark which are put on-shell by their scattering off the nuclear target. 

For the present purposes, the nucleus can be described in the spirit of the Color Glass Condensate, as a random color background field whose correlations reproduce the nuclear gluon distribution. Due to the high-energy kinematics, this background field is viewed by the projectile as a shockwave and the scattering can be computed in the eikonal approximation: when crossing the shockwave, a parton from the projectile does not get deflected, but merely acquires a color rotation represented by a Wilson line (see below for details). We shall work in a Lorentz frame similar to the laboratory frame at RHIC or the LHC:  the proton is an energetic right mover, with large longitudinal momentum $P^+$, whereas the nucleus is an energetic left mover, so that the shockwave is localised near $x^+=0$.

We shall use light-cone (LC) perturbation theory, which is the time-dependent (``old-fashioned'') perturbation theory of quantum mechanics adapted to ultrarelativistic kinematics. The ingredients of the formalism, in particular, the LC Hamiltonian of QCD, are summarized in the Appendices (see also  \cite{Venugopalan:1998zd} for a pedagogical introduction and more details). In this section, we shall remind some general features of this formalism, which are useful for understanding the subsequent manipulations. 

Namely, we shall explain how to build the outgoing ({\it out}) state at LC time $x^+\to \infty$, which is the state from which we shall compute particle production, starting with an incoming ({\it in}) state at $x^+\to -\infty$ which represents an on-shell, bare, quark. We shall assume, as usual, that interactions are adiabatically switched off when $|x^+|\to \infty$, so that the eigenstates of the LC Hamiltonian at asymptotically large (negative and positive) times are the Fock states built with on-shell bare partons. The outgoing state is obtained by dressing the bare incoming state with the QCD interactions, as encoded in the time evolution operator
\beq\label{U}
U(x^+,\,x^+_0)\,=\,{\rm e}^{-i {H}(x^+-x^+_0)}\,=\,U_0(x^+,\,x_0^+)\, U_I(x^+,\,x_0^+)\,,\eeq
and also by the scattering with the shockwave (the nucleus), whose treatment will be shortly specified. In \eqn{U}, $H$ denotes the LC QCD Hamiltonian, $H_0$ is its free (non-interacting) part, and
 \beq
U_I(x^+,\,x_0^+) = {\rm T}\exp\left\{-i\int_{x_0^+}^{x^+}\rmd y^+ H_{I}(y^+)\right\}
\eeq
is the evolution operator in the interaction picture, built with the respective version of the interaction  piece of the Hamiltonian, that is (below, $H_{\rm int}\equiv H-H_0$),
\beq
H_{I}(x^+)=U_0^\dagger (x^+,\,x_0^+)\,H_{\rm int}\,U_0(x^+,\,x_0^+)\,.
\eeq
In what follows we shall systematically use the interaction picture, which is well suited for the purposes of the perturbation theory. This in particular means that $H_{\rm int}$, $H_{I}(x^+)$ and hence the evolution operator are expressed in terms of Fock space creation and annihilation operators for bare quarks and gluons.

Consider now the scattering between the (dressed) quark and the nucleus. In the eikonal approximation appropriate at high energy, one can assume that the shockwave representing the nucleus has an infinitesimal extent in $x^+$. (The corrections to this picture are suppressed by inverse powers of the center-of-mass energy.) This in turn has two main consequences: \texttt{(i)} The transverse coordinates of the partons from the projectile are not modified by the scattering; the only effect of the latter is a color procession described by Wilson lines (see below). \texttt{(ii)} One can ignore parton branchings occurring during the scattering, i.e. within the support of the shockwave: the quantum evolution of the projectile occurs either well before, or well after, its scattering off the target\footnote{The choice of a gauge is important too for this sharp separation in time: it holds in the projectile light-cone gauge, as inherent in the LC perturbation theory, but not also in other gauges like the covariant one; see e.g.  the discussion in \cite{Blaizot:2004wv,Gelis:2005pt}.}. This is tantamount to saying that the scattering and the evolution can be factorized from each other in $x^+$. Accordingly,  the outgoing state produced by the scattering between a quark from the projectile and the shockwave can be computed as follows:
\beq\label{qout}
\left|q_{\lambda}^{\alpha}(p^{+},\,\bp)\right\rangle^{out}\,=\,U_{I}(\infty,\,0) \,\hat S \, U_{I}(0,\,-\infty)\,
\left|q_{\lambda}^{\alpha}(p^{+},\,\bp)\right\rangle,
\eeq
where in the r.h.s. $\left|q_{\lambda}^{\alpha}(p^{+},\,\bm{p})\right\rangle \,\equiv\,{b_{\lambda}^{\alpha\dagger}(p^{+},\,\bm{p})}\left|0\right\rangle$ represents an on-shell bare quark state with 4-momentum $p^\mu=(p^+,p^-,\bp)$, with $p^+>0$ and $p^-=\bp^2/2p^+$, helicity $\lambda=\pm 1/2$ and color index $\alpha=1,2,\dots,N_c$.  This state is obtained by acting with the respective Fock space creation operator $b_{\lambda}^{\alpha\dagger}(p^{+},\,\bm{p})$ on the bare vacuum state $\left|0\right\rangle$. (Our conventions for the bare Fock space are summarized in Appendix \ref{fieldef}.) The unitary operator $U_{I}(0,\,-\infty)$ describes the QCD evolution of this state prior to scattering (``initial state evolution''), $U_{I}(\infty,\,0)$ similarly refers to the final state evolution, and $\hat S $ is the collision $S$-matrix in the eikonal approximation and in the  interaction representation. The differential cross-section for inclusive multi-jet production is finally computed as the average of a product of Fock space number density operators for {\it bare} partons over the {\em out} state (see \eqn{locrosdefin} for an example for the case of dijet production and Eq.~\eqref{qqqcross} for the production of three jets). This procedure is different from, but equivalent to, the one that was most often in the recent literature and which consists in counting the number of {\it dressed} partons ``at time $0+$'', i.e. immediately after the scattering \cite{Baier:2005dv,Kovner:2006wr,Kovner:2006ge,Marquet:2007vb,Altinoluk:2014eka,Altinoluk:2018uax}.

For what follows it is important to notice that in the absence of scattering, i.e. in the limit  $\hat S \to 1$, the  {\it in} and {\it out} states in \eqn{qout} coincide with each other, up to a phase. Indeed, by energy-momentum conservation, an on-shell quark cannot decay into a set of on-shell partons; hence, the partonic fluctuations that can develop at intermediate times via the QCD evolution are necessarily virtual and must close back at $x^+\to\infty$. This implies that there is no particle production in the absence of scattering, as expected on physical grounds.

In order to describe the scattering, it is convenient to use the transverse coordinate representation, where the $S$-matrix becomes diagonal. This is defined via the usual Fourier transform, e.g.
\begin{equation}\label{FTdef}
\left|q_{\lambda}^{\alpha}(p^{+},\,\bx)\right\rangle\,=\,\int \,\frac{d^{2}\bm{p}}{(2\pi)^{2}}\,
\rme^{-i\bm{x}\cdot\bm{p}}\left|q_{\lambda}^{\alpha}(p^{+},\,\bm{p})\right\rangle,
\end{equation}
and similarly for the {\it out} state.
By using the perturbative expansion of the evolution operators in \eqn{qout}, to be shortly explained, the state in the r.h.s. will be explicitly constructed as a superposition of bare Fock states including several partons (up to 3 in the approximations of interest). Via the  Fourier transform \eqref{FTdef}, each of the partons in this dressed state will have a well defined  transverse coordinate at the time scattering. The effect of the scattering on any such a bare parton is a multiplication by a Wilson line, in the appropriate representation of the color group:
\begin{align}\label{qscat}
\hat S  \left|q_{\lambda}^{\alpha}(p^{+},\,\bx)\right\rangle
 & =V_{\beta\alpha}(\bm{x})\,\left|q_{\lambda}^{\beta}(p^{+},\,\bx)\right\rangle,\quad
 \hat S \left|\bar q_{\lambda}^{\alpha}(p^{+},\,\bx)\right\rangle=
 V^{\dagger}_{\alpha\beta}(\bm{x})\,\left|\bar q_{\lambda}^{\beta}(p^{+},\,\bx)\right\rangle,  
 \\[0.3cm]
 \label{gscat}
\hat S \left|g_{i}^{a}(p^{+},\,\bx)\right\rangle
 & =U_{ba}(\bm{x})\,\left|g_i^b(p^{+},\,\bx)\right\rangle, \end{align}
 where  $\left|q \right\rangle$, $\left|\bar q \right\rangle$, $\left|g \right\rangle$ denote bare quark, antiquark, and gluon states respectively, and $U_{ba}(\bm{x})$ and $V_{\beta\alpha}(\bm{x})$ are matrix elements of Wilson lines in the adjoint and respectively fundamental representation:
\beq\label{SA}
U(\bm{x})\,=\,{\rm T}\exp\left\{ ig\int dx^{+}\,T^{a}A_{a}^{-}(x^{+},\,\bm{x})\right\} ,
\qquad   
  V(\bm{x})\,=\,{\rm T}\exp\left\{ ig\int dx^{+}\, t^{a}A^{-}_a(x^{+},\, \bm{x})\right\}.
   \eeq
These Wilson lines involve the ``minus'' component $A^{-}_a$ of the color field representing the (small-$x$) gluons in the target. As anticipated, this field is random and must be averaged out at the level of the cross-section, in order to reconstruct the correlations in the gluon distribution of the target.

As mentioned in the Introduction, in this paper we shall compute the leading-order cross-section for producing three partons in the final state. One of these three partons is the original valence quark and the other two (two gluons or a quark-antiquark pair) are produced via two successive parton branchings and are put on-shell by the scattering with the shockwave.  In preparation for this and for the sake of the presentation, we shall first compute the production of a pair of partons, by either an incoming (bare) quark, or a bare gluon. To construct all the relevant Fock space configurations, one needs the perturbative expansion of the evolution operators in \eqn{qout} up to second order. In this expansion, one can discard second-order contributions which represent one-loop corrections to the production of a single quark, as they do not contribute to the final states of interest. 

Our subsequent developments in this section will be rather schematic: they will be transformed into more explicit formulae in the subsequent sections and also in the Appendices. In particular, we shall use generic notations like $\left|i\right\rangle$, $\left|j\right\rangle$, $\left|f\right\rangle$ ... for the energy-momentum eigenstates of the free QCD Hamiltonian $H_0$, i.e. the Fock states built with bare partons. These states are assumed to be normalized and to form a complete set obeying
\beq
\sum_j|j\ran \lan j| =1,\qquad \lan j | H_I(x^+) | i\ran =\rme^{i(E_j-E_i)x^+} \lan j | H_{\rm int} |i\ran\,,
\eeq
where $E_j$ is the LC energy of the bare state $|j\ran $, i.e. the sum of the LC energies of the bare partons composing that state  (an eigenvalue of $H_0$). 

To the order of interest,  the perturbative expansions of $U_I(\infty,0)$ and $U_I(0,-\infty)$ read as follows
\beq\label{pertUin}
U_I(0,-\infty)=1 - i\int_{-\infty}^0\,dx^+\,H_I(x^+)-\int_{-\infty}^0\,dy^+\int_{-\infty}^{y^+}dx^+ \,H_I(y^+) H_I(x^+)+\,\cdots,
\eeq
and respectively 
\beq\label{pertUfin}
U_I(\infty,0)=1 - i\int^{\infty}_0\,dx^+\,H_I(x^+)-\int^{\infty}_0\,dx^+\int^{\infty}_{x^+}dy^+
\, H_I(y^+) H_I(x^+)+\,\cdots.
\eeq
Using these equations, it is easy to deduce the (formal) action of these evolution operators on some generic state $ | i\ran$. One finds\footnote{When performing the time integrations, it is convenient to insert a convergency factor $\rme^{-\epsilon |x^+|}$, with $\epsilon$ infinitesimal and positive. But for the present purposes, we can safely take the limit $\epsilon\to 0$, since the various states that we shall be interested in are not degenerate with each other.}\begin{align}\label{UinWV}
U_I(0,-\infty)\,|i\ran&=\,|i\ran -\sum_f\frac{\lan f| H_{\rm int} |i\ran}{E_f-E_i} |f\ran +
\sum_{f,j} \frac{\lan f| H_{\rm int} |j\ran \,\lan j| H_{\rm int} |i\ran
}{(E_f-E_i)(E_j-E_i)}\,|f\ran +\,\cdots \nonumber
 \\*[0.2cm] 
U_I(\infty,0)\,|i\ran&=\,|i\ran +\sum_f\frac{\lan f| H_{\rm int} |i\ran}{E_f-E_i} |f\ran +
\sum_{f,j} \frac{\lan f| H_{\rm int} |j\ran \,\lan j| H_{\rm int} |i\ran
}{(E_f-E_j)(E_f-E_i)}\,|f\ran +\,\cdots\,.
\end{align}
The energy denominators in these expressions have been generated by integrating over the interaction times (the time arguments of the interaction Hamiltonian). Note an important difference between the initial state evolution (as generated by  $U_I(0,-\infty)$) and, respectively, the final state evolution (described by $U_I(\infty,0)$), which becomes manifest starting with the second order: for the initial state evolution, the energy denominators involve the difference between the energy of the intermediate state under consideration and that ($E_i$) of the {\it initial} state; on the contrary, for the final state evolution, the energy  of the intermediate state is mesured w.r.t. to that of the {\em final} state ($E_f$).  

Here, however, we are more interested in the scattering state \eqref{qout}. To second order, one can write (using generic notations, once again)
\begin{align}\label{outexp}
|out\ran &=\,U_{I}(\infty,\,0) \,\hat S \, U_{I}(0,\,-\infty)\,|in\ran =\,
|in\ran \,+\,|out\ran^{(1)} \,+\,|out\ran^{(2)}+\,\cdots\,,
\end{align}
where $|out\ran^{(1)}$ denotes the first-order correction, that takes the following form (with $\lan f| S|j\ran$ denoting the matrix element of $\hat S $ between the respective bare Fock states)
\begin{align}\label{out1}
|out\ran^{(1)}  &=- \sum_{f,j,i}|f\ran\lan f| S|j\ran\frac{\lan j|H_{\rm int} |in\ran}{E_j-E_{in}}
+\sum_{f,j,i}|f\ran\,\frac{\lan f|H_{\rm int} |j\ran}{E_f-E_{j}}\,\lan j| S|in\ran\,,
\end{align}
whereas $|out\ran^{(2)}$ is the second-order correction and involves 3 pieces:
\begin{align}\label{out2}
|out\ran^{(2)}  = &\sum_{f,j,i} |f\ran\lan f| S|j\ran \frac{\lan j| H_{\rm int} |i\ran \,\lan i| H_{\rm int} |in\ran
}{(E_j-E_{in})(E_i-E_{in})} \,+\sum_{f,j,i} |f\ran \frac{\lan f| H_{\rm int} |j\ran \,\lan j| H_{\rm int} |i\ran
}{(E_f-E_{j})(E_f-E_{i})}\,\lan i| S|in\ran\nn
&- \sum_{f,j,i} |f\ran \,\frac{\lan f| H_{\rm int} |j\ran}{E_f-E_j}\,\lan j| S|i\ran\,
\frac{\lan i| H_{\rm int} |in\ran}{E_i-E_{in}}
\,.
\end{align}
Clearly, the action of $H_{\rm int} $ upon a Fock state built with bare partons is to generate one additional parton splitting. Within the first-order correction $|out\ran^{(1)}$, this splitting can occur either prior to the scattering off the shockwave (the first term in the r.h.s. of \eqn{out1}), or after that scattering (the second term in \eqn{out1}). Similarly, the second order correction $|out\ran^{(2)}$ involves two additional splittings, which can both occur prior to the scattering, or both after the scattering, or one before and the other one after (corresponding to the 3 terms in the r.h.s. of \eqn{out2}). As a matter of facts, the QCD interaction Hamiltonian in the light-cone gauge $H_{\rm int}$ also contains some pieces of order $g^2$ which formally describe {\em two} successive parton branchings connected by the instantaneous piece of either the gluon, or the quark, propagator. (These pieces are generated when using the operator equations of motion to eliminate the constrained components of the field in terms of the dynamical ones  \cite{Venugopalan:1998zd}.) Precisely because they occur simultaneously, these two branchings must occur on the same side of the shockwave --- that is, either before, or after the collision. This discussion also shows that the 1st order correction $|out\ran^{(1)}$, which is linear $H_{\rm int} $, involves not only contributions of order $g$, but also of order $g^2$.

At this level, one can explicitly check that the 1st order and 2nd order corrections to the outgoing state indeed cancel in the absence of scattering, i.e. when $\lan f| S|j\ran=\delta_{fj}$. For the 1st order piece \eqref{out1} this is quite obvious. For the 2nd order piece, one can write
\begin{align}\label{out2S1}
|out\ran^{(2)}\Big |_{S=1}  = &\sum_{f,j,i}|f\ran\,\lan f|H_{{\rm int}}|j\ran\,\lan j|H_{{\rm int}}|in\ran\\
&\times \left\{\frac{1}{(E_f-E_{in})(E_j-E_{in})}  + \frac{1}{(E_f-E_{j})(E_f-E_{in})} - \frac{1}{(E_f-E_{j})(E_j-E_{in})}\right\},\nonumber
\end{align}
where the sum of the energy denominators inside the braces is indeed equal to zero. Notice that, even in the absence of scattering ($\hat S =1$), the 3 second-order contributions are still different from each other (they involve different energy denominators), it is only their sum which vanishes.

 \section{The cross-section for forward dijets}
 \label{locros}
 
 As a warm up, and also with the purpose of introducing some notations and explaining our general method, we shall first compute the cross-section for dijet production in the scattering between either a quark, or a gluon, and the shockwave.  The main ingredient in that sense will be the two-parton Fock space component in the light-cone wave-function (LCWF) of the incoming parton (quark or gluon). This represents the leading-order (LO) perturbative correction to the respective bare state, so for brevity we shall refer to it as ``the LO quark (or gluon) wave-function''. With reference to the outgoing state, this is the same as the first-order correction $|out\ran^{(1)}$ defined in \eqn{out1} {\em except} for the fact that one keeps only the pieces of $\order{g}$ in the interaction Hamiltonian.
 
 \comment{ dressed by one parton branching. For brevity, the LCWF of the quark dressed by one gluon will be referred to as the ``LO quark wave-function''. As it will shortly become clear, the coefficient of the quark-gluon component in the LO quark wave-function (as is the case for the coefficient of the gluon-gluon and quark-antiquark component in the LO gluon wave-function) is needed to $\order{g}$, whereas the normalization of the bare quark component (describing the reduction in the probability for the bare quark or gluon to exist) is needed to $\order{g^2}$. In the second part of this section, we shall use the quark and gluon LO WFs in order to compute the (leading-order) cross-sections for producing a quark-gluon, quark-antiquark, and gluon-gluon dijets at forward rapidities and thus recover the expected result \cite{Marquet:2007vb}.
 }
 
\subsection{Quark production: transverse momentum broadening}  
\label{ptBroad}

To start with, we shall consider an even simpler process, with the purpose of checking the normalization of our general formula for the cross-section for particle production. Namely, we shall study the transverse momentum broadening of a quark which scatters off the shockwave.  This broadening is introduced by the multiple scattering between the quark and the gluons in the nuclear wave function and is therefore of the order of the nuclear saturation momentum $Q_s(x_g)$, with $x_g$ the longitudinal momentum fraction of the gluon from the target involved in the collision.

To LO in pQCD, we can neglect the evolution of the quark LCWF via radiation, hence the outgoing state differs from the incoming one (a bare quark with definite momentum) merely via the color precession introduced by the collision (cf. Eqs.~\eqref{FTdef} and \eqref{qscat}):
\beq
\left|q_{\lambda}^{\alpha}(q^{+},\,\bm{q})\right\rangle^{out}  =
\hat S  \left|q_{\lambda}^{\alpha}(q^{+},\,\bq)\right\rangle
  = \int d^2{\bm{x}}\,\rme^{i\bm{x}\cdot\bm{q}}\,
V_{\beta\alpha}(\bm{x})\,\big|q_{\lambda}^{\beta}(p^{+},\,\bx)\big\rangle.\quad
\eeq
Using this result, we now compute the cross-section for the inclusive quark production at forward rapidities and to LO. As we shall shortly check, this is correctly given (including normalization) by the following formula
\begin{equation}\begin{split}\label{1qcrosdefin}
\frac{d\sigma_{\lo}^{qA\rightarrow q+X}}{dp^{+}d^{2}\bm{p}}\,(2\pi)\delta(p^{+}-q^{+})\equiv\,\frac{1}{2N_{c}}\!\!{}_{\ \ }\!^{out}\left\langle q_{\lambda}^{\alpha}(q^{+},\,\bm{q})\right|\,\hat{\mathcal{N}}_{q}(p^{+},\,\bm{p})\,\left|q_{\lambda}^{\alpha}(q^{+},\,\bm{q})\right\rangle ^{out},
\end{split}\end{equation}
with $\hat{\mathcal{N}}_{q}(p^{+},\,\bm{p})$ the bare quark number density operator, which with our present conventions reads
\begin{equation}
\label{Nq}
\hat{\mathcal{N}}_{q}(p^{+},\,\bm{p})\,\equiv\,\frac{1}{(2\pi)^{3}}\,b_{\lambda}^{\alpha\dagger}(p^{+},\,\bm{p})\,b_{\lambda}^{\alpha}(p^{+},\,\bm{p})\,=\,\frac{1}{(2\pi)^{3}}\,\int_{\overline{\bm{x}},\,\bm{x}}e^{i\bm{p}\cdot(\overline{\bm{x}}-\bm{x})}\,b_{\lambda}^{\alpha\dagger}(p^{+},\overline{\bm{x}})\,b_{\lambda}^{\alpha}(p^{+},\,\bm{x}).
\end{equation}
The factor ${1}/(2N_{c})$ in the r.h.s. of \eqn{1qcrosdefin}
 accounts for the average over the color and helicity states of the initial quark. 
When no confusion is possible, we shall also use the simplified notation $p\equiv (p^+,\bp)$ for the three-momentum vector. The normalization in \eqn{Nq} is such that, when evaluating the total number of bare quarks, $\hat {n}_q\equiv \int d^3p \,\hat{\mathcal{N}}_{q}(p)$  for a properly normalized wavepacket representing a single bare quark, one finds indeed an average value $\langle \hat{n}_q\rangle =1$; see the discussion in App.~\ref{fieldef}, around \eqn{bqwp}.

Using the definition of the bare state and the commutation relations for the Fock space operators shown in Appendix \ref{fieldef}, it is easy to deduce 
\beq\label{sigma1q}
\frac{d\sigma_{\lo}^{qA\rightarrow q+X}}{dp^{+}d^{2}\bm{p}}\,=\,\delta(p^+-q^+)\frac{1}{(2\pi)^2}
\int d^2{\bm{x}} d^2{\bm{y}}\,\rme^{i(\bm{x}-\bm{y})\cdot\bm{p}}\,\frac{1}{N_{c}}\,\left\langle\mathrm{tr}\big(V^{\dagger}({\bm{y}})\,V(\bm{x})\big)\right\rangle,
\eeq
where the trace of the Wilson lines has been generated by the sum/average over the color indices in the final/initial state and the brackets encompassing the trace denote the average over the color fields in the target. This average, to be understood in the sense of the CGC effective theory \cite{Iancu:2003xm,Gelis:2010nm}, generates all the target multi-gluon correlations that are probed by the projectile quark via multiple scattering.

As anticipated, \eqn{sigma1q} is indeed the correct result, as well known in the literature. Its normalization can also be checked on physical grounds: if one integrates the cross-section \eqref{sigma1q} over $p^+$ and $\bp$, one must find the total surface of the target (since the probability for the quark to emerge with {\em any} value of the transverse momentum is one). And indeed, the integral over $\bp$ yields $(2\pi)^2\delta^{(2)}(\bm{x}-\bm{y})$, so the two Wilson lines multiply to unity and one is left with
\beq
\int dp^+ d^2\bp\,\frac{d\sigma_{\lo}^{qA\rightarrow q+X}}{d^{3}p}\,=\,\int d^2\bx,
\eeq
 where the integral in the r.h.s. covers the surface of the  target.
 
\subsection{The outgoing states for dijet production} 
\label{looutgoing}

\begin{figure}[t]\center
  \includegraphics[scale=0.7]{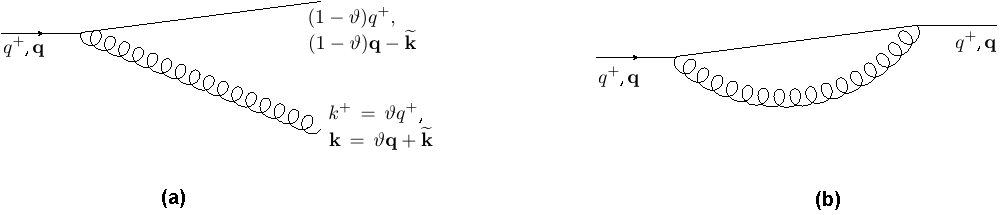}
  \caption{Left: One ``real'' gluon emission with 3-momentum $(k^+,\bk)=(\vartheta q^{+},\,\vartheta\bm{q}+\bm{\widetilde{k}})$ from an initial quark with 3-momentum $\left(q^{+},\,\bm{q}\right)$. We are using a non-eikonal vertex, in which the recoil of the quark will not be neglected. Right:  One ``virtual'' gluon emission contributing to the  normalization of the WF.
  \label{lofig}}
\end{figure}

The quark-gluon component of the quark LCWF is illustrated in Fig.~\ref{lofig}.a, which also shows our notations for the kinematics; such a process, where the gluon fluctuation is emitted in the final state, is conventionally referred to as ``real''. By contrast, the gluon loop shown in Fig.~\ref{lofig}.b, which contributes to the normalization of the LO  LCWF, is referred to as ``virtual''. As it should be clear by inspection of these graphs, the coefficient of the quark-gluon component is a quantity of $\order{g}$, whereas the normalization of the bare quark component (describing the reduction in the probability for the bare quark to exist) is an effect of $\order{g^2}$. 

Before we construct the full outgoing state (including the effects of the scattering), let us first consider the quark state at the time of scattering, as obtained by acting with $U_{I}(0,\,-\infty)$ on the bare ({\it in}) quark state.
To LO in perturbative QCD, this reads
\begin{equation}\begin{split}\label{loexp}
&\left|q_{\lambda}^{\alpha}(q^{+},\,\bm{q})\right\rangle _{\lo}\,\equiv\,U_{I}(0,\,-\infty) |_{\lo}\,\left|q_{\lambda}^{\alpha}(q^{+},\,\bm{q})\right\rangle \,=\,\mathcal{Z}_{\lo}\left|q_{\lambda}^{\alpha}(q)\right\rangle \,+\,\left|q_{\lambda}^{\alpha}(q)\right\rangle _{qg}\,,
\end{split}\end{equation}
with:
\begin{equation}\label{qgpart}
\left|q_{\lambda}^{\alpha}(q)\right\rangle _{qg}\,\equiv\,-\int_{0}^{\infty}\frac{ds^{+}}{2\pi}\,
\frac{dk^{+}}{2\pi}\int \frac{d^2\bm{s}}{(2\pi)^2}\,\frac{d^2\bm{k}}{(2\pi)^2}\,\left|q_{\lambda_{1}}^{\beta}(s)\,g_{i}^{a}(k)\right\rangle \frac{\left\langle q_{\lambda_{1}}^{\beta}(s)\,g_{i}^{a}(k)\left|\mathsf{H}_{q\rightarrow qg}\right|q_{\lambda}^{\alpha}(q)\right\rangle }{\Delta(q;\,s,\,k)}\,,
\end{equation}
where it is understood that the repeated discrete indices $i,\,a$ (for the gluon) and $\lambda_1,\,\beta$ (for the quark) must be summed over. The 3-momentum of the incoming quark has been denoted as $q=(q^{+},\,\bm{q})$, whereas 
$s\equiv (s^+,\bm{s})$  and  $k\equiv (k^+,\bk)$ similarly refer to the final quark and gluon. The constraint of momentum conservation is implicitly included in the transition matrix element of the interaction Hamiltonian.  $\alpha,\beta$ are color indices for the quark, i.e. for the fundamental representation of SU$(N_c)$, and $a,\,b$ similarly refer to the gluon (adjoint representation). Finally, $\lambda,\,\lambda_1$ represent quark helicity states, whereas $i=1,2$ denotes the transverse polarisation of the on-shell gluon.

 The first term in the r.h.s of \eqn{loexp} is the modified bare quark state. The normalization coefficient  $\mathcal{Z}_\lo$ is in general complex and its modulus --- as determined by the condition that the full wave function be normalized to unity\footnote{More precisely, this normalization strictly holds only for a physical wavepacket built as a linear superposition of ``plane waves'' (momentum eigenstates); see \eqn{bqwp} and the subsequent discussion.}
  (see \eqn{ZLO} below) --- measures the bare quark survival  probability to the order of interest. Alternatively, this coefficient can be computed by evaluating the one-loop graph in Fig.~\ref{lofig}.b.
The second term in the r.h.s. of \eqn{loexp} corresponds to the ``real'' graph in Fig.~\ref{lofig}.a and describes a bare quark-gluon pair. 

All the bare parton states are obtained by acting with the respective creation operators on the bare vacuum (see Appendix \ref{fieldef} for a summary of our conventions). Furthermore, the matrix elements of the interaction piece of  the LC Hamiltonian are listed in Appendix \ref{mateleapp}, see notably Eqs.~\eqref{gqq} and \eqref{hami_start} for the piece describing the quark splitting into a quark-gluon pair. Besides the relevant matrix element, \eqn{qgpart} also involves the following energy denominator --- the difference between the on-shell LC energies of the final quark-gluon state and the initial quark state:
\begin{equation}
\Delta(q;\,s,\,k)\,\equiv\,E(s,\,k)\,-\,E(q)\,=\,\frac{\bm{k}^{2}}{2k^{+}}\,+\,\frac{\bm{s}^{2}}{2s^{+}}\,-\,\frac{\bm{q}^{2}}{2q^{+}}\,.
\end{equation}

The condition that the LO WF be normalized to unity implies 
\begin{equation}\label{ZLO}\begin{split}
\left\Vert \mathcal{Z}_{\lo}\right\Vert \,=\,1\,-\,\int_{0}^{\infty}\frac{ds^{+}}{2\pi}\,\frac{dk^{+}}{2\pi}\int\frac{d^{2}\bm{s}}{(2\pi)^{2}}\,\frac{d^{2}\bm{k}}{(2\pi)^{2}}\,\frac{\left|\left\langle q_{\lambda_{1}}^{\beta}(s)\,g_{i}^{a}(k)\left|\mathsf{H}_{q\rightarrow qg}\right|q_{\lambda}^{\alpha}(q)\right\rangle \right|^{2}}{2\left[\Delta(q;\,s,\,k)\right]^{2}}\,.
\end{split}\end{equation}

To proceed with the calculation of (\ref{loexp}), it is convenient to introduce new variables according to
\begin{equation}\label{newvar}
\vartheta\,\equiv\,\frac{k^{+}}{q^{+}},\qquad\qquad\qquad\bm{\widetilde{k}}\,\equiv\,\bm{k}-\vartheta\bm{q}.
\end{equation}
$\vartheta$ is the longitudinal momentum fraction of the gluon while $\bm{\widetilde{k}}$ is proportional to the relative transverse velocity of the emitted gluon with respect to the parent quark. The matrix element which appears in (\ref{qgpart}) is given in (\ref{qqg}). By using the Pauli matrix identity $\sigma^{i}\sigma^{j}=i\varepsilon^{ij}\sigma^{3}+\delta^{ij}I$ we deduce
\begin{equation}
\frac{2\bm{k}^{i}}{k^{+}}-\frac{\sigma\cdot(\bm{q}-\bm{k})}{q^{+}-k^{+}}\sigma^{i}-\sigma^{i}\frac{\sigma\cdot\bm{q}}{q^{+}}\,=\,\frac{(2-\vartheta)\delta^{ij}-i\vartheta\varepsilon^{ij}\sigma^{3}}{\vartheta(1-\vartheta)q^{+}}\bm{\widetilde{k}}^{j},
\end{equation}
which helps us to simplify the matrix element (\ref{gqq1}) as follows:
\begin{equation}\label{qgqbasic}
\left\langle q_{\lambda_{1}}^{\beta}(s^{+},\,\bm{s})\,g_{i}^{a}(\vartheta q^{+},\,\vartheta\bm{q}+\bm{\widetilde{k}})\left|\mathsf{H}_{q\rightarrow qg}\right|q_{\lambda}^{\alpha}(q^{+},\,\bm{q})\right\rangle =\frac{gt_{\beta\alpha}^{a}\phi_{\lambda_{1}\lambda}^{ij}(\vartheta)\,\bm{\widetilde{k}}^{j}}{(1-\vartheta)(2\vartheta q^{+})^{3/2}}(2\pi)^{3}\delta^{(3)}(k+s-q).
\end{equation}
with (the 2-dimensional spinors $\chi_{\lambda}$ are defined in (\ref{chis})):
  \begin{equation}
\phi_{\lambda_{1}\lambda}^{ij}(\vartheta)\,\equiv\,\chi_{\lambda_{1}}^{\dagger}\left[(2-\vartheta)\delta^{ij}-i\vartheta\varepsilon^{ij}\sigma^{3}\right]\chi_{\lambda}\,=\,\delta_{\lambda\lambda_{1}}\left[(2-\vartheta)\delta^{ij}-2i\vartheta\varepsilon^{ij}\lambda\right].
\end{equation}
In terms of these new variables (\ref{newvar}), the energy denominator takes a simple form:
\begin{equation}\begin{split}\label{enede1}
&\Delta(q;\,k,\,q-k)\,=\,\frac{\bm{k}^{2}}{2k^{+}}\,+\,\frac{(\bm{q}-\bm{k})^{2}}{2(q^{+}-k^{+})}\,-\,\frac{\bm{q}^{2}}{2q^{+}}\,=\,\frac{(\bm{k}-\vartheta\bm{q})^{2}}{2\vartheta(1-\vartheta)q^{+}}\,=\,\frac{\bm{\widetilde{k}}^{2}}{2\vartheta(1-\vartheta)q^{+}}\,,
\end{split}\end{equation}
After inserting (\ref{qgqbasic}) and (\ref{enede1}) into expression (\ref{qgpart}), one finds
\begin{equation}\begin{split}\label{lobeg}
&\left|q_{\lambda}^{\alpha}(q^{+},\,\bm{q})\right\rangle _{qg}\\
&=\,-\,\int_{0}^{1}d\vartheta\,\int d^{2}\bm{\widetilde{k}}\,\frac{gt_{\beta\alpha}^{a}\phi_{\lambda_{1}\lambda}^{ij}(\vartheta)\,\sqrt{q^{+}}\,\bm{\widetilde{k}}^{j}}{8\sqrt{2\vartheta}\pi^{3}\bm{\widetilde{k}}^{2}}\,\left|q_{\lambda_{1}}^{\beta}\big((1-\vartheta)q^{+},\,(1-\vartheta)\bm{q}-\bm{\widetilde{k}}\big)\,g_{i}^{a}\big(\vartheta q^{+},\,\vartheta\bm{q}+\bm{\widetilde{k}}\big)\right\rangle .
\end{split}\end{equation}
To compute the scattering with the shockwave, we will need the Fourier transform of this wave-function to
the transverse coordinate representation, as introduced in \eqn{FTdef}; specifically,
\begin{equation}\label{chanrep}
\left|q_{\lambda}^{\alpha}(q^{+},\,\bm{w})\right\rangle_{qg} \equiv\int 
\frac{d^{2}\bm{q}}{(2\pi)^{2}}\,\rme^{-i\bm{w}\cdot\bm{q}}\left|q_{\lambda}^{\alpha}(q^{+},\,\bm{q})\right\rangle_{qg}.
\end{equation}
Then, with the aid of (\ref{delta}) and (\ref{fourier.1}), we arrive at
\begin{eqnarray}\label{lobegon}
&&\left|q_{\lambda}^{\alpha}(q^{+},\,\bm{w})\right\rangle _{qg}\\
&&=\,\int_{\bm{x},\,\bm{z}}\,\int_{0}^{1}d\vartheta\,\frac{igt_{\beta\alpha}^{a}\phi_{\lambda_{1}\lambda}^{ij}(\vartheta)\sqrt{q^{+}}\,\bm{X}^{j}}{4\sqrt{2\vartheta}\pi^{2}\bm{X}^{2}}\,\delta^{(2)}(\bm{w}-(1-\vartheta)\bm{x}-\vartheta\bm{z})\left|q_{\lambda_{1}}^{\beta}((1-\vartheta)q^{+},\,\bm{x})\,g_{i}^{a}(\vartheta q^{+},\,\bm{z})\right\rangle ,\nonumber
\end{eqnarray}
with the shorthand notations $\int_{\bm{z}}\,\equiv\,\int d^{2}\bm{z}$ and $\bm{X}\equiv\bm{x}-\bm{z}$. The transverse coordinate $\bm{w}=(1-\vartheta)\bm{x}+\vartheta\bm{z}$ is the center of mass of the quark-gluon pair.  
The bare quark-gluon component in the mixed representation is defined similarly to \eqn{chanrep} and is explicitly shown in \eqn{qgchanrep}.

\begin{figure}[t]\center
  \includegraphics[scale=0.85]{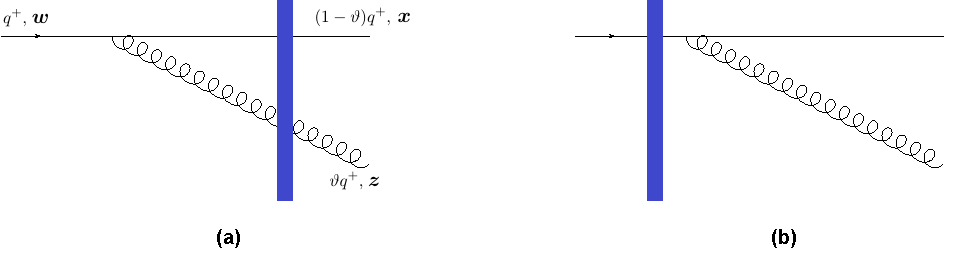}
  \caption{The quark-gluon component of the outgoing wave-function for an incoming quark. There are two ways to insert the shockwave, before and after the gluons emission, represented in Fig.~a and b, respectively. The transverse position of the quark is deflected by the emission of the gluon, from $\bm{w}$ to $\bm{x}=\frac{\bm{w}-\vartheta\bm{z}}{1-\vartheta}$.
  \label{emiglu}}
\end{figure}

At this level, it is straightforward to add the effect of the scattering (in coordinate space, this amounts to multiplication by Wilson lines, cf. Eqs.~(\ref{gscat})--(\ref{qscat})) and those of the final evolution (computed to leading order, once again). In displaying the outgoing state in what follows, we shall omit its bare quark component, since this does not contribute to dijet (here, quark-gluon) production. In suggestive but schematic notations to be often used in what follows, the quark-gluon component of the outgoing quark state reads
\begin{equation}\begin{split}\label{qqgout}
\left|q_{\lambda}^{\alpha}\right\rangle _{qg}^{out}\,=\,\left|q_{\lambda_{2}}^{\gamma}\,g_{j}^{b}\right\rangle \left\{-\left\langle q_{\lambda_{2}}^{\gamma}\,g_{j}^{b}\right|\hat{S}\left|q_{\lambda_{1}}^{\beta}\,g_{i}^{a}\right\rangle \frac{\left\langle q_{\lambda_{1}}^{\beta}\,g_{i}^{a}\left|\mathsf{H}_{q\rightarrow qg}\right|q_{\lambda}^{\alpha}\right\rangle }{E_{qg}-E_{q}}+\frac{\left\langle q_{\lambda_{2}}^{\gamma}\,g_{j}^{b}\right|\mathsf{H}_{q\rightarrow qg}\left|q_{\lambda_{1}}^{\beta}\right\rangle }{E_{qg}-E_{q}}\left\langle q_{\lambda_{1}}^{\beta}\left|\hat{S}\right|q_{\lambda}^{\alpha}\right\rangle \right\},
\end{split}\end{equation}
where the intermediate states are implicitly summed over. The first term inside the brackets corresponds to the initial-state evolution (gluon emission before the scattering), as explicitly computed in \eqn{lobeg} or \eqref{lobegon}. The second term describes the final-state evolution (gluon emission after the scattering).  The transverse coordinates or momenta are not explicitly shown in \eqn{qqgout}, which formally holds in any representation. But, clearly, the effects
of the scattering are easiest to compute in the  transverse coordinate representation, where the matrix elements of $\hat S$ are diagonal. One finally obtains (we recall that $\bm{X}\equiv\bm{x}-\bm{z}$):
\begin{equation}\begin{split}\label{asilo}
&\left|q_{\lambda}^{\alpha}(q^{+},\,\bm{w})\right\rangle _{qg}^{out}\,=\,-\int_{\bm{x},\,\bm{z}}\,\int_{0}^{1}d\vartheta\,\frac{ig\phi_{\lambda_{1}\lambda}^{ij}(\vartheta)\sqrt{q^{+}}\,\bm{X}^{j}}{4\sqrt{2\vartheta}\pi^{2}\,\bm{X}^{2}}\,\delta^{(2)}(\bm{w}-(1-\vartheta)\bm{x}-\vartheta\bm{z})\\
&\times\left[V^{\gamma\beta}(\bm{x})\,U^{ba}(\bm{z})\,t_{\beta\alpha}^{a}\,-\,t_{\gamma\beta}^{b}\,V^{\beta\alpha}(\bm{w})\right]\,\left|q_{\lambda_{1}}^{\gamma}((1-\vartheta)q^{+},\,\bm{x})\,g_{i}^{b}(\vartheta q^{+},\,\bm{z})\right\rangle .
\end{split}\end{equation}
The two terms inside the square bracket refer to the situation when the scattering with the shockwave occurs before and respectively after the emission of the gluon (see Fig.~\ref{emiglu}).

Notice an important difference w.r.t. to the  LO WF in the absence of scattering, cf. \eqn{lobeg}:
in the latter, the total transverse momentum of the quark-gluon pair is equal to that of the original quark;
in particular, when $\bq=0$, the final quark and gluon propagate back-to-back in the transverse plane. By contrast, in \eqn{asilo} there is generally an imbalance in transverse momentum between the produced quark-gluon pair and the original quark. This imbalance (which would become apparent when constructing the Fourier transform of \eqn{asilo}  back to the transverse momentum space) represents the momentum transferred by the target via multiple scattering, similarly to the transverse momentum broadening studied in Sect.~\ref{ptBroad}.

\begin{figure}[!h]
 \center \includegraphics[scale=0.78]{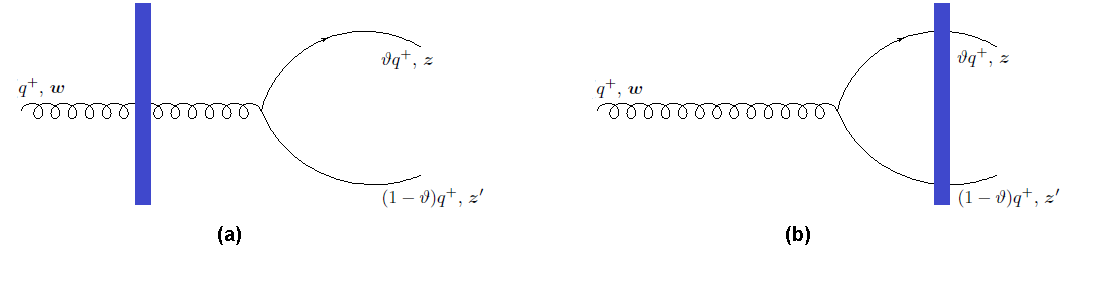}
\caption{The quark-antiquark  contributions to the gluon outgoing state.  \label{fig:gqq}}
\end{figure}

\begin{figure}[!h] \center \includegraphics[scale=0.78]{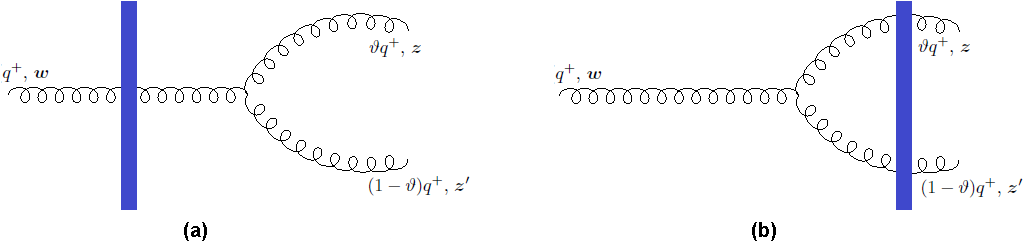}
\caption{The two-gluon contributions to the gluon outgoing state. \label{fig:ggg}}
\end{figure}

Following the same logic and using the matrix elements of the LC Hamiltonian  listed in Appendix \ref{mateleapp},
it is straightforward to construct the LO LCWF for an incoming {\em gluon}: starting with its general definition, that is,
\begin{align}
\label{scatgluo}
\left|g_{i}^{a}(q^{+},\,\bm{q})\right\rangle^{out} &\,\equiv\,U(\infty,\,0)\,\hat{S}\,U(0,\,-\infty)\,\left|g_{i}^{a}(q^{+},\,\bm{q})\right\rangle,
\end{align}
and expanding the two evolution operators to linear order in the interaction Hamiltonian, one finds that the outgoing state may involve two different two-parton Fock space components: gluon-gluon and quark-antiquark (see Figs.~\ref{fig:gqq} and \ref{fig:ggg} for illustrations). Similarly to \eqn{qqgout}, one can write
\begin{equation}\begin{split}\label{qqcompo}
\left|g_{i}^{a}\right\rangle _{q\overline{q}}^{out}\,=\,\left|\overline{q}_{\lambda_{2}}^{\delta}\:q_{\lambda_{1}}^{\gamma}\right\rangle \left\{-\left\langle \overline{q}_{\lambda_{2}}^{\delta}\:q_{\lambda_{1}}^{\gamma}\left|\hat{S}\right|\overline{q}_{\lambda_{4}}^{\alpha}\,q_{\lambda_{3}}^{\beta}\right\rangle \frac{\left\langle \overline{q}_{\lambda_{4}}^{\alpha}\,q_{\lambda_{3}}^{\beta}\left|\mathsf{H}_{g\rightarrow q\overline{q}}\right|g_{i}^{a}\right\rangle }{E_{gg}-E_{g}}+\frac{\left\langle \overline{q}_{\lambda_{2}}^{\delta}\:q_{\lambda_{1}}^{\gamma}\left|\mathsf{H}_{g\rightarrow q\overline{q}}\right|g_{j}^{b}\right\rangle }{E_{gg}-E_{g}}\left\langle g_{j}^{b}\left|\hat{S}\right|g_{i}^{a}\right\rangle \right\},
\end{split}\end{equation}
for the quark-antiquark component of the outgoing state and 
\begin{equation}\begin{split}\label{ggcompo}
\left|g_{i}^{a}\right\rangle _{gg}^{out}\,=\,\frac{1}{2}\left|g_{l}^{c}\,g_{j}^{e}\right\rangle \left\{-\left\langle g_{l}^{c}\,g_{j}^{e}\left|\hat{S}\right|g_{m}^{b}\,g_{n}^{d}\right\rangle \frac{\left\langle g_{m}^{b}\,g_{n}^{d}\left|\mathsf{H}_{g\rightarrow gg}\right|g_{i}^{a}\right\rangle }{E_{gg}-E_{g}}+\frac{\left\langle g_{l}^{c}\,g_{j}^{e}\left|\mathsf{H}_{g\rightarrow gg}\right|g_{k}^{b}\right\rangle }{E_{gg}-E_{g}}\left\langle g_{k}^{b}\left|\hat{S}\right|g_{i}^{a}\right\rangle \right\},
\end{split}\end{equation}
for the two-gluon component.

Once again, more explicit expressions can be deduced by working in the  transverse coordinate space. Some of the intermediate manipulations are presented in Appendix~\ref{glucha}; one eventually finds (with the notation $\bm{Z}\equiv \bm{z}-\bm{z}^{\prime}$)
\begin{equation}\begin{split}\label{qqasistat}
&\left|g_{i}^{a}(q^{+},\,\bm{w})\right\rangle _{q\bar{q}}^{out}\,=\,\int_{\bm{z},\,\bm{z}^{\prime}}\,\int_{0}^{1}d\vartheta\,\frac{igt_{\alpha\beta}^{a}\varphi_{\lambda_{2}\lambda_{1}}^{ij}(\vartheta)\sqrt{q^{+}}\,\bm{Z}^{j}}{4\sqrt{2}\pi^{2}\bm{Z}^{2}}\,\delta^{(2)}(\bm{w}-(1-\vartheta)\bm{z}^{\prime}-\vartheta\bm{z})\\
&\times\left[V^{\gamma\beta}(\bm{z}^{\prime})\,t_{\beta\alpha}^{a}\,V^{\dagger\alpha\delta}(\bm{z})\,-\,t_{\gamma\delta}^{b}\,U^{ba}(\bm{w})\right]\,\left|\bar{q}_{\lambda_{2}}^{\delta}((1-\vartheta)q^{+},\,\bm{z}^{\prime})\,q_{\lambda_{1}}^{\gamma}(\vartheta q^{+},\,\bm{z})\right\rangle,
\end{split}\end{equation}
and respectively
\begin{equation}\begin{split}\label{ggasistat}
&\left|g_{i}^{a}(q^{+},\,\bm{w})\right\rangle _{gg}^{out}\,=\,\int_{\bm{z},\,\bm{z}^{\prime}}\,\int_{0}^{1}d\vartheta\,\frac{igf^{abc}\,\sqrt{\vartheta(1-\vartheta)q^{+}}}{4\sqrt{2}\pi^{2}\bm{Z}^{2}}\left(\bm{Z}^{i}\delta_{jl}\,-\,\frac{1}{1-\vartheta}\bm{Z}^{j}\delta_{il}\,-\,\frac{1}{\vartheta}\bm{Z}^{l}\delta_{ij}\right)\\
&\times\delta^{(2)}(\bm{w}-(1-\vartheta)\bm{z}^{\prime}-\vartheta\bm{z})\,\left[U^{ed}(\bm{z}^{\prime})\,U^{cb}(\bm{z})\,f^{abd}\,-\,f^{bce}\,U^{ba}(\bm{w})\right]\,\left|g_{l}^{e}((1-\vartheta)q^{+},\,\bm{z}^{\prime})\,g_{j}^{c}(\vartheta q^{+},\,\bm{z})\right\rangle .
\end{split}\end{equation}

\subsection{The cross-sections for forward dijets}

Using the above results, we shall now compute the LO cross section for producing a quark-gluon pair at forward rapidities in a quark-nucleus scattering. The number density operator for (bare) quarks has already been introduced in \eqn{Nq}. The corresponding operators for antiquarks and gluons read  (we recall the simplified notation $p\equiv (p^+,\bp)$)
\begin{equation}\label{Nqbar}
\hat{\mathcal{N}}_{\overline{q}}(p)\,\equiv\,\frac{1}{(2\pi)^{3}}\,d_{\lambda}^{\alpha\dagger}(p)\,d_{\lambda}^{\alpha}(p)\,=\,\frac{1}{(2\pi)^{3}}\,\int_{\overline{\bm{x}},\,\bm{x}}e^{i\bm{p}\cdot(\overline{\bm{x}}-\bm{x})}\,d_{\lambda}^{\alpha\dagger}(p^{+},\overline{\bm{x}})\,d_{\lambda}^{\alpha}(p^{+},\,\bm{x}),
\end{equation}
\begin{equation}\label{Ng}
\hat{\mathcal{N}}_{g}(k)\,\equiv\,\frac{1}{(2\pi)^{3}}\,a_{i}^{a\dagger}(k)\,a_{i}^{a}(k)\,=\,\frac{1}{(2\pi)^{3}}\,\int_{\overline{\bm{z}},\,\bm{z}}e^{i\bm{k}\cdot(\overline{\bm{z}}-\bm{z})}\,a_{i}^{a\dagger}(k^{+},\,\overline{\bm{z}})\,a_{i}^{a}(k^{+},\,\bm{z}).
\end{equation}
If there are several quark flavors, then the sum over flavors is implicitly understood in Eqs.~\eqref{Nq} and \eqref{Nqbar}. The mixed-space representations shown in the r.h.s. of the above equations will be useful when evaluating the action of the Fock-space operators on the outgoing states (which are themselves computed in such a mixed representation; see e.g. \eqn{asilo}).

\subsubsection{The $qA\rightarrow qg+X$ channel}

The inclusive cross section for forward quark-gluon production in the scattering between a bare quark and the nucleus can be computed as the average of the respective product of number density operators over the momentum-space outgoing state, as obtained via inverse Fourier transform from  \eqn{asilo} :
\beq\label{backF}
\left|q_{\lambda}^{\alpha}(q^{+},\,\bm{q})\right\rangle _{qg}^{out}\,=\int_{\bm{w}}\,\,\rme^{i\bm{w}\cdot\bm{q}}
\left|q_{\lambda}^{\alpha}(q^{+},\,\bm{w})\right\rangle _{qg}^{out}\,.\eeq
Specifically, with our present conventions one can write (compare to \eqn{1qcrosdefin})
\begin{equation}\begin{split}\label{locrosdefin}
\frac{d\sigma_{\lo}^{qA\rightarrow qg+X}}{d^{3}k\,d^{3}p}\,(2\pi)\delta(k^++p^+-q^+)
\equiv\,\frac{1}{2N_{c}}\!\!{}_{\ \ \ qg}^{\ \ out}\!\left\langle q_{\lambda}^{\alpha}(q^{+},\,\bm{q})\right|\,\hat{\mathcal{N}}_{q}(p)\,\hat{\mathcal{N}}_{g}(k)\,\left|q_{\lambda}^{\alpha}(q^{+},\,\bm{q})\right\rangle _{qg}^{out}.
\end{split}\end{equation}

\begin{figure}[t]\center
  \includegraphics[scale=0.85]{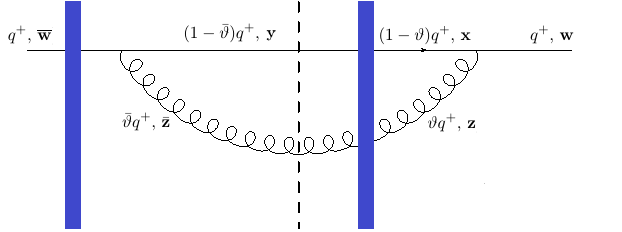}
  \caption{One of the contributions (the one proportional to $S_{q\bar{q}g}\left(\overline{\bm{w}},\,\bm{x},\,\bm{z}\right)$) to the quark-gluon production cross section in Eq.~(\ref{LOfinal}). There are four such contributions, in which the shockwave can be located either before or after the gluon emission, in both the direct amplitude and the complex conjugate amplitude. The dashed line represents the final state at $x^{+}\rightarrow\infty$. \label{loopfig} }
\end{figure}

After inserting the outgoing state given by Eqs.~(\ref{backF}) and (\ref{asilo}), together with the expressions for the quark and gluon number density operators in the mixed representation, we are led to the matrix element evaluated in (\ref{matqg}).  Using the identity $\varepsilon^{il}\varepsilon^{jk}=\delta^{ij}\delta^{kl}-\delta^{ik}\delta^{jl}$, one can perform the sums over the quark helicities and the gluon polarizations:
\begin{equation}\begin{split}\label{simptheta}
&\phi_{\lambda_{1}\lambda}^{ij\dagger}(\vartheta)\,\phi_{\lambda_{1}\lambda}^{ik}(\vartheta)\,=\,\chi_{\lambda}^{\dagger}\left((2-\vartheta)\delta^{ij}+i\vartheta\varepsilon^{ij}\sigma^{3}\right)\,I\,\big((2-\vartheta)\delta^{ik}-i\vartheta\varepsilon^{ik}\sigma^{3}\big)\chi_{\lambda}\\
&=\,4\delta^{jk}\,\left(1+(1-\vartheta)^{2}\right).
\end{split}\end{equation}
After also performing the integration over $\bm{w}$ in \eqn{backF} with the help of the $\delta$-function in Eq.~(\ref{asilo}), we obtain
\begin{equation}\begin{split}
&\frac{d\sigma_{\lo}^{qA\rightarrow qg+X}}{d^{3}k\,d^{3}p}\,=\,\frac{2\alpha_{s}\left(1+(1-\vartheta)^{2}\right)}{(2\pi)^{6}\vartheta N_{c}q^{+}}\,\int_{\bm{x},\,\overline{\bm{x}},\,\bm{z},\,\overline{\bm{z}}}\,\frac{\bm{X}\cdot\overline{\bm{X}}}{\bm{X}^{2}\,\overline{\bm{X}}^{2}}\,\rme^{-i(\bm{p}-(1-\vartheta)\bm{q})\cdot(\bm{x}-\overline{\bm{x}})-i(\bm{k}-\vartheta\bm{q})\cdot(\bm{z}-\bar{\bm{z}})}\\
&\times\left[t_{\alpha\delta}^{c}\,U^{\dagger cb}(\overline{\bm{z}})\,V^{\dagger\delta\gamma}(\overline{\bm{x}})\,-\,V^{\alpha\delta\dagger}(\overline{\bm{w}})\,t_{\delta\gamma}^{b}\right]\left[V^{\gamma\beta}(\bm{x})\,U^{ba}(\bm{z})\,t_{\beta\alpha}^{a}\,-\,t_{\gamma\beta}^{b}\,V^{\beta\alpha}(\bm{w})\right]\\
&\times\,\delta(q^{+}-k^{+}-p^{+}),
\end{split}\end{equation}
with $\bm{X}\equiv\bm{x}-\bm{z}$, $\overline{\bm{X}}\equiv\overline{\bm{x}}-\overline{\bm{z}}$, $\bm{w}=(1-\vartheta)\bm{x}+\vartheta\bm{z}$ and $\overline{\bm{w}}=(1-\vartheta)\overline{\bm{x}}+\vartheta\overline{\bm{z}}$. 

Without loss of generality, we can set $\bm{q}=0$ (this amounts to choosing the longitudinal axis parallel with the direction of the momentum of the initial quark). After some simple color algebra, one recovers the expected result for the forward quark-gluon production cross section, originally presented  in \cite{Marquet:2007vb}:
\begin{equation}\begin{split}\label{LOfinal}
\frac{d\sigma_{\lo}^{qA\rightarrow qg+X}}{dk^{+}\,d^{2}\bm{k}\,dp^{+}\,d^{2}\bm{p}}\,&=\,\frac{2\alpha_{s}C_{F}\left(1+(1-\vartheta)^{2}\right)}{(2\pi)^{6}\vartheta q^{+}}\,\delta(q^{+}-k^{+}-p^{+})
\\&\times\,\int_{\bm{x},\,\overline{\bm{x}},\,\bm{z},\,\overline{\bm{z}}}\,\frac{\bm{X}\cdot\overline{\bm{X}}}{\bm{X}^{2}\,\overline{\bm{X}}^{2}}\,\rme^{-i\bm{p}\cdot(\bm{x}-\overline{\bm{x}})-i\bm{k}\cdot(\bm{z}-\bar{\bm{z}})}\\
&\times\left[S_{qgqg}\left(\overline{\bm{x}},\,\overline{\bm{z}},\,\bm{x},\,\bm{z}\right)\,-\,S_{qqg}\left(\overline{\bm{w}},\,\bm{x},\,\bm{z}\right)\,-\,S_{qqg}\left(\overline{\bm{x}},\,\bm{w},\,\overline{\bm{z}}\right)\,+\,\mathcal{S}\left(\overline{\bm{w}},\,\bm{w}\right)\right],
\end{split}\end{equation}
with $C_{F}\equiv t^{a}t^{a}=\frac{N_{c}^{2}-1}{2N_{c}}$. Above, we have introduced the following $S$-matrices describing the forward scattering of colorless systems made with up to four partons: a quark-antiquark {\em dipole},
\begin{equation}\label{lowils2}
\mathcal{S}\left(\overline{\bm{w}},\,\bm{w}\right)\,\equiv\,\frac{1}{N_{c}}\,\left\langle \mathrm{tr}\big(V^{\dagger}(\overline{\bm{w}})\,V(\bm{w})\big)\right\rangle,
\end{equation}
a quark-antiquark-gluon triplet (this is illustrated in Fig.~\ref{loopfig}),
\begin{equation}\begin{split}\label{lowils3}
&S_{qqg}\left(\overline{\bm{w}},\,\bm{x},\,\bm{z}\right)\,\equiv\,\frac{1}{C_{F}N_{c}}\,\left\langle \mathrm{tr}\left(V^{\dagger}(\overline{\bm{w}})\,t^{b}\,V(\bm{x})\,t^{a}\right)\,U^{ba}(\bm{z})\right\rangle \\
&=\,\frac{1}{2C_{F}N_{c}}\,\left(N_{c}^{2}\,\mathcal{S}(\overline{\bm{w}},\,\bm{z})\,\mathcal{S}(\bm{z},\,\bm{x})\,-\,\mathcal{S}(\overline{\bm{w}},\,\bm{x})\right)\,\simeq\,\mathcal{S}(\overline{\bm{w}},\,\bm{z})\,\mathcal{S}(\bm{z},\,\bm{x})\,,
\end{split}\end{equation}
and finally a quark-antiquark pair accompanied by two gluons:
\begin{equation}\begin{split}\label{lowils1}
&S_{qgqg}\left(\overline{\bm{x}},\,\overline{\bm{z}},\,\bm{x},\,\bm{z}\right)\,\equiv\,\frac{1}{C_{F}N_{c}}\,\left\langle \mathrm{tr}\left(V^{\dagger}(\overline{\bm{x}})\,V(\bm{x})\,t^{a}\,t^{c}\right)\,\left[U^{\dagger}(\overline{\bm{z}})\,U(\bm{z})\right]^{ca}\right\rangle \\
&=\,\frac{1}{2C_{F}N_{c}}\,\left(N_{c}^{2}\,\mathcal{Q}(\overline{\bm{x}},\,\bm{x},\,\bm{z},\,\overline{\bm{z}})\,\mathcal{S}(\overline{\bm{z}},\,\bm{z})\,-\,\mathcal{S}(\overline{\bm{x}},\,\bm{x})\right)\,\simeq\,\mathcal{Q}(\overline{\bm{x}},\,\bm{x},\,\bm{z},\,\overline{\bm{z}})\,\mathcal{S}(\overline{\bm{z}},\,\bm{z})\,.
\end{split}\end{equation}
These expressions also involve the Wilson line in the adjoint representation of the color group, related to that in the fundamental representation  via
\begin{equation}\label{Uadj}
U^{ab}(\bm{z})\,=\,2\,\mathrm{tr}\left(t^{a}\,V(\bm{z})\,t^{b}\,V^{\dagger}(\bm{z})\right).
\end{equation}
The second equalities in the r.h.s. of Eqs.~(\ref{lowils1}) and (\ref{lowils3}) are obtained after using \eqn{Uadj} together with Fiertz identities. They show that all the $S$-matrices of interest for the process at hand can be computed in terms of two basic gauge-invariant correlators, $\mathcal{S}$ and $\mathcal{Q}$, which refer to (colorless) projectiles made with quarks and antiquarks alone: the dipole already introduced in \eqn{lowils2} and the {\em quadrupole},
\begin{equation}
\mathcal{Q}\,(\overline{\bm{x}},\,\bm{x},\,\bm{z},\,\overline{\bm{z}})\,\equiv\,\frac{1}{N_{c}}\,\left\langle
\mathrm{tr}\big(V^{\dagger}(\overline{\bm{x}})\,V(\bm{x})\,V^{\dagger}(\bm{z})\,V(\overline{\bm{z}})\big)\right\rangle.
\end{equation}
Techniques for computing these basic correlators under various approximations --- in particular, the corresponding B-JIMWLK equations for their high-energy evolution --- are discussed at length in the literature and will be briefly referred to in the concluding section. The final, approximate equalities in Eqs.~(\ref{lowils1}) and (\ref{lowils3}) hold in the multi-color limit $N_c\to\infty$, which allows for important simplifications (see the discussion in Sect.~\ref{Conc}).

\subsubsection{The $gA\rightarrow q\bar{q}+X$ channel}
\begin{figure}[!t]
\center \includegraphics[scale=0.91]{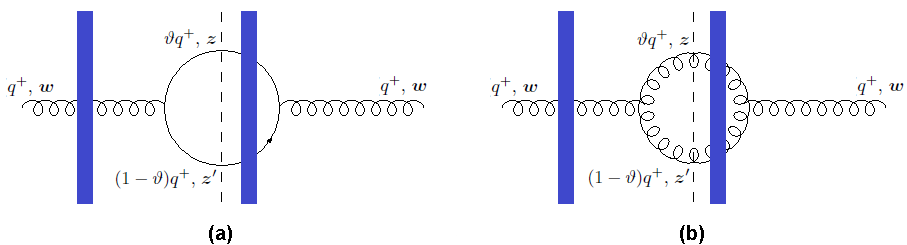}
\caption{One of the four contributions to the cross section for the case of $(a)$ $gA\rightarrow q\bar{q}+X$, $(b)$ $gA\rightarrow gg+X$.}
\label{fig:g2partons}
\end{figure}

The relation between the outgoing state (\ref{qqasistat}) and the $q\bar q$ inclusive cross section is given by the following formula (we take $\bq={\bm 0}$)
\begin{equation}\begin{split}\label{locrosdefg1}
\hspace*{-0.5cm}
\frac{d\sigma_{\lo}^{gA\rightarrow q\bar{q}+X}}{d^{3}k\,d^{3}p}\,(2\pi)\delta(k^++p^+-q^+)
=&\,\frac{1}{2(N_{c}^{2}-1)}
{}^{\  out}\!\left\langle g_{i}^{a}(q^{+},\,\bm{q}={\bm 0})\right|\,\hat{\mathcal{N}}_{q}(p)\,\hat{\mathcal{N}}_{\bar{q}}(k)\,\left|g_{i}^{a}(q^{+},\,\bm{q}={\bm 0})\right\rangle {}^{out}\\
=&\,\frac{1}{2(N_{c}^{2}-1)}\int_{\bm{w},\,\overline{\bm{w}}}\!\!{}_{\ \ \ q\bar q}^{\ \ out}\!\left\langle g_{i}^{a}(q^{+},\,\overline{\bm{w}})\right|\,\hat{\mathcal{N}}_{q}(p)\,\hat{\mathcal{N}}_{\bar{q}}(k)\,\left|g_{i}^{a}(q^{+},\,\bm{w})\right\rangle _{q\bar{q}}^{out}.
\end{split}\end{equation}
The factor ${1}/{2(N_{c}^{2}-1)}$ accounts for the average over the colors and polarizations of the initial gluon. After using (\ref{qqasistat}) together with the matrix element (\ref{matqq}), one finds
\begin{equation}\label{gqbarq}
\begin{split}
\frac{d\sigma_{\lo}^{gA\rightarrow q\bar{q}+X}}{dk^{+}\,d^{2}\bm{k}\,dp^{+}\,d^{2}\bm{p}}\,&=\,\frac{\alpha_{s}\left(1+(1-2\vartheta)^{2}\right)}{2(2\pi)^{6}q^{+}}\,\delta(q^{+}-k^{+}-p^{+})
\\&\times\,\int_{\overline{\bm{z}},\,\overline{\bm{z}}^{\prime},\,\bm{z},\,\bm{z}^{\prime}}\,\frac{\bm{Z}\cdot\overline{\bm{Z}}}{\bm{Z}^{2}\,\overline{\bm{Z}}^{2}}\,\rme^{-i\bm{p}\cdot(\bm{z}^{\prime}-\overline{\bm{z}}^{\prime})-i\bm{k}\cdot(\bm{z}-\overline{\bm{z}})}\\
&\times\left[S_{q\overline{q}q\overline{q}}\left(\overline{\bm{z}},\,\bm{\overline{z}}^{\prime},\,\bm{z},\,\bm{z}^{\prime}\right)\,-\,S_{q\overline{q}g}\left(\bm{z},\,\bm{z}^{\prime},\,\overline{\bm{w}}\right)-S_{q\overline{q}g}\left(\bm{\overline{z}}^{\prime},\,\overline{\bm{z}},\,\bm{w}\right)\,+\,\mathcal{\widetilde{S}}\left(\overline{\bm{w}},\,\bm{w}\right)\right],
\end{split}\end{equation}
where $\bm{Z}= \bm{z}^{\prime}-\bm{z}$, $\overline{\bm{Z}}= \overline{\bm{z}}^{\prime}-\overline{\bm{z}}$, and
we have introduced the new $S$-matrices
\begin{equation}\begin{split}
&S_{q\bar{q}q\bar{q}}\left(\overline{\bm{z}},\,\overline{\bm{z}}^{\prime},\,\bm{z},\,\bm{z}^{\prime}\right)\equiv\,\frac{1}{C_{F}\,N_{c}}\,\left\langle\mathrm{tr}\left(V(\bm{\bar{z}})\,t^{a}\,V^{\dagger}(\bm{\bar{z}}^{\prime})\,V(\bm{z}^{\prime})\,t^{a}\,V^{\dagger}(\bm{z})\right)\right\rangle\\
&=\,\frac{1}{2C_{F}N_{c}}\,\left(N_{c}^{2}\,\mathcal{S}(\bm{z},\,\overline{\bm{z}})\,\mathcal{S}(\overline{\bm{z}}^{\prime},\,\bm{z}^{\prime})\,-\,\mathcal{Q}(\bm{z},\,\overline{\bm{z}},\,\overline{\bm{z}}^{\prime},\,\bm{z}^{\prime})\right)\,\simeq\,\mathcal{S}(\bm{z},\,\overline{\bm{z}})\,\mathcal{S}(\overline{\bm{z}}^{\prime},\,\bm{z}^{\prime}),
\end{split}\end{equation}
for a colorless system made with 2 quarks and 2 antiquarks and respectively
\begin{equation}\begin{split}
\mathcal{\widetilde{S}}\left(\overline{\bm{w}},\,\bm{w}\right)\,\equiv\,&\frac{1}{2C_{F}N_{c}}\,
\left\langle\mathrm{Tr}\left(U^{\dagger}(\overline{\bm{w}})\,U(\bm{w})\right)\right\rangle \\ \,=\,&\frac{1}{2C_{F}N_c}\left(N_{c}^{2}\,\mathcal{S}\left(\overline{\bm{w}},\,\bm{w}\right)\,\mathcal{S}\left(\bm{w},\,\overline{\bm{w}}\right)-1\right)\,\simeq\,\mathcal{S}\left(\overline{\bm{w}},\,\bm{w}\right)\,\mathcal{S}\left(\bm{w},\,\overline{\bm{w}}\right)
\end{split}\end{equation}
for a gluon-gluon dipole. (We recall that $S_{q\overline{q}g}$ was already defined in Eq.~(\ref{lowils3}).) As before, the final, approximate, equalities hold in the large $N_c$ limit. One of the processes contributing to the cross-section \eqref{gqbarq} --- actually, the third piece inside the square brackets, which is an interference term --- is pictorially represented in Fig.~\ref{fig:g2partons}.a.

\subsubsection{The $gA\rightarrow gg+X$ channel}

The cross-section for inclusive two gluons production is similarly computed as 
\begin{equation}\begin{split}\label{defcrosgg2}
\frac{d\sigma_{\lo}^{gA\rightarrow gg+X}}{d^{3}k\,d^{3}p}
\,(2\pi)\delta(k^++p^+-q^+)
=\frac{1}{2(N_{c}^{2}-1)}\,\int_{\bm{w},\,\overline{\bm{w}}}\!\!{}_{\ \ \ gg}^{\ \ out}\!\left\langle g_{i}^{a}(q^{+},\,\overline{\bm{w}})\right|\,\hat{\mathcal{N}}_{g}(p)\,\hat{\mathcal{N}}_{g}(k)\,\left|g_{i}^{a}(q^{+},\,\bm{w})\right\rangle _{gg}^{out}.
\end{split}\end{equation}
After using Eq.~(\ref{ggasistat}) together with the matrix element (\ref{matgg}), one finds
\begin{equation}\label{crossggg}
\begin{split}
\frac{d\sigma_{\lo}^{gA\rightarrow gg+X}}{dk^{+}\,d^{2}\bm{k}\,dp^{+}\,d^{2}\bm{p}}\,&=\,\frac{4\alpha_{s}}{(2\pi)^{6}q^{+}}\left(\vartheta(1-\vartheta)\,+\,\frac{\vartheta}{(1-\vartheta)}\,+\,\frac{1-\vartheta}{\vartheta}\right)\,\delta(q^{+}-k^{+}-p^{+})
\\&\times\,\int_{\overline{\bm{z}},\,\overline{\bm{z}}^{\prime},\,\bm{z},\,\bm{z}^{\prime}}\,\frac{\bm{Z}\cdot\overline{\bm{Z}}}{\bm{Z}^{2}\,\overline{\bm{Z}}^{2}}\,\left(e^{-i\bm{p}\cdot(\bm{z}^{\prime}-\overline{\bm{z}}^{\prime})-i\bm{k}\cdot(\bm{z}-\overline{\bm{z}})}+e^{-i\bm{k}\cdot(\bm{z}^{\prime}-\overline{\bm{z}}^{\prime})-i\bm{p}\cdot(\bm{z}-\overline{\bm{z}})}\right)\\
&\times\left[S_{gggg}\left(\overline{\bm{z}},\,\bm{\overline{z}}^{\prime},\,\bm{z},\,\bm{z}^{\prime}\right)\,-\,S_{ggg}\left(\overline{\bm{w}},\,\bm{z},\,\bm{z}^{\prime}\right)\,-\,S_{ggg}\left(\bm{w},\,\bm{\overline{z}}^{\prime},\,\overline{\bm{z}}\right)\,+\,\mathcal{\widetilde{S}}\left(\overline{\bm{w}},\,\bm{w}\right)\right],\\
\end{split}\end{equation}
where the new $S$-matrices (exclusively built with adjoint Wilson lines) are
\begin{equation}\begin{split}
&S_{gggg}\left(\overline{\bm{z}},\,\bm{\overline{z}}^{\prime},\,\bm{z},\,\bm{z}^{\prime}\right)\equiv\,\frac{1}{2C_{F}\,N_{c}^{2}}\,\left\langle
f^{amn}\,f^{abc}\,\left[U^{\dagger}(\overline{\bm{z}}^{\prime})\,U(\bm{z}^{\prime})\right]^{nc}\,\left[U^{\dagger}(\overline{\bm{z}})\,U(\bm{z})\right]^{mb}\right\rangle\\
&\simeq\,\frac{1}{2}\left(\mathcal{Q}\left(\overline{\bm{z}}^{\prime},\,\bm{z}^{\prime},\,\bm{z},\,\overline{\bm{z}}\right)\,\mathcal{S}\left(\bm{z}^{\prime},\,\overline{\bm{z}}^{\prime}\right)\,\mathcal{S}\left(\overline{\bm{z}},\,\bm{z}\right)+\mathcal{Q}\left(\bm{z}^{\prime},\,\overline{\bm{z}}^{\prime},\,\overline{\bm{z}},\,\bm{z}\right)\,\mathcal{S}\left(\bm{z},\,\overline{\bm{z}}\right)\,\mathcal{S}\left(\overline{\bm{z}}^{\prime},\,\bm{z}^{\prime}\right)\right),
\end{split}\end{equation}
for a 4-gluon system and respectively (see Fig.~\ref{fig:g2partons}.b)
\begin{equation}\begin{split}
S_{ggg}\left(\overline{\bm{w}},\,\bm{z},\,\bm{z}^{\prime}\right) &\,\equiv\,\frac{1}{2C_{F}\,N_{c}^{2}}\,
\left\langle f^{gce}\,f^{abd}\,U^{ga}(\overline{\bm{w}})\,U^{cb}(\bm{z})\,U^{ed}(\bm{z}^{\prime})\right\rangle\\
&\simeq\,\frac{1}{2}\left(\mathcal{S}\left(\overline{\bm{w}},\,\bm{z}^{\prime}\right)\,\mathcal{S}\left(\bm{z}^{\prime},\,\bm{z}\right)\,\mathcal{S}\left(\bm{z},\,\overline{\bm{w}}\right)+\mathcal{S}\left(\overline{\bm{w}},\,\bm{z}\right)\,\mathcal{S}\left(\bm{z}^{\prime},\,\overline{\bm{w}}\right)\,\mathcal{S}\left(\bm{z},\,\bm{z}^{\prime}\right)\right)
\end{split}\end{equation}
for a colorless system of 3 gluons. Once again, one has relied on \eqn{Uadj} together with appropriate Fiertz identities (summarized in Appendix~\ref{wilsdict}) in order to reexpress all the $S$-matrices in terms of gauge-invariant correlators involving fundamental Wilson lines alone. To simplify writing, we have shown the corresponding results only at large $N_c$, where they reduce to products of dipoles and quadrupoles. This property --- that, in the multi-color limit, the cross-section for inclusive multi-particle production can be fully expressed in terms of dipoles and quadrupoles --- is expected to be a general result, i.e. to hold for an arbitrary final state (at least, for cross-sections computed to LO in pQCD) \cite{Kovner:2006wr,Kovner:2006ge,Dominguez:2011wm,Dominguez:2012ad}. Our final formulae for the cross-sections in Eqs.~\eqref{gqbarq} and \eqref{crossggg} agree with the corresponding results in the literature \cite{Dominguez:2011wm,Iancu:2013dta}.

\section{Three parton components of the quark outgoing state}
\label{sec:NLOprod}

After the warm-up calculation of the dijet production in Sect.~\ref{locros}, we now turn to our main objective in this paper: the calculation of the cross-section for producing three forward ``jets'' (actually, partons) in the quark-nucleus scattering. As in the case of dijets, we shall divide our calculation into two parts, for presentation purposes: first, we 
shall construct the three parton Fock space components of the quark outgoing state (in this section); then, in the next section, we shall compute the actual cross-sections.

\subsection{The general structure}

Our starting point is, once again, the exact but formal expression, Eq.~(\ref{qout}), for the outgoing state produced by an incoming quark. To generate a final state involving three partons, meaning two additional bare partons (two gluons or a quark-antiquark pair) besides the original quark, one needs to expand the evolution operators Eq.~(\ref{qout}) up to order $g^{2}$. Given the familiar structure of the QCD Lagragian, the following three types of contributions are easy to anticipate: \texttt{(i)} two initial state emissions, as generated by the second order terms in the expansion 
of $U_I(0,-\infty)$ in the first line of \eqref{UinWV};  \texttt{(ii)} two final state emissions, from the second order terms in the expansion of $U_I(\infty,0)$ (cf. the second line in  \eqref{UinWV});  \texttt{(iii)} one emission in the initial state and another one in the final state, corresponding to the product of the first-order terms in both evolution operators. Additionally, due to the peculiar structure of the LC-gauge QCD Hamiltonian which also involves $1\to 3$ splitting vertices (corresponding to instantaneous interactions), there is a fourth type of contributions, namely \texttt{(iv)} one instantaneous splitting in either the initial state, or the final state, as generated by the respective pieces in the first order terms in the expansion of either $U_I(0,-\infty)$, or $U_I(\infty,0)$.

As in Sect.~\ref{looutgoing}, we shall only display those Fock space components which matter for the cross-sections to be computed in the next section --- that is, those involving three bare partons. These are naturally denoted as $\left|q_{\lambda}^{\alpha}\right\rangle^{out}_{qq\overline{q}}$ for the three-fermion state and, respectively, $\left|q_{\lambda}^{\alpha}\right\rangle^{out} _{qgg}$ for the state built with the original quark plus two gluons. In turn, each of these three-parton states receives two types of contributions: {\em regular} contributions, which group together the terms of types \texttt{(i)}, \texttt{(ii)} and \texttt{(iii)} alluded to above, and {\em instantaneous} contributions, which refer to type \texttt{(iv)}.

Specifically, for the final state with three fermions (two quarks and an antiquark), one can write (using a generic representation, like in \eqn{qqgout})
\begin{equation}\begin{split}\label{nloasi}
\left|q_{\lambda}^{\alpha}\right\rangle^{out}_{qq\overline{q}}\,=\,\left|q_{\lambda}^{\alpha}\right\rangle _{qq\overline{q}}^{reg}\,+\,\left|q_{\lambda}^{\alpha}\right\rangle _{qq\overline{q}}^{inst},
\end{split}\end{equation}
where (the 3 terms with the accolades in the subsequent equation correspond to contributions of types \texttt{(i)}, \texttt{(ii)} and \texttt{(iii)}, respectively)
\begin{equation}\begin{split}\label{out.qqq1}
\left|q_{\lambda}^{\alpha}\right\rangle _{qq\overline{q}}^{reg}\,\equiv \,\frac{1}{2}\left|\overline{q}_{\lambda_{3}}^{\rho}\,q_{\lambda_{2}}^{\varrho}\,q_{\lambda_{1}}^{\sigma}\right\rangle & \,\left\{ \frac{\left\langle \overline{q}_{\lambda_{3}}^{\rho}\,q_{\lambda_{2}}^{\varrho}\,q_{\lambda_{1}}^{\sigma}\left|\hat{S}\right|\overline{q}_{\lambda_{7}}^{\delta}\,q_{\lambda_{6}}^{\epsilon}\,q_{\lambda_{5}}^{\kappa}\right\rangle \left\langle \overline{q}_{\lambda_{7}}^{\delta}\,q_{\lambda_{6}}^{\epsilon}\,q_{\lambda_{5}}^{\kappa}\left|\mathsf{H}_{g\rightarrow q\overline{q}}\right|q_{\lambda_{4}}^{\beta}\,g_{i}^{a}\right\rangle \left\langle q_{\lambda_{4}}^{\beta}\,g_{i}^{a}\left|\mathsf{H}_{q\rightarrow qg}\right|q_{\lambda}^{\alpha}\right\rangle }{(E_{qq\overline{q}}-E_{q})\,(E_{qg}-E_{q})}\right.\\
&\,+\,\frac{\left\langle \overline{q}_{\lambda_{3}}^{\rho}\,q_{\lambda_{2}}^{\varrho}\,q_{\lambda_{1}}^{\sigma}\right|\mathsf{H}_{g\rightarrow q\overline{q}}\left|q_{\lambda_{5}}^{\gamma}\,g_{i}^{a}\right\rangle \left\langle q_{\lambda_{5}}^{\gamma}\,g_{i}^{a}\right|\mathsf{H}_{q\rightarrow qg}\left|q_{\lambda_{4}}^{\beta}\right\rangle \left\langle q_{\lambda_{4}}^{\beta}\left|\hat{S}\right|q_{\lambda}^{\alpha}\right\rangle }{(E_{qq\overline{q}}-E_{qg})(E_{qq\overline{q}}-E_{q})}\\
&\left.\,-\,\frac{\left\langle \overline{q}_{\lambda_{3}}^{\rho}\,q_{\lambda_{2}}^{\varrho}\,q_{\lambda_{1}}^{\sigma}\right|\mathsf{H}_{g\rightarrow q\overline{q}}\left|q_{\lambda_{5}}^{\gamma}\,g^{j}\right\rangle \left\langle q_{\lambda_{5}}^{\gamma}\,g^{j}\right|\hat{S}\left|q_{\lambda_{4}}^{\beta}\,g_{i}^{a}\right\rangle \left\langle q_{\lambda_{4}}^{\beta}\,g_{i}^{a}\left|\mathsf{H}_{q\rightarrow qg}\right|q_{\lambda}^{\alpha}\right\rangle }{(E_{qq\overline{q}}-E_{qg})(E_{qg}-E_{q})}\right\} ,
\end{split}\end{equation}
and (the two terms below describe initial-state and respectively final-state $q\rightarrow qq\overline{q}$ splittings)
\begin{align}\label{out.qqq2}
\left|q_{\lambda}^{\alpha}\right\rangle _{qq\overline{q}}^{inst}\,\equiv\,\frac{1}{2}\left|\overline{q}_{\lambda_{3}}^{\rho}\,q_{\lambda_{2}}^{\varrho}\,q_{\lambda_{1}}^{\sigma}\right\rangle & \,\left\{ -\,\frac{\left\langle \overline{q}_{\lambda_{3}}^{\rho}\,q_{\lambda_{2}}^{\varrho}\,q_{\lambda_{1}}^{\sigma}\left|\hat{S}\right|\overline{q}_{\lambda_{6}}^{\epsilon}\,q_{\lambda_{5}}^{\delta}\,q_{\lambda_{4}}^{\beta}\right\rangle \,\left\langle \overline{q}_{\lambda_{6}}^{\epsilon}\,q_{\lambda_{5}}^{\delta}\,q_{\lambda_{4}}^{\beta}\left|\mathsf{H}_{q\rightarrow qq\overline{q}}\right|q_{\lambda}^{\alpha}\right\rangle }{E_{qq\overline{q}}-E_{q}}
\right.\nn
&\left. \ \ +\,
\frac{\left\langle \overline{q}_{\lambda_{3}}^{\rho}\,q_{\lambda_{2}}^{\varrho}\,q_{\lambda_{1}}^{\sigma}\left|\mathsf{H}_{q\rightarrow qq\overline{q}}\right|q_{\lambda_{4}}^{\beta}\right\rangle \,\left\langle q_{\lambda_{4}}^{\beta}\left|\hat{S}\right|q_{\lambda}^{\alpha}\right\rangle }{E_{qq\overline{q}}-E_{q}}
\right\},
\end{align}
 
Concerning the $qgg$ final state, it is convenient (for the purpose of the calculation) to separately consider the contributions which involve the  $g\to gg$ splitting vertex and those which do not. Such a separation is quite obvious for the regular contributions, but it also makes sense for the instantaneous ones: indeed, the instantaneous vertices $q\to qgg$ receive two types of contributions, represented by two pieces in the interaction Hamiltonian --- the ``gluon'' piece $\mathsf{H}_{q\rightarrow qgg}^{g}$ and the ``quark'' piece $\mathsf{H}_{q\rightarrow qgg}^{q}$ ---, corresponding to the situations where the instantaneous propagation refers to an intermediate gluon, or to an intermediate quark, respectively. It is therefore natural to decompose $\left|q_{\lambda}^{\alpha}\right\rangle^{out} _{qgg}$ as follows:
\beq\label{qggsplit}
\left|q_{\lambda}^{\alpha}\right\rangle^{out}_{qgg}\,=\,\sum_{i=1,2}\left(\left|q_{\lambda}^{\alpha}\right\rangle _{qgg}^{reg,\,i}\,+\,\left|q_{\lambda}^{\alpha}\right\rangle _{qgg}^{inst,\,i}\right),
\eeq
where $i=1$ corresponds to regular pieces built with the $q\to qg$ vertex alone and to the ``quark'' piece $\mathsf{H}_{q\rightarrow qgg}^{q}$ of the instantaneous $q\to qgg$ vertex,
\begin{align}
\label{out.qgg3}
\left|q_{\lambda}^{\alpha}\right\rangle _{qgg}^{reg,\,1}\,\equiv\,\frac{1}{2}\left|q_{\lambda_{3}}^{\delta}\,g_{m}^{d}\,g_{n}^{c}\right\rangle &\,\left\{ \frac{\left\langle q_{\lambda_{3}}^{\delta}\,g_{m}^{d}\,g_{n}^{c}\right|\hat{S}\left|q_{\lambda_{2}}^{\gamma}\,g_{j}^{b}\,g_{l}^{c}\right\rangle \left\langle q_{\lambda_{2}}^{\gamma}\,g_{j}^{b}\,g_{l}^{c}\right|\mathsf{H}_{q\rightarrow qg}\left|q_{\lambda_{1}}^{\beta}\,g_{i}^{a}\right\rangle \left\langle q_{\lambda_{1}}^{\beta}\,g_{i}^{a}\left|\mathsf{H}_{q\rightarrow qg}\right|q_{\lambda}^{\alpha}\right\rangle }{(E_{qgg}-E_{q})(E_{qg}-E_{q})}\right. \nn
&+\,\frac{\left\langle q_{\lambda_{3}}^{\delta}\,g_{m}^{d}\,g_{n}^{c}\right|\mathsf{H}_{q\rightarrow qg}\left|q_{\lambda_{2}}^{\gamma}\,g_{i}^{c}\right\rangle \left\langle q_{\lambda_{2}}^{\gamma}\,g_{i}^{c}\right|\mathsf{H}_{q\rightarrow qg}\left|q_{\lambda_{1}}^{\beta}\right\rangle \left\langle q_{\lambda_{1}}^{\beta}\left|\hat{S}\right|q_{\lambda}^{\alpha}\right\rangle }{(E_{qgg}-E_{qg})(E_{qgg}-E_{q})}\nn
&\left. -\,\frac{\left\langle q_{\lambda_{3}}^{\delta}\,g_{m}^{d}\,g_{n}^{c}\right|\mathsf{H}_{q\rightarrow qg}\left|q_{\lambda_{2}}^{\gamma}\,g_{j}^{b}\right\rangle \left\langle q_{\lambda_{2}}^{\gamma}\,g_{j}^{b}\right|\hat{S}\left|q_{\lambda_{1}}^{\beta}\,g_{i}^{a}\right\rangle \left\langle q_{\lambda_{1}}^{\beta}\,g_{i}^{a}\left|\mathsf{H}_{q\rightarrow qg}\right|q_{\lambda}^{\alpha}\right\rangle }{(E_{qgg}-E_{qg})(E_{qg}-E_{q})}\right\},
\end{align}
\begin{align}\label{out.qgg4}
\left|q_{\lambda}^{\alpha}\right\rangle _{qgg}^{inst,\,1}\,\equiv\,\frac{1}{2}\left|q_{\lambda_{1}}^{\gamma}\,g_{i}^{d}\,g_{j}^{e}\right\rangle & \,\left\{ 
\,-\,\frac{\left\langle q_{\lambda_{1}}^{\gamma}\,g_{i}^{d}\,g_{j}^{e}\right|\hat{S}\left|q_{\lambda_{2}}^{\beta}\,g_{m}^{b}\,g_{n}^{c}\right\rangle \left\langle q_{\lambda_{2}}^{\beta}\,g_{m}^{b}\,g_{n}^{c}\left|\mathsf{H}_{q\rightarrow qgg}^{q}\right|q_{\lambda}^{\alpha}\right\rangle }{E_{qgg}-E_{q}}\right. \nn
&\left. \ \ +\,
 \frac{\left\langle q_{\lambda_{1}}^{\gamma}\,g_{i}^{d}\,g_{j}^{e}\right|\mathsf{H}_{q\rightarrow qgg}^{q}\left|q_{\lambda_{2}}^{\beta}\right\rangle \left\langle q_{\lambda_{2}}^{\beta}\left|\hat{S}\right|q_{\lambda}^{\alpha}\right\rangle }{E_{qgg}-E_{q}}\right\},
\end{align}
whereas $i=2$ corresponds to regular pieces involving the 3-gluon vertex $g\to gg$ and to instantaneous pieces computed with the ``gluon'' part $\mathsf{H}_{q\rightarrow qgg}^{g}$ of the instantaneous interaction Hamiltonian:
\begin{align}
\label{out.qgg1}
\left|q_{\lambda}^{\alpha}\right\rangle _{qgg}^{reg,\,2}\,\equiv \,\frac{1}{2}\left|q_{\lambda_{1}}^{\gamma}\,g_{i}^{d}\,g_{j}^{e}\right\rangle &\,\left\{ \frac{\left\langle q_{\lambda_{1}}^{\gamma}\,g_{i}^{d}\,g_{j}^{e}\left|\hat{S}\right|q_{\lambda_{2}}^{\delta}\,g_{m}^{b}\,g_{n}^{c}\right\rangle \left\langle q_{\lambda_{2}}^{\delta}\,g_{m}^{b}\,g_{n}^{c}\left|\mathsf{H}_{g\rightarrow gg}\right|q_{\lambda_{3}}^{\beta}\,g_{l}^{a}\right\rangle \left\langle q_{\lambda_{3}}^{\beta}\,g_{l}^{a}\left|\mathsf{H}_{q\rightarrow qg}\right|q_{\lambda}^{\alpha}\right\rangle }{(E_{qgg}-E_{q})(E_{qg}-E_{q})}
\right.\nn
&+\,\frac{\left\langle q_{\lambda_{1}}^{\gamma}\,g_{i}^{d}\,g_{j}^{e}\left|\mathsf{H}_{g\rightarrow gg}\right|q_{\lambda_{2}}^{\delta}\,g_{l}^{a}\right\rangle \left\langle q_{\lambda_{2}}^{\delta}\,g_{l}^{a}\left|\mathsf{H}_{q\rightarrow qg}\right|q_{\lambda_{3}}^{\beta}\right\rangle \left\langle q_{\lambda_{3}}^{\beta}\left|\hat{S}\right|q_{\lambda}^{\alpha}\right\rangle }{(E_{qgg}-E_{qg})(E_{qgg}-E_{q})}\nn
&\left.-\,\frac{\left\langle q_{\lambda_{1}}^{\gamma}\,g_{i}^{d}\,g_{j}^{e}\left|\mathsf{H}_{g\rightarrow gg}\right|q_{\lambda_{2}}^{\delta}\,g_{l}^{b}\right\rangle \left\langle q_{\lambda_{2}}^{\delta}\,g_{l}^{b}\left|\hat{S}\right|q_{\lambda_{3}}^{\beta}\,g_{m}^{a}\right\rangle \left\langle q_{\lambda_{3}}^{\beta}\,g_{m}^{a}\left|\mathsf{H}_{q\rightarrow qg}\right|q_{\lambda}^{\alpha}\right\rangle }{(E_{qgg}-E_{qg})(E_{qg}-E_{q})}\right\}, 
\end{align}
\begin{align}\label{out.qgg2}
\left|q_{\lambda}^{\alpha}\right\rangle _{qgg}^{inst,\,2}\,\equiv\,\frac{1}{2}\left|q_{\lambda_{1}}^{\delta}\,g_{i}^{d}\,g_{j}^{c}\right\rangle & \,\left\{- \frac{\left\langle q_{\lambda_{1}}^{\delta}\,g_{i}^{d}\,g_{j}^{c}\left|\hat{S}\right|q^{\beta}\,g_{m}^{b}\,g_{n}^{a}\right\rangle \left\langle q^{\beta}\,g_{m}^{b}\,g_{n}^{a}\left|\mathsf{H}_{q\rightarrow qgg}^{g}\right|q_{\lambda}^{\alpha}\right\rangle }{E_{qgg}-E_{q}}\right.  \nn 
&\left. \ +\,\frac{\left\langle q_{\lambda_{1}}^{\delta}\,g_{i}^{d}\,g_{j}^{c}\right|\mathsf{H}_{q\rightarrow qgg}^{g}\left|q_{\lambda_{2}}^{\beta}\right\rangle \left\langle q_{\lambda_{2}}^{\beta}\left|\hat{S}\right|q_{\lambda}^{\alpha}\right\rangle }{E_{qgg}-E_{q}}\right\}.
\end{align}
In the remaining part of this section we shall present explicit results for these three-parton states. The details of the calculations are deferred to Appendix~\ref{details}, on the example of the initial-state evolution --- that is, we explicitly construct there the three-parton components of the quark LCWF at the time of scattering. The corresponding calculations for the final-state evolution are entirely similar, as they differ only in their energy denominators (compare the first to the second line in \eqn{UinWV}).

\subsection{The $qq\overline{q}$ component}
\begin{figure}[!t]
\includegraphics[scale=0.65]{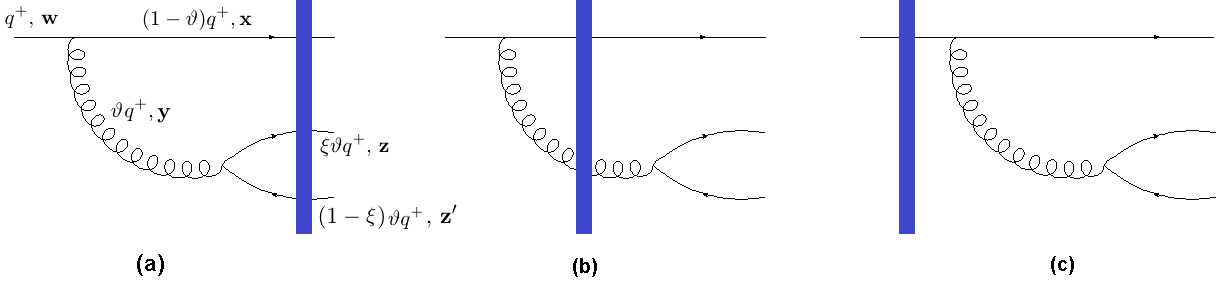}
\caption{The three possible configurations for the interplay between the quark evolution and the scattering, for a final partonic state built with three quarks and a propagating intermediate gluon: (a) initial-state evolution (the first term within the square brackets in Eq.~(\ref{fin.qqq1})), (b) mixed evolution: the gluon is emitted prior to the scattering, but it splits after the scattering  (the third term in Eq.~(\ref{fin.qqq1})), (c) final-state evolution (the second term in Eq.~(\ref{fin.qqq1})). 
}
\label{quarinsert}
\end{figure}

The ``regular'' contribution to this state is defined in Eq.~(\ref{out.qqq1}) and explicitly constructed in subsection \ref{qqqsubs} of Appendix~\ref{details}. The final result in transverse-coordinate space reads (cf. \eqn{qqqreg})
\begin{eqnarray} \label{fin.qqq1}
&&\left|q_{\lambda}^{\alpha}(q^{+},\,\bm{w})\right\rangle _{qq\bar{q}}^{reg}\,=\,-\,\int_{\bm{x},\,\bm{z},\,\bm{z}^{\prime}}\,\int_{0}^{1}d\vartheta\,d\xi\:\frac{g^{2}\,\varphi_{\lambda_{2}\lambda_{3}}^{il}(\xi)\,\phi_{\lambda_{1}\lambda}^{ij}(\vartheta)\,\bm{Z}^{l}\,\left(\bm{X}^{j}+\xi\bm{Z}^{j}\right)\,q^{+}}{2(2\pi)^{4}\left(\bm{X}+\xi\bm{Z}\right)^{2}\,\bm{Z}^{2}}\nonumber\\
&&\times\left[\Theta_{1}(\bm{x},\,\bm{z},\,\bm{z}^{\prime})\,V^{\varrho\delta}(\bm{z}^{\prime})\,t_{\delta\epsilon}^{a}\,V^{\dagger\epsilon\rho}(\bm{z})\,V^{\sigma\beta}(\bm{x})\,t_{\beta\alpha}^{a}\,+\,\Theta_{2}(\bm{x},\,\bm{z},\,\bm{z}^{\prime})\,t_{\varrho\rho}^{a}\,t_{\sigma\beta}^{a}\,V^{\beta\alpha}(\bm{w})\right.\\
&&\left.-\,t_{\varrho\rho}^{b}\,V^{\sigma\beta}(\bm{x})\,U^{ba}(\bm{y})\,t_{\beta\alpha}^{a}\right]\delta^{(2)}\left(\bm{w}-\bm{C}\right)\left|\bar{q}_{\lambda_{3}}^{\rho}((1-\xi)\vartheta q^{+},\,\bm{z})\,q_{\lambda_{2}}^{\varrho}(\xi\vartheta q^{+},\,\bm{z}^{\prime})\,q_{\lambda_{1}}^{\sigma}((1-\vartheta)q^{+},\bm{x})\right\rangle ,\nonumber
 \end{eqnarray}
 with the following notations:
\begin{equation}
\Theta_{1}(\bm{x},\,\bm{z},\,\bm{z}^{\prime})\,\equiv\,\frac{(1-\vartheta)\left(\bm{X}+\xi\bm{Z}\right)^{2}}{(1-\vartheta)\left(\bm{X}+\xi\bm{Z}\right)^{2}\,+\,\xi(1-\xi)\bm{Z}^{2}}\,,
\end{equation}
\begin{equation}
\Theta_{2}(\bm{x},\,\bm{z},\,\bm{z}^{\prime})\,=\,1-\Theta_{1}(\bm{x},\,\bm{z},\,\bm{z}^{\prime})\,\equiv\,\frac{\xi(1-\xi)\bm{Z}^{2}}{(1-\vartheta)\left(\bm{X}+\xi\bm{Z}\right)^{2}\,+\,\xi(1-\xi)\bm{Z}^{2}}\,,
\end{equation}
where  $\bx$ and $\bz'$ denote the transverse coordinates of two final quarks, while $\bz$ is the transverse coordinate of the antiquark. In terms of these, the transverse position $\by$ of the intermediate gluon and the corresponding position $\bw$ of the incoming quark are given by (see also see Fig.~\ref{quarinsert})
\begin{equation}\label{defyc}
\bm{y}\,\equiv\,\xi\bm{z}^{\prime}+(1-\xi)\bm{z}\,;\qquad\qquad\bm{w}\,=\,\bm{C}\,\equiv\,(1-\vartheta)\bm{x}\,+\,\xi\vartheta\bm{z}^{\prime}\,+\,(1-\xi)\vartheta\bm{z}.
\end{equation}
We have further written $\bm{Z}\equiv\bm{z}-\bm{z}^{\prime}$ and $\bm{X}\equiv\bm{x}-\bm{z}$.

The three terms inside the square brackets of Eq.~(\ref{fin.qqq1}), which carry the information about the scattering, correspond to the three terms in the r.h.s. of Eq.~(\ref{out.qqq1}). They exhibit different color structures corresponding to the three possibilities for the insertion of the shockwave, as depicted on Fig.~\ref{quarinsert}.  In particular, the first term inside the square brackets in (\ref{fin.qqq1}), which involves three fundamental Wilson lines and corresponds to Fig.~\ref{quarinsert}.a, has been obtained simply by acting with the $S$-matrix on the $qq\overline{q}$ Fock component shown in  (\ref{qqqreg}). 

The other factors within the integrand of Eq.~(\ref{fin.qqq1}) can be understood as follows: the overall factor outside the brackets was obtained as the product of the two emission vertices times the energy denominators corresponding to one emission prior to the scattering (initial-state) and a second one after the scattering (final-state); that is, these are the  energy denominators associated with the third term in Eq.~(\ref{out.qqq1}). This explains why the corresponding (third) term within the brackets of Eq.~(\ref{fin.qqq1}) has a unit weight. The factors $\Theta_{1}$ and $\Theta_{2}$  multiplying the other color structures restore the proper energy denominators for having both emissions in the initial state and, respectively, in the final state. The fact that $\Theta_{1}$ and $\Theta_{2}$ sum up to unity is in agreement with the fact that the three terms inside the braces in \eqn{out2S1} mutually cancel. In other terms, the r.h.s. of Eq.~(\ref{fin.qqq1}) would vanish in the absence of scattering, in agreement with the general arguments in 
Sect.~\ref{sec:prod}. A similar discussion applies to all the other ``regular'' Fock components to be exhibited in what follows. 

\begin{figure}[!t]
\center \includegraphics[scale=0.68]{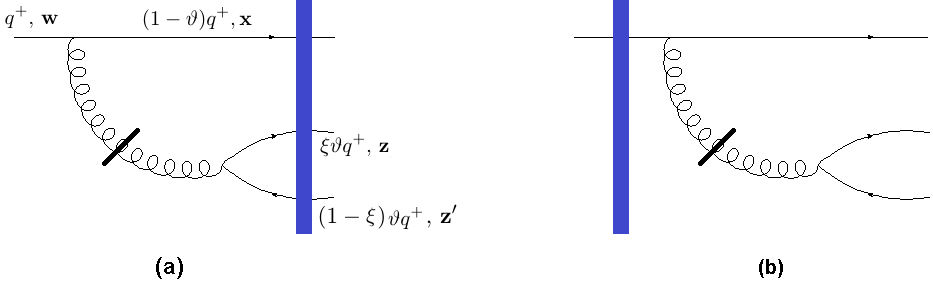}
\caption{The two possible insertions of the shockwave in the case where the final quark antiquark pair has been produced via an instantaneous gluon interaction: (a) initial-state evolution, (b)  final-state evolution.}
\label{quarkinst}
\end{figure}

The instantaneous gluon contribution to the $qq\overline{q}$ final state is defined in Eq.~(\ref{out.qqq2}) and illustrated in Fig.~\ref{quarkinst}, in which the ``cut'' gluon line represents an instantaneous gluon exchange, which cannot scatter off the nucleus. Accordingly, there is no diagram in which the  ``cut'' gluon crosses the shockwave: the two diagrams in Fig.~\ref{quarkinst}.a and b correspond to the two terms in the r.h.s. of Eq.~(\ref{out.qqq2}): the first term, involving three fundamental Wilson lines, refers to initial-state evolution, where the second term with a single Wilson line, to the final-state evolution. Both contributions are easily evaluated starting with the expression \eqref{qq_cont2} for the appropriate Fock space component. One finds (cf. \eqn{qqqinst})
\begin{align}\label{fin.qqq2}
\left|q_{\lambda}^{\alpha}(q^{+},\,\bm{w})\right\rangle _{qq\bar{q}}^{inst}&\,=-\,\int_{\bm{x},\,\bm{z},\,\bm{z}^{\prime}}\,\int_{0}^{1}d\vartheta\,d\xi\:\frac{g^{2}\,(1-\vartheta)\xi(1-\xi)q^{+}}{(2\pi)^{4}\left((1-\vartheta)\left(\bm{X}+\xi\bm{Z}\right)^{2}\,+\,\xi(1-\xi)\bm{Z}^{2}\right)}\nn
&\times\,\left[V^{\varrho\delta}(\bm{z}^{\prime})\,t_{\delta\epsilon}^{a}\,V^{\dagger\epsilon\rho}(\bm{z})\,V^{\sigma\beta}(\bm{x})\,t_{\beta\alpha}^{a}-\,t_{\varrho\rho}^{a}\,t_{\sigma\beta}^{a}\,V^{\beta\alpha}(\bm{w})\right]\nn
&\times\,\delta^{(2)}\left(\bm{w}-\bm{C}\right)\left|\bar{q}_{\lambda_{1}}^{\rho}((1-\xi)\vartheta q^{+},\,\bm{z})\,q_{\lambda_{1}}^{\varrho}(\xi\vartheta q^{+},\,\bm{z}^{\prime})\,q_{\lambda}^{\sigma}((1-\vartheta)q^{+},\bm{x})\right\rangle ,
 \end{align}
 with the same notations for the various transverse coordinates as in Eq.~(\ref{fin.qqq1}) and Eq.~(\ref{defyc}).
 
\subsection{The $qgg$ component}

We now turn to the four Fock space components which involve a gluon pair in the final state, together with the incoming quark. The respective calculations are detailed in  Appendix~\ref{qggsubs}.

 \begin{figure}[hb]
    \includegraphics[scale=0.59]{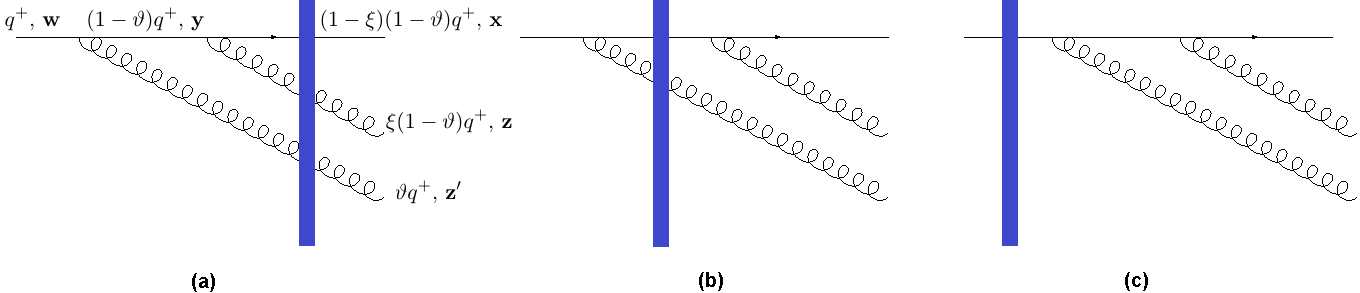}
  \caption{Two gluon emissions by the original quark. There are three possibilities for the insertion of the shockwave: (a) initial-state evolution, (b) mixed (the first gluon emission occurs prior to scattering and the second after the scattering), (c) final-state evolution. 
\label{fig:qggreg1}}
\end{figure}

Consider first the ``type $i=1$'' topology, which does not involve the triple gluon vertex at all (recall \eqn{qggsplit}).
The ``regular'' contribution to this state is defined in (\ref{out.qgg3}) and illustrated in Fig~\ref{fig:qggreg1}: both gluons are emitted by the incoming quark, either before, or after, its scattering off the shockwave.  Thes processes yield (cf. \eqn{qggreg1})
  \begin{eqnarray}\label{fin.qgg1}
  \left|q_{\lambda}^{\alpha}(q^{+},\,\bm{w})\right\rangle _{qgg}^{reg,\,1}&&\,=\,\int_{\bm{x},\,\bm{z},\,\bm{z}^{\prime}}\int_{0}^{1}d\xi\,d\vartheta\,\frac{g^{2}\phi_{\lambda_{2}\lambda_{1}}^{jm}(\xi)\,\phi_{\lambda_{1}\lambda}^{il}(\vartheta)\,\left(\xi\bm{X}-\bm{X^{\prime}}\right)^{l}\,\bm{X}^{m}\,\delta^{(2)}\left(\bm{w}-\bm{D}\right)\,\sqrt{(1-\vartheta)}q^{+}}{2(2\pi)^{4}\sqrt{\xi}\vartheta\,\bm{X}^{2}\,\left(\xi\bm{X}-\bm{X^{\prime}}\right)^{2}}\nonumber\\
&&\times\left[\Theta_{3}(\bm{x},\,\bm{z},\,\bm{z}^{\prime})\,V^{\delta\gamma}(\bm{x})\,t_{\gamma\beta}^{b}\,U^{db}(\bm{z})\,U^{ca}(\bm{z}^{\prime})\,t_{\beta\alpha}^{a}\,+\,\Theta_{4}(\bm{x},\,\bm{z},\,\bm{z}^{\prime})\,t_{\delta\gamma}^{d}\,t_{\gamma\beta}^{c}\,V^{\beta\alpha}(\bm{w})\right. \nn
&&\left.-\,t_{\delta\gamma}^{d}\,V^{\gamma\beta}(\bm{y})\,t_{\beta\alpha}^{a}\,U^{ca}(\bm{z}^{\prime})\right]\left|q_{\lambda_{2}}^{\delta}((1-\xi)(1-\vartheta)q^{+},\,\bm{x})\,g_{j}^{d}(\xi(1-\vartheta)q^{+},\,\bm{z})\,g_{i}^{c}(\vartheta q^{+},\,\bm{z}^{\prime})\right\rangle,\nn
  \end{eqnarray}
with $\bm{X}\equiv\bm{x}-\bm{z}$, $\bm{X}^{\prime}\,\equiv\,\bm{x}\,-\,\bm{z}^{\prime}$ and
\begin{equation}
\Theta_{3}(\bm{x},\,\bm{z},\,\bm{z}^{\prime})\,\equiv\,\frac{\vartheta\left(\xi\bm{X}-\bm{X^{\prime}}\right)^{2}}{\vartheta\left(\xi\bm{X}-\bm{X^{\prime}}\right)^{2}+\xi(1-\xi)\bm{\bm{X}}^{2}}\,,
\end{equation}
\begin{equation}
\Theta_{4}(\bm{x},\,\bm{z},\,\bm{z}^{\prime})\,\equiv\,1-\Theta_{3}(\bm{x},\,\bm{z},\,\bm{z}^{\prime})=\,\frac{\xi(1-\xi)\bm{\bm{X}}^{2}}{\vartheta\left(\xi\bm{X}-\bm{X^{\prime}}\right)^{2}+\xi(1-\xi)\bm{\bm{X}}^{2}}\,.
\end{equation}
In these expressions, $\bx$ represents the transverse coordinate of the final quark, while $\bz'$ and $\bz$ denote the transverse coordinates of the first and second emitted gluon. The transverse position of the intermediate gluon $\by$ and the corresponding position $\bw$ of the incoming quarks are given by
 \begin{equation}\label{defyd}
\bm{y}\,=\,(1-\xi)\bm{x}\,+\,\xi\bm{z}\,;\qquad\qquad\bm{w}=\,\bm{D}\,\equiv\,(1-\xi)(1-\vartheta)\bm{x}\,+\,\xi(1-\vartheta)\bm{z}\,+\,\vartheta\bm{z}^{\prime}.
\end{equation}

 \begin{figure}[!h]\center
    \includegraphics[scale=0.575]{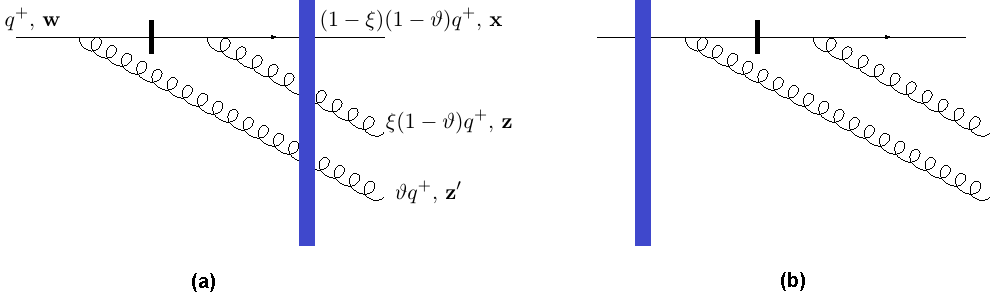}
  \caption{The two possible insertions of the shockwave in the case where the intermediate quark exchange is instantaneous: (a) initial-state evolution, (b)  final-state evolution.
  \label{fig:qgginst1}}
\end{figure}

Consider now the $qgg$ state involving one instantaneous quark propagator, as defined in (\ref{out.qgg4}) and illustrated in Fig.~\ref{fig:qgginst1}; the calculation in Appendix~\ref{qggsubs} yields (cf. \eqn{qgginst1})
  \begin{align}\label{fin.qgg2}
\left|q_{\lambda}^{\alpha}(q^{+},\,\bm{w})\right\rangle _{qgg}^{inst,\,1}&\,=\,-\int_{\bm{x},\,\bm{z},\,\bm{z}^{\prime}}\int_{0}^{1}d\xi\,d\vartheta\,\frac{2g^{2}\sqrt{\xi\vartheta}(1-\xi)\,\chi_{\lambda_{1}}^{\dagger}\left(\delta^{ij}-i\varepsilon^{ij}\sigma^{3}\right)\chi_{\lambda}\,q^{+}}{(2\pi)^{4}\sqrt{1-\vartheta}\,\left(\vartheta\left(\xi\bm{X}-\bm{X^{\prime}}\right)^{2}+\xi(1-\xi)\bm{\bm{X}}^{2}\right)}\nn
&\times\left[V^{\delta\gamma}(\bm{x})\,t_{\gamma\beta}^{b}\,U^{db}(\bm{z})\,U^{ca}(\bm{z}^{\prime})\,t_{\beta\alpha}^{a}\,-\,t_{\delta\gamma}^{d}\,t_{\gamma\beta}^{c}\,V^{\beta\alpha}(\bm{w})\right]\nn
&\times\delta^{(2)}\left(\bm{w}-\bm{D}\right)\left|q_{\lambda_{1}}^{\delta}((1-\xi)(1-\vartheta)q^{+},\,\bm{x})\,g_{j}^{d}(\xi(1-\vartheta)q^{+},\,\bm{z})\,g_{i}^{c}(\vartheta q^{+},\,\bm{z}^{\prime})\right\rangle,
\end{align}
 with the same notations for the various transverse coordinates as in Eq.~(\ref{fin.qgg1}) and Eq.~(\ref{defyd}).

 \begin{figure}[!h]
    \includegraphics[scale=0.65]{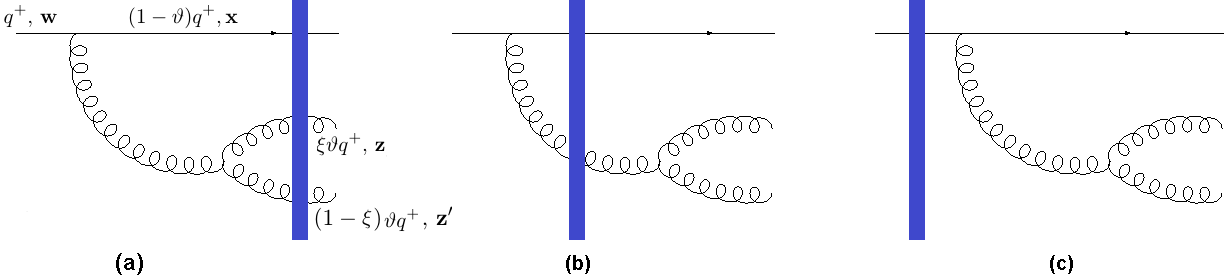}
  \caption{The three possible configurations for the interplay between the quark evolution and the scattering, for a final partonic state built with a quark and two gluons which were produced from a propagating intermediate gluon: (a) initial-state evolution, (b) mixed (the gluon emission occurs prior to scattering, but its splitting happens after the scattering), (c) final-state evolution.
  \label{fig:qggreg2}}
\end{figure}

We finally turn to the $qgg$ states which also involve the triple gluon vertex. The ``regular'' one was defined in (\ref{out.qgg1}) and is illustrated in Fig.~\ref{fig:qggreg2}. The calculation in Appendix~\ref{qggsubs} yields (cf. \eqn{qggreg2})
\begin{align}\label{fin.qgg3}
\left|q_{\lambda}^{\alpha}(q^{+},\,\bm{w})\right\rangle _{qgg}^{reg,\,2}&\,=\,-\,\int_{\bm{x},\,\bm{z},\,\bm{z}^{\prime}}\,\int_{0}^{1}d\vartheta\,d\xi\,\frac{ig^{2}\sqrt{\xi(1-\xi)}\,\phi_{\lambda_{1}\lambda}^{im}(\vartheta)\,\delta^{(2)}\left(\bm{w}-\bm{C}\right)\,q^{+}}{2(2\pi)^{4}\,\bm{Z}^{2}\,\left(\bm{X}+\xi\bm{Z}\right)^{2}}\nn
&\times\left(\bm{X}^{m}\,+\,\xi\bm{Z}^{m}\right)\,\left(\bm{Z}^{i}\delta_{jl}\,-\,\frac{1}{\xi}\bm{Z}^{j}\delta_{il}\,-\,\frac{1}{1-\xi}\bm{Z}^{l}\delta_{ij}\right)\nn
&\times\left[\Theta_{1}(\bm{x},\,\bm{z},\,\bm{z}^{\prime})\,f^{abc}\,V^{\gamma\beta}(\bm{x})\,U^{ec}(\bm{z})\,U^{db}(\bm{z}^{\prime})\,t_{\beta\alpha}^{a}\,+\,\Theta_{2}(\bm{x},\,\bm{z},\,\bm{z}^{\prime})\,f^{ade}\,t_{\gamma\beta}^{a}\,V^{\beta\alpha}(\bm{w})\right.\nn
&\left.-\,f^{bde}\,V^{\gamma\beta}(\bm{x})\,U^{ba}(\bm{y})\,t_{\beta\alpha}^{a}\right]\left|q_{\lambda_{1}}^{\gamma}((1-\vartheta)q^{+},\,\bm{x})\,g_{j}^{d}(\vartheta\xi q^{+},\,\bm{z}^{\prime})\,g_{l}^{e}(\vartheta(1-\xi)q^{+},\,\bm{z})\right\rangle .
\end{align}
where $\bx$ in the transverse coordinate of the final quark, while $\bz'$ and $\bz$ denote the transverse coordinates of the final gluons after the original gluon splitted. The transverse position of the intermediate gluon $\by$ and the corresponding position $\bw$ of the incoming quarks are given by (\ref{defyc}).  Like before, we used the notations $\bm{Z}\equiv\bm{z}-\bm{z}^{\prime}$ and $\bm{X}\equiv\bm{x}-\bm{z}$.

\begin{figure}[!h]\center
    \includegraphics[scale=0.65]{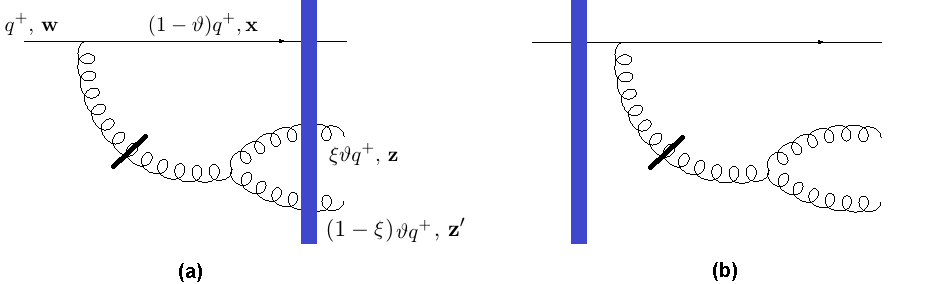}
  \caption{The two possible insertions of the shockwave in the case where a pair of gluons has been produced via an instantaneous gluon interaction: (a) initial-state evolution, (b)  final-state evolution.
  \label{fig:qgginst2}}
\end{figure}

As for the state involving an instantaneous propagator, as defined in (\ref{out.qgg2}) and illustrated in Fig.~\ref{fig:qgginst2}, this yields  (cf. \eqn{qgginst2})
 \begin{align}\label{fin.qgg4}
\left|q_{\lambda}^{\alpha}(q^{+},\,\bm{w})\right\rangle _{qgg}^{inst,\,2}&\,=-\,\int_{\bm{x},\,\bm{z},\,\bm{z}^{\prime}}\,\int_{0}^{1}d\vartheta\,d\xi\,\frac{ig^{2}(1-2\xi\vartheta)(1-\vartheta)\sqrt{\xi(1-\xi)}q^{+}}{2(2\pi)^{4}\left((1-\vartheta)\left(\bm{X}+\xi\bm{Z}\right)^{2}\,+\,\xi(1-\xi)\bm{Z}^{2}\right)}\nn
&\times\left[f^{abc}\,V^{\gamma\beta}(\bm{x})\,U^{ec}(\bm{z})\,U^{db}(\bm{z}^{\prime})\,t_{\beta\alpha}^{a}\,-\,f^{ade}\,t_{\gamma\beta}^{a}\,V^{\beta\alpha}(\bm{w})\right]\nn
&\times\delta^{(2)}\left(\bm{w}-\bm{C}\right)\left|q_{\lambda}^{\gamma}((1-\vartheta)q^{+},\,\bm{x})\,g_{i}^{d}(\vartheta\xi q^{+},\,\bm{z}^{\prime})\,g_{i}^{e}(\vartheta(1-\xi)q^{+},\,\bm{z})\right\rangle,
 \end{align}
  with the same notations for the various transverse coordinates as in Eq.~(\ref{fin.qgg3}) and Eq.~(\ref{defyc}).

\section{The trijet cross section\label{trijetfinal}}

We are now in a position to explicitly compute the physical quantity of interest for us in this work, namely the cross-section for the production of three `jets' (actually, partons) at forward rapidities in the scattering between a quark `taken' (within the collinear factorization) from a dilute projectile (a proton) and a shockwave describing a dense target (a large nucleus) within the CGC effective theory. The calculation is conceptually straightforward --- it merely amounts to taking the expectation value of the product of three parton number density operators over the quark outgoing state constructed in the previous section ---, yet this is quite tedious in practice, because of the complexity of the final state, which receives contributions from Feynman graphs with many possible topologies.

Specifically, as discussed in Sect.~\ref{sec:NLOprod}, for an incoming quark there are two possible final states containing three partons: $qq\bar q$ and $qgg$. Hence, the LO cross-section for trijet production in the quark channel can be written as 
\begin{equation}
\frac{d\sigma^{pA\to3jet+X}}{d^{3}q_{1}\,d^{3}q_{2}\,d^{3}q_{3}}\,=\,\int dx_{p}\,q(x_{p},\mu^{2})\left(\frac{d\sigma^{qA\rightarrow qgg+X}}{d^{3}q_{1}\,d^{3}q_{2}\,d^{3}q_{3}}\,+\,\frac{d\sigma^{qA\rightarrow qq\overline{q}+X}}{d^{3}q_{1}\,d^{3}q_{2}\,d^{3}q_{3}}\right),
 \end{equation}
with $q(x_p,\mu^2)$ is the quark distribution function of the proton evaluated for a longitudinal momentum fraction $x_p=q^+/Q^+$ (with $Q^+$ the proton longitudinal momentum and $q^+$ the respective momentum of the participating quark) and for a transverse resolution scale $\mu^2$ (of the order of the hardest among the transverse momenta of the produced jets).  In the kinematic of interest, the ``plus'' momentum transferred by the target is negligible, hence the conservation of longitudinal momentum requires  $q^{+}=q_{1}^{+}+q_{2}^{+}+q_{3}^{+}$, which in turn fixes the value of $x_p$.

Consider the partonic  cross-section for the ${qA\rightarrow qgg+X}$ channel, for definiteness. This is computed similarly to \eqn{locrosdefin}, that is (a factor $2\pi\delta(q_{1}^{+}+q_{2}^{+}+q_{3}^{+}-q^{+})$ is implicit in the l.h.s.)
 \begin{equation}\begin{split}\label{qqqcross}
\frac{d\sigma^{qA\rightarrow qq\overline{q}+X}}{d^{3}q_{1}\,d^{3}q_{2}\,d^{3}q_{3}}&\,\equiv\,\frac{1}{2N_{c}}\ {}^{out}\!\left\langle q_{\lambda}^{\alpha}(q^{+},\,\bm{q}=0_{\perp})\right|\,\hat{\mathcal{N}}_{q}(q_{1})\,\hat{\mathcal{N}}_{q}(q_{2})\,\hat{\mathcal{N}}_{\overline{q}}(q_{3})\,\left|q_{\lambda}^{\alpha}(q^{+},\,\bm{q}=0_{\perp})\right\rangle ^{out}\\
&=\,\frac{1}{2N_{c}}\,\int_{\bm{w},\,\overline{\bm{w}}}
\!\!{}_{\ \ \ qg \bar{q}}^{\ \ \ out}\!\left\langle q_{\lambda}^{\alpha}(q^{+},\,\overline{\bm{w}})\right|\,\hat{\mathcal{N}}_{q}(q_{1})\,\hat{\mathcal{N}}_{q}(q_{2})\,\hat{\mathcal{N}}_{\overline{q}}(q_{3})\,\left|q_{\lambda}^{\alpha}(q^{+},\,\bm{w})\right\rangle _{qq\bar{q}}^{out},
 \end{split}\end{equation}
 which involves the number-density Fock space operators for bare quarks and antiquarks previously introduced in \eqref{Nq} and \eqref{Nqbar}.

\comment{and respectively
 \begin{equation}\begin{split}\label{qggcross}
&\frac{d\sigma^{qA\rightarrow qgg+X}}{d^{3}q_{1}\,d^{3}q_{2}\,d^{3}q_{3}}\,\equiv\,\frac{1}{2N_{c}\,L}\,^{out}\left\langle q_{\lambda}^{\alpha}(q^{+},\,\bm{q}=0_{\perp})\right|\,\hat{\mathcal{N}}_{q}(q_{1})\,\hat{\mathcal{N}}_{g}(q_{2})\,\hat{\mathcal{N}}_{g}(q_{3})\,\left|q_{\lambda}^{\alpha}(q^{+},\,\bm{q}=0_{\perp})\right\rangle ^{out}\\
&=\,\frac{1}{2N_{c}\,L}\int_{\bm{w},\,\overline{\bm{w}}}\,{}_{qgg}\left\langle Q_{\lambda}^{\alpha}(q^{+},\,\overline{\bm{w}})\right|\,\hat{\mathcal{N}}_{q}(q_{1})\,\hat{\mathcal{N}}_{g}(q_{2})\,\hat{\mathcal{N}}_{g}(q_{3})\,\left|Q_{\lambda}^{\alpha}(q^{+},\,\bm{w})\right\rangle _{qgg},
\end{split}\end{equation}
}

\begin{figure}[!h]\center
    \includegraphics[scale=0.85]{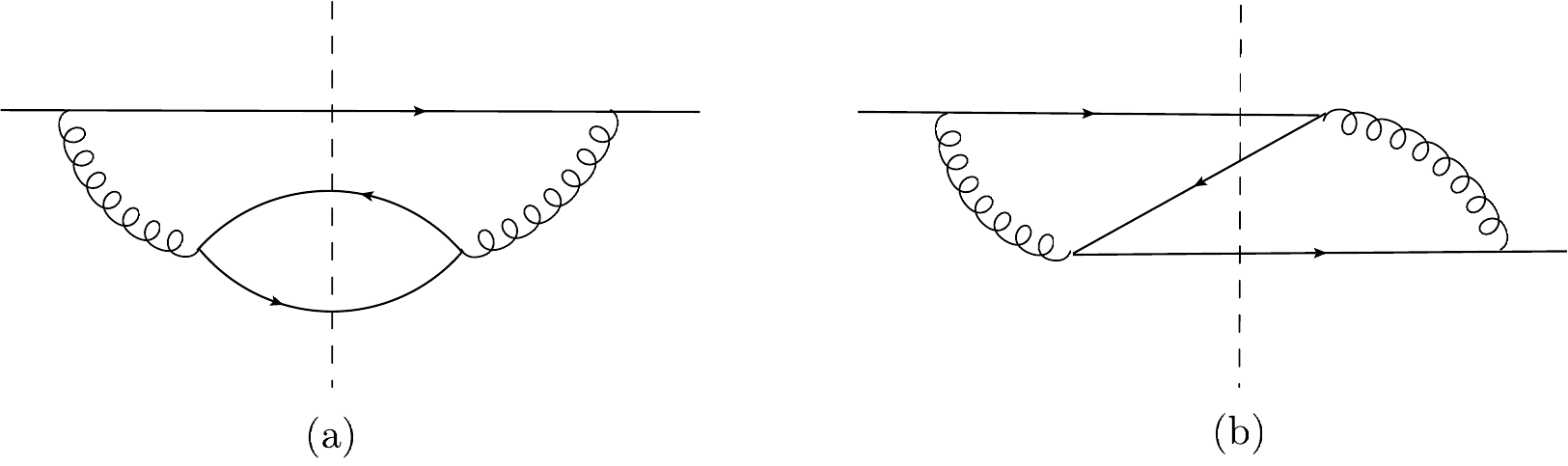}
  \caption{Two examples of graphs for $qq\bar q$ production. The graph on the left  contributes in the large $N_{c}$ limit and will be indeed included in our calculation. The  one on the right is suppressed at large $N_{c}$ and will be ignored. The dashed vertical lines indicates the final state at time $x^+\to\infty$. \label{croqqq}}
\end{figure}

\begin{figure}[!h]\center
    \includegraphics[scale=0.75]{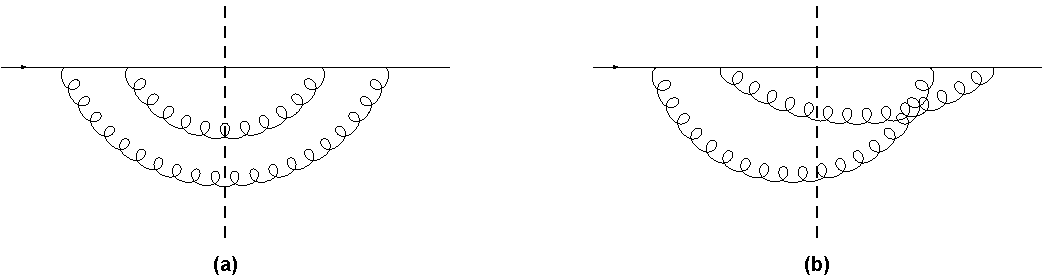}
  \caption{Two examples of graphs for $qgg$ production. The planar graph on the left contributes in the large $N_{c}$ limit and will be indeed included in our calculation. The non-planar one on the right is suppressed at large $N_{c}$ and will be ignored.\label{croqgg}}
\end{figure}

As explained in Sect.~\ref{sec:NLOprod},  the $qq\bar q$ component of the outgoing quark state contains two pieces, a  ``regular'' piece with three contributing topologies (cf. Fig.~\ref{quarinsert}) and an ``instantaneous'' one, with two
possible topologies (cf. Fig.~\ref{quarkinst}). When ``squaring'' the wave function to compute the cross-section \eqref{qqqcross}, one therefore generates 25 different topologies: 9 topologies involving the ``regular'' pieces of the LCWF alone, 12 interference topologies between ``regular'' and ``instantaneous'' graphs, and 4 built with the ``instantaneous''  graphs alone. More precisely, there are $2\times 25=50$ additional topologies in which the 2 quarks in the final state are differently connected between the direct amplitude (DA) and the complex conjugate amplitude (CCA): e.g., the original quark in the DA gets connected with the daughter quark in the CCA; see  Fig.~\ref{croqqq}.b. However, such ``crossed'' diagrams are suppressed in the multi-color limit $N_c\to\infty$ to which we shall eventually restrict in what follows (since the results look already complicated in that limit). Hence, from now on we shall neglect all such ``crossed'' topologies. A similar observation applies to  the diagrams involving gluons in the final state: we shall keep only the planar graphs, like that Fig.~\ref{croqgg}.a, but ignore the non-planar ones
like that in Fig.~\ref{croqgg}.b, which are suppressed at large $N_c$.

\comment{
Each of the final results presented below have the same structure: the first square brackets contain the contributions from color structure when the emission of the two partons, both in the amplitude and the conjugate amplitude, were created from an intermediate gluon. The second and third square brackets contain the contribution from the color structure when the the two partons emitted directly (instantaneous gluon). The fourth square brackets correspond to the case in which both the partons in the amplitude as well as the conjugate amplitude were emitted instantaneously. Since our intention was to keep only the contributions which are not suppressed at the $N_{c}$ limit, we dropped the contributions which contain crossing of partons (such as appear in Fig.~\ref{croqgg}.b). Each of the results below requires the introduction of a new set of Wilson lines in addition to those that were defined in the LO part, (\ref{lowils1}), (\ref{lowils3}), and (\ref{lowils2}). The various definitions of these structures and their large $N_{c}$ limit is available in App. \ref{wilsdict}. As expected, at the large $N_{c}$ limit all the results can be expressed exclusively in terms of dipole and quadropole.
}

\subsection{The $qq\overline{q}$ final state\label{qqqtrij}}

The starting point is Eq.~(\ref{qqqcross}) together with Eqs.~(\ref{fin.qqq1}) and (\ref{fin.qqq2}) for the $qq\overline{q}$ Fock space components of the quark outgoing state. It is then straightforward to evaluate the expectation value of the three-parton particle number densities (the matrix element shown in \eqn{matqqq} is useful in that respect), with the following result:
   \begin{equation}\begin{split}\label{qqcr}
&\frac{d\sigma^{qA\rightarrow qq\overline{q}+X}}{d^{3}q_{1}\,d^{3}q_{2}\,d^{3}q_{3}}\equiv\,\frac{\alpha_{s}^{2}\,C_{F}\,N_{f}}{2(2\pi)^{10}(q^{+})^{2}}\,\delta(q^{+}-q_{1}^{+}-q_{2}^{+}-q_{3}^{+})\,\int_{\bm{\overline{x}},\,\bm{\overline{z}},\,\bm{\overline{z}}^{\prime},\,\bm{x},\,\bm{z},\,\bm{z}^{\prime}}\,\rme^{-i\bm{q}_{1}\cdot(\bm{x}-\bm{\overline{x}})-i\bm{q}_{2}\cdot(\bm{z}-\bm{\bar{z}})-i\bm{q}_{3}\cdot(\bm{z}^{\prime}-\bm{\bar{z}}^{\prime})}\\
&\times\left\{ K_{qq\overline{q}}^{1}\left(\bm{\overline{x}},\,\bm{\overline{z}},\,\bm{\overline{z}}^{\prime},\,\bm{x},\,\bm{z},\,\bm{z}^{\prime}\right)\left[\overline{\Theta}_{1}\Theta_{1}S_{q\overline{q}qq\overline{q}q}\left(\bm{\overline{x}},\,\bm{\overline{z}},\,\bm{\overline{z}}^{\prime},\,\bm{x},\,\bm{z},\,\bm{z}^{\prime}\right)\,-\,\overline{\Theta}_{1}S_{q\overline{q}qqg}\left(\bm{\overline{x}},\,\bm{\overline{z}},\,\bm{\overline{z}}^{\prime},\,\bm{x},\,\bm{y}\right)\right.\right.\\
&-\,\Theta_{1}S_{qgq\overline{q}q}\left(\bm{\overline{x}},\,\bm{\overline{y}},\,\bm{x},\,\bm{z},\,\bm{z}^{\prime}\right)\,+\,\overline{\Theta}_{2}\Theta_{1}S_{qq\overline{q}q}\left(\bm{\overline{w}},\,\bm{x},\,\bm{z},\,\bm{z}^{\prime}\right)\,+\,\overline{\Theta}_{1}\Theta_{2}S_{qq\overline{q}q}\left(\bm{\overline{x}},\,\bm{w},\,\bm{\overline{z}}^{\prime},\,\bm{\overline{z}}\right)\\
&\left.\,+\,S_{qgqg}\left(\bm{\overline{x}},\,\bm{\overline{y}},\,\bm{x},\,\bm{y}\right)-\,\overline{\Theta}_{2}S_{qqg}\left(\overline{\bm{w}},\,\bm{y},\,\bm{x}\right)\,-\,\Theta_{2}S_{qqg}\left(\bm{\overline{x}},\,\bm{w},\,\bm{\overline{y}}\right)\,+\,\overline{\Theta}_{2}\Theta_{2}\mathcal{S}\left(\bm{\overline{w}},\,\bm{w}\right)\right]\\
&+\,K_{qq\overline{q}}^{2}\left(\bm{\overline{x}},\,\bm{\overline{z}},\,\bm{\overline{z}}^{\prime},\,\bm{x},\,\bm{z},\,\bm{z}^{\prime}\right)\left[\Theta_{1}S_{q\overline{q}qq\overline{q}q}\left(\bm{\overline{x}},\,\bm{\overline{z}},\,\bm{\overline{z}}^{\prime},\,\bm{x},\,\bm{z},\,\bm{z}^{\prime}\right)\,-\,S_{q\overline{q}qqg}\left(\bm{\overline{x}},\,\bm{\overline{z}},\,\bm{\overline{z}}^{\prime},\,\bm{x},\,\bm{y}\right)\right.\\
&\left.-\,\Theta_{1}S_{qq\overline{q}q}\left(\bm{\overline{w}},\,\bm{x},\,\bm{z},\,\bm{z}^{\prime}\right)\,+\,\Theta_{2}S_{qq\overline{q}q}\left(\bm{\overline{x}},\,\bm{w},\,\bm{\overline{z}}^{\prime},\,\bm{\overline{z}}\right)\,+\,S_{qqg}\left(\overline{\bm{w}},\,\bm{y},\,\bm{x}\right)\,-\,\Theta_{2}\mathcal{S}\left(\bm{\overline{w}},\,\bm{w}\right)\right]\\
&+\,K_{qq\overline{q}}^{2}\left(\bm{x},\,\bm{z},\,\bm{z}^{\prime},\,\bm{\overline{x}},\,\bm{\overline{z}},\,\bm{\overline{z}}^{\prime}\right)\left[\overline{\Theta}_{1}S_{q\overline{q}qq\overline{q}q}\left(\bm{\overline{x}},\,\bm{\overline{z}},\,\bm{\overline{z}}^{\prime},\,\bm{x},\,\bm{z},\,\bm{z}^{\prime}\right)\,-\,S_{qgq\overline{q}q}\left(\bm{\overline{x}},\,\bm{\overline{y}},\,\bm{x},\,\bm{z},\,\bm{z}^{\prime}\right)\right.\\
&\left.+\,\overline{\Theta}_{2}S_{qq\overline{q}q}\left(\bm{\overline{w}},\,\bm{x},\,\bm{z},\,\bm{z}^{\prime}\right)\,-\,\overline{\Theta}_{1}S_{qq\overline{q}q}\left(\bm{\overline{x}},\,\bm{w},\,\bm{\overline{z}}^{\prime},\,\bm{\overline{z}}\right)\,+\,S_{qqg}\left(\bm{\overline{x}},\,\bm{w},\,\bm{\overline{y}}\right)\,-\,\overline{\Theta}_{2}\mathcal{S}\left(\bm{\overline{w}},\,\bm{w}\right)\right]\\
&+\,K_{qq\overline{q}}^{3}\left(\bm{\overline{x}},\,\bm{\overline{z}},\,\bm{\overline{z}}^{\prime},\,\bm{x},\,\bm{z},\,\bm{z}^{\prime}\right)\left[S_{q\overline{q}qq\overline{q}q}\left(\bm{\overline{x}},\,\bm{\overline{z}},\,\bm{\overline{z}}^{\prime},\,\bm{x},\,\bm{z},\,\bm{z}^{\prime}\right)\,-\,S_{qq\overline{q}q}\left(\bm{\overline{w}},\,\bm{x},\,\bm{z},\,\bm{z}^{\prime}\right)\right.\\
&\left.\left.-\,S_{qq\overline{q}q}\left(\bm{\overline{x}},\,\bm{w},\,\bm{\overline{z}}^{\prime},\,\bm{\overline{z}}\right)\,+\,\mathcal{S}\left(\bm{\overline{w}},\,\bm{w}\right)\right]\right\} \,+\,\left(q_{1}^{+}\leftrightarrow q_{2}^{+},\:\bm{q}_{1}\leftrightarrow\bm{q}_{2}\right).
\end{split}\end{equation}
The transverse coordinates $\bx$ and $\bz'$ refer to the two quarks in the DA, while $\bz$ refers to the antiquark. The respective coordinates in the CCA are denoted as $\overline\bx$, ${\overline \bz}'$, ${\overline \bz}'$; e.g. the fermion at $\overline\bx$ effectively acts as an {\em anti}quark for the purpose of the collision with the shockwave. The transverse coordinates which do not appear under the integration sign are defined as follows:
\begin{align}\label{defyw}
\bm{y}&\,\equiv\,\xi\bm{z}^{\prime}+(1-\xi)\bm{z}\,;\qquad\qquad\bm{w}\,\equiv\,(1-\vartheta)\bm{x}\,+\,\xi\vartheta\bm{z}^{\prime}\,+\,(1-\xi)\vartheta\bm{z};\nn
\overline{\bm{y}}&\,\equiv\,\xi\overline{\bm{z}}^{\prime}+(1-\xi)\overline{\bm{z}}\,;\qquad\qquad\overline{\bm{w}}\,\equiv\,(1-\vartheta)\overline{\bm{x}}\,+\,\xi\vartheta\overline{\bm{z}}^{\prime}\,+\,(1-\xi)\vartheta\overline{\bm{z}}.
\end{align}
The longitudinal momentum fractions are fixed by the external momenta as follows:
\begin{equation}
\vartheta\,=\,\frac{q_{2}^{+}+q_{3}^{+}}{q^{+}},\qquad\qquad\xi\,=\,\frac{q_{3}^{+}}{q_{2}^{+}+q_{3}^{+}}.
\end{equation}
In deriving the three kernels $K_{qq\overline{q}}^{1}$, $K_{qq\overline{q}}^{2}$  and $K_{qq\overline{q}}^{3}$, we have performed the sums over polarizations and helicities with the help of the following identities:
\begin{align}
\varphi_{\lambda_{1}\lambda}^{mn\dagger}(\xi)\,\varphi_{\lambda_{1}\lambda}^{il}(\xi)&\,=\,2\left((2\xi-1)^{2}\delta^{mn}\delta^{il}\,+\,\delta^{mi}\delta^{nl}\,-\,\delta^{ni}\delta^{ml}\right),\nn
\phi_{\lambda_{1}\lambda}^{mp\dagger}(\vartheta)\,\phi_{\lambda_{1}\lambda}^{ij}(\vartheta)&\,=\,2\left((\vartheta-2)^{2}\delta^{mp}\delta^{ij}\,+\,\vartheta^{2}\left(\delta^{im}\delta^{pj}\,-\,\delta^{ip}\delta^{mj}\right)\right),
\end{align}
and
 \begin{eqnarray}
&&\phi_{\lambda_{1}\lambda}^{mp\dagger}(\vartheta)\,\varphi_{\lambda_{2}\lambda_{3}}^{mn\dagger}(\xi)\,\varphi_{\lambda_{2}\lambda_{3}}^{il}(\xi)\,\phi_{\lambda_{1}\lambda}^{ij}(\vartheta)\\
&&=\,2\left[\left((2\xi-1)^{2}(\vartheta-2)^{2}+\vartheta^{2}\right)\delta^{np}\delta^{lj}+\left((\vartheta-2)^{2}+2\vartheta^{2}\right)\delta^{pj}\delta^{nl}-\left((2\xi-1)^{2}\vartheta^{2}+(\vartheta-2)^{2}\right)\delta^{lp}\delta^{nj}\right].\nonumber
\end{eqnarray}

 \begin{figure}[!h]
  \includegraphics[scale=1]{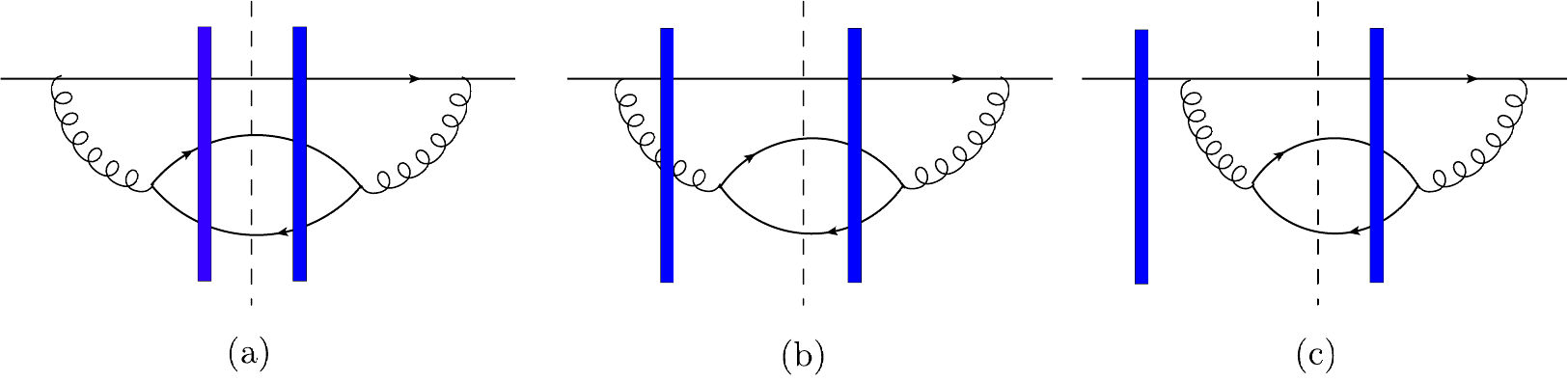}
  \caption{Three of the nine diagrams of the first type contributing to the cross-section for the $qq\bar q$ final state. The 
 intermediate gluon is non-local in time (``propagating'') in both the DA and the CCA. All such graphs enter in the final result with the kernel $K_{qq\overline{q}}^{1}$. The $S$-matrices corresponding to these particular graphs are $S_{q\overline{q}q\overline{q}q\overline{q}}$ for graph $(a)$, $S_{q\overline{q}q\overline{q}g}^{(2)}$ for graph $(b)$ and, finally, $S_{q\overline{q}q\overline{q}}^{(1)}$ for graph $(c)$.\label{qqqfig1}}
\end{figure}

We recognize in the r.h.s. of \eqn{qqcr} the four types of contributions anticipated at the beginning of this section. All the terms of a given type have the same branching pattern (in both the DA and the CCA) --- hence they are multiplied by the same emission kernel ---, but differ from each other at the level of the shockwave insertions --- hence they involve $S$-matrices with different color structures. As in the previous sections, we label the $S$-matrices via lower indices which indicate the partons involved in the scattering and also (when needed) via upper indices $(1)$ or $(2)$, to distinguish between different color configurations with the same partonic content. To facilitate the reading of this section, we have collected the definitions of the various $S$--matrices in Appendix~\ref{wilsdict}, see notably Eqs.~(\ref{wils1})--(\ref{wils2}). 
But their simplified versions valid in the large-$N_c$ limit will be exhibited later in this section.

 Specifically, the nine terms multiplying the kernel $K_{qq\overline{q}}^{1}$, with (we use the notations $\bm{Z}\equiv\bm{z}-\bm{z}^{\prime}$, $\bm{X}\equiv\bm{x}-\bm{z}$,
$\overline{\bm{Z}}\equiv\overline{\bm{z}}-\overline{\bm{z}}^{\prime}$ and $\overline{\bm{X}}\equiv\overline{\bm{x}}-\overline{\bm{z}}$)
\begin{equation}\begin{split}
&K_{qq\overline{q}}^{1}\left(\bm{\overline{x}},\,\bm{\overline{z}},\,\bm{\overline{z}}^{\prime},\,\bm{x},\,\bm{z},\,\bm{z}^{\prime}\right)\equiv\,\frac{\bm{\overline{Z}}^{n}\,\left(\bm{\overline{X}}^{p}+\xi\bm{\overline{Z}}^{p}\right)\,\bm{Z}^{l}\,\left(\bm{X}^{j}+\xi\bm{Z}^{j}\right)}{\bm{\overline{Z}}^{2}\,\left(\overline{\bm{X}}+\xi\overline{\bm{Z}}\right)^{2}\,\bm{Z}^{2}\,\left(\bm{X}+\xi\bm{Z}\right)^{2}}\\
&\times\left(\left((2\xi-1)^{2}(\vartheta-2)^{2}+\vartheta^{2}\right)\delta^{np}\delta^{lj}+\left((\vartheta-2)^{2}+2\vartheta^{2}\right)\delta^{pj}\delta^{nl}-\left((2\xi-1)^{2}\vartheta^{2}+(\vartheta-2)^{2}\right)\delta^{lp}\delta^{nj}\right),
\end{split}\end{equation}
corresponds to diagrams where the intermediate gluon propagator is non-local in time, i.e. it shows a genuine propagation. Three of the nine associated diagrams are shown in Fig.~ \ref{qqqfig1}.

\begin{figure}[t]
    \includegraphics[scale=.9]{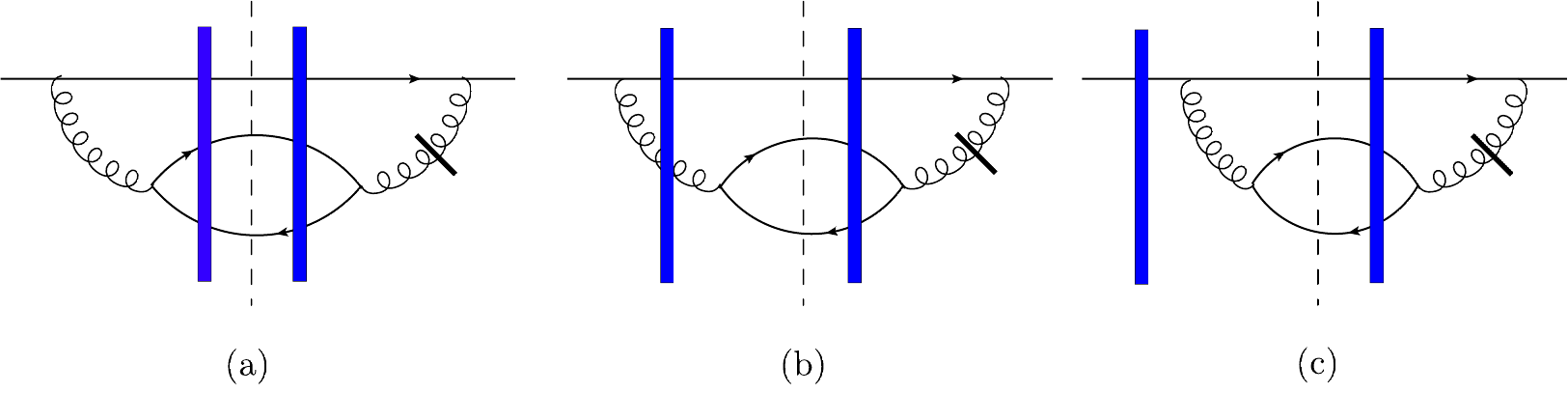}
  \caption{{Three among the 12 interference graphs between a propagating intermediate gluon in the DA and an instantaneous one in the CCA, or vice-versa. All such diagrams yield contributions proportional to $K_{qq\overline{q}}^{2}$. \label{qqqfig2}}}
\end{figure}

Furthermore, the two subsequent types of contributions, containing 6 terms each and proportional to the same  kernel $K_{qq\overline{q}}^{2}$ (but with different transverse coordinates for the two types), represent interference terms between ``propagating'' gluons in the DA and ``instantaneous'' gluons in the CCA, or vice-versa (see Fig.~\ref{qqqfig2}).
The corresponding kernel reads:
 \begin{equation}\begin{split}
 &K_{qq\overline{q}}^{2}\left(\bm{\overline{x}},\,\bm{\overline{z}},\,\bm{\overline{z}}^{\prime},\,\bm{x},\,\bm{z},\,\bm{z}^{\prime}\right)\equiv\,\frac{4(2\xi-1)(2-\vartheta)(1-\vartheta)\xi(1-\xi)\,\bm{Z}\cdot\left(\bm{X}+\xi\bm{Z}\right)}{\left(\xi(1-\xi)\overline{\bm{Z}}^{2}\,+\,(1-\vartheta)\left(\overline{\bm{X}}+\xi\overline{\bm{Z}}\right)^{2}\right)\,\bm{Z}^{2}\,\left(\bm{X}+\xi\bm{Z}\right)^{2}},
\end{split}\end{equation}

\begin{figure}[h]\center
    \includegraphics[scale=0.7]{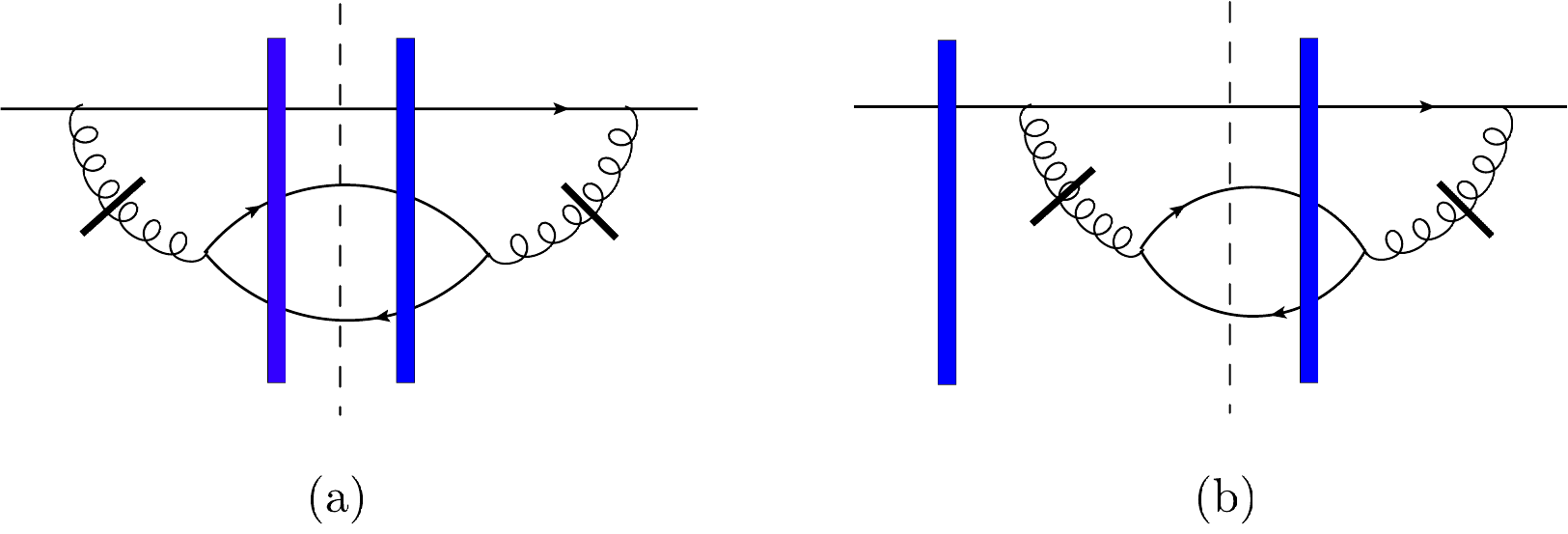}
  \caption{Two among the four diagrams in which the intermediate gluon is instantaneous in both the DA and the CCA. Such diagrams enter in the final result with the kernel $K_{qq\overline{q}}^{3}$. \label{qqqfig3}}
\end{figure}

Finally, the four terms multiplying $K_{qq\overline{q}}^{3}$, with
 \begin{equation}\begin{split}
&K_{qq\overline{q}}^{3}\left(\bm{\overline{x}},\,\bm{\overline{z}},\,\bm{\overline{z}}^{\prime},\,\bm{x},\,\bm{z},\,\bm{z}^{\prime}\right)\\
&\equiv\,\frac{2(1-\vartheta)^{2}\xi^{2}(1-\xi)^{2}}{\left(\xi(1-\xi)\overline{\bm{Z}}^{2}\,+\,(1-\vartheta)\left(\overline{\bm{X}}+\xi\overline{\bm{Z}}\right)^{2}\right)\,\left(\xi(1-\xi)\bm{Z}^{2}\,+\,(1-\vartheta)\left(\bm{X}+\xi\bm{Z}\right)^{2}\right)},
\end{split}\end{equation}
represent diagrams in which the intermediate gluon is instantaneous in both the DA and the CCA. cf. Fig.~\ref{qqqfig3}.
 
 As promised, the color structure of the various $S$-matrices becomes more transparent when taking the large-$N_c$ limit (see App. \ref{wilsdict} for details).  After retaining only the structures which are left at the large $N_{c}$ limit and replacing $C_{F}\,\rightarrow\,N_{c}/2$, the result (\ref{qqcr}) can be fully expressed in terms of dipoles and quadrupoles, as expected on general grounds \cite{Kovner:2006wr,Kovner:2006ge,Dominguez:2011wm,Dominguez:2012ad}. Specifically, one finds
 \begin{equation}\label{qqqNc}
 \begin{split}
&\frac{d\sigma^{qA\rightarrow qq\overline{q}+X}}{d^{3}q_{1}\,d^{3}q_{2}\,d^{3}q_{3}}\,\equiv\,\frac{\alpha_{s}^{2}\,N_{c}\,N_{f}}{4(2\pi)^{10}(q^{+})^{2}}\,\delta(q^{+}-q_{1}^{+}-q_{2}^{+}-q_{3}^{+})\,\int_{\bm{\overline{x}},\,\bm{\overline{z}},\,\bm{\overline{z}}^{\prime},\,\bm{x},\,\bm{z},\,\bm{z}^{\prime}}\,\rme^{-i\bm{q}_{1}\cdot(\bm{x}-\bm{\overline{x}})-i\bm{q}_{2}\cdot(\bm{z}-\bm{\bar{z}})-i\bm{q}_{3}\cdot(\bm{z}^{\prime}-\bm{\bar{z}}^{\prime})}\\
&\times\left\{ K_{qq\overline{q}}^{1}\left(\bm{\overline{x}},\,\bm{\overline{z}},\,\bm{\overline{z}}^{\prime},\,\bm{x},\,\bm{z},\,\bm{z}^{\prime}\right)\left[\overline{\Theta}_{1}\,\Theta_{1}\,{\mathcal{Q}(\overline{\bm{x}},\,\bm{x},\,\bm{z},\,\overline{\bm{z}})\,\mathcal{S}(\overline{\bm{z}}^{\prime},\,\bm{z}^{\prime})}\,-\,\overline{\Theta}_{1}\,\mathcal{Q}(\overline{\bm{x}},\,\bm{x},\,\bm{y},\,\overline{\bm{z}})\,\mathcal{S}(\overline{\bm{z}}^{\prime},\,\bm{y})\right.\right.\\
&-\,\Theta_{1}\,\mathcal{Q}(\overline{\bm{x}},\,\bm{x},\,\bm{z},\,\overline{\bm{y}})\,\mathcal{S}(\overline{\bm{y}},\,\bm{z}^{\prime})\,+\,\overline{\Theta}_{2}\,\Theta_{1}\,\mathcal{S}(\overline{\bm{w}},\,\bm{z}^{\prime})\,\mathcal{S}(\bm{z},\,\bm{x})\,+\,\overline{\Theta}_{1}\,\Theta_{2}\,\mathcal{S}(\overline{\bm{x}},\,\bm{\overline{z}})\,\mathcal{S}(\bm{\overline{z}}^{\prime},\,\bm{w})\\
&\left.\,+\,\mathcal{Q}(\overline{\bm{x}},\,\bm{x},\,\bm{y},\,\overline{\bm{y}})\,\mathcal{S}(\overline{\bm{y}},\,\bm{y})-\,\overline{\Theta}_{2}\,\mathcal{S}(\overline{\bm{w}},\,\bm{x})\,\mathcal{S}(\bm{x},\,\bm{y})\,-\,\Theta_{2}\,\mathcal{S}(\overline{\bm{x}},\,\bm{\overline{y}})\,\mathcal{S}(\bm{\overline{y}},\,\bm{w})\,+\,\overline{\Theta}_{2}\,\Theta_{2}\mathcal{S}\left(\bm{\overline{w}},\,\bm{w}\right)\right]\\
&+\,K_{qq\overline{q}}^{2}\left(\bm{\overline{x}},\,\bm{\overline{z}},\,\bm{\overline{z}}^{\prime},\,\bm{x},\,\bm{z},\,\bm{z}^{\prime}\right)\left[\Theta_{1}\,\mathcal{Q}(\overline{\bm{x}},\,\bm{x},\,\bm{z},\,\overline{\bm{z}})\,\mathcal{S}(\overline{\bm{z}}^{\prime},\,\bm{z}^{\prime})\,-\,\mathcal{Q}(\overline{\bm{x}},\,\bm{x},\,\bm{y},\,\overline{\bm{z}})\,\mathcal{S}(\overline{\bm{z}}^{\prime},\,\bm{y})\right.\\
&\left.-\,\Theta_{1}\,\mathcal{S}(\overline{\bm{w}},\,\bm{z}^{\prime})\,\mathcal{S}(\bm{z},\,\bm{x})\,+\,\Theta_{2}\,\mathcal{S}(\overline{\bm{x}},\,\bm{\overline{z}})\,\mathcal{S}(\bm{\overline{z}}^{\prime},\,\bm{w})\,+\,\mathcal{S}(\overline{\bm{w}},\,\bm{x})\,\mathcal{S}(\bm{x},\,\bm{y})\,-\,\Theta_{2}\,\mathcal{S}\left(\bm{\overline{w}},\,\bm{w}\right)\right]\\
&+\,K_{qq\overline{q}}^{2}\left(\bm{x},\,\bm{z},\,\bm{z}^{\prime},\,\bm{\overline{x}},\,\bm{\overline{z}},\,\bm{\overline{z}}^{\prime}\right)\left[\overline{\Theta}_{1}\,\mathcal{Q}(\overline{\bm{x}},\,\bm{x},\,\bm{z},\,\overline{\bm{z}})\,\mathcal{S}(\overline{\bm{z}}^{\prime},\,\bm{z}^{\prime})\,-\,\mathcal{Q}(\overline{\bm{x}},\,\bm{x},\,\bm{z},\,\overline{\bm{y}})\,\mathcal{S}(\overline{\bm{y}},\,\bm{z}^{\prime})\right.\\
&\left.+\,\overline{\Theta}_{2}\,\mathcal{S}(\overline{\bm{w}},\,\bm{z}^{\prime})\,\mathcal{S}(\bm{z},\,\bm{x})\,-\,\overline{\Theta}_{1}\,\mathcal{S}(\overline{\bm{x}},\,\bm{\overline{z}})\,\mathcal{S}(\bm{\overline{z}}^{\prime},\,\bm{w})\,+\,\mathcal{S}(\overline{\bm{x}},\,\bm{\overline{y}})\,\mathcal{S}(\bm{\overline{y}},\,\bm{w})\,-\,\overline{\Theta}_{2}\,\mathcal{S}\left(\bm{\overline{w}},\,\bm{w}\right)\right]\\
&+\,K_{qq\overline{q}}^{3}\left(\bm{\overline{x}},\,\bm{\overline{z}},\,\bm{\overline{z}}^{\prime},\,\bm{x},\,\bm{z},\,\bm{z}^{\prime}\right)\left[\mathcal{Q}(\overline{\bm{x}},\,\bm{x},\,\bm{z},\,\overline{\bm{z}})\,\mathcal{S}(\overline{\bm{z}}^{\prime},\,\bm{z}^{\prime})\,-\,\mathcal{S}(\overline{\bm{w}},\,\bm{z}^{\prime})\,\mathcal{S}(\bm{z},\,\bm{x})\right.\\
&\left.\left.-\,\mathcal{S}(\overline{\bm{x}},\,\bm{\overline{z}})\,\mathcal{S}(\bm{\overline{z}}^{\prime},\,\bm{w})\,+\,\mathcal{S}\left(\bm{\overline{w}},\,\bm{w}\right)\right]\right\} \,+\,\left(q_{1}^{+}\leftrightarrow q_{2}^{+},\:\bm{q}_{1}\leftrightarrow\bm{q}_{2}\right).
\end{split}\end{equation}

\subsection{The $qgg$ final state}

The calculation of the cross section for the final state involving the original quark plus two gluons is similar to the previous one for the case where the quark was accompanied by a quark-antiquark pair, but a little more cumbersome, due to the fact that there are twice as many contributions to the $qgg$ Fock space component of the quark LCWF (cf. Eqs.~\eqref{qggsplit}--\eqref{out.qgg2}). Specifically, besides the (regular and instantaneous) contributions involving an intermediate gluon, labelled with a subscript $i=2$ in Sect.~\ref{sec:NLOprod} (see Figs.~\ref{fig:qggreg2} and \ref{fig:qgginst2}), there are also contributions, labelled with $i=1$, involving an intermediate quark  (see Figs.~\ref{fig:qggreg1} and \ref{fig:qgginst1}). Hence, when computing the cross-section for $qgg$ production, we shall also meet with interference contributions between ``intermediate gluon'' and ``intermediate quark''.

\subsubsection{Contributions with an intermediate quark\label{qggtrijaa}}

This calculation and the physical interpretation of its results are entirely similar to that for the $qq\bar q$ final state presented in Sect.~\ref{qqqtrij}, so here we shall only quote the final results. Namely, starting with Eqs.~\eqref{fin.qgg1} and \eqref{fin.qgg2} and using the matrix element (\ref{matqgg}), one finds 
  \begin{equation}\begin{split}\label{gpar1}
&\frac{d\sigma_{QQ}^{qA\rightarrow qgg+X}}{d^{3}q_{1}\,d^{3}q_{2}\,d^{3}q_{3}}\equiv\,\frac{2\alpha_{s}^{2}\,C_{F}^{2}}{(2\pi)^{10}(q^{+})^{2}}\,\delta(q^{+}-q_{1}^{+}-q_{2}^{+}-q_{3}^{+})\,\int_{\bm{\overline{x}},\,\bm{\overline{\bm{z}}},\,\bm{\overline{\bm{z}}}^{\prime},\,\bm{x},\,\bm{\bm{z}},\,\bm{\bm{z}}^{\prime}}\:e^{-i\bm{q}_{1}\cdot(\bm{x}-\bm{\overline{x}})-i\bm{q}_{2}\cdot(\bm{\bm{z}}-\bm{\overline{\bm{z}}})-i\bm{q}_{3}\cdot(\bm{\bm{z}}^{\prime}-\bm{\overline{\bm{z}}}^{\prime})}\\
&\times\left\{ K_{qgg}^{1}\left(\bm{\overline{x}},\,\bm{\overline{z}},\,\bm{\overline{z}}^{\prime},\,\bm{x},\,\bm{z},\,\bm{z}^{\prime}\right)\left[\overline{\Theta}_{3}\,\Theta_{3}\,S_{qggqgg}^{(1)}\left(\bm{\overline{x}},\,\overline{\bm{z}},\,\overline{\bm{z}}^{\prime},\,\bm{x},\,\bm{z},\,\bm{z}^{\prime}\right)\,-\,\overline{\Theta}_{3}\,S_{qgqgg}^{(1)}\left(\bm{\overline{y}},\,\overline{\bm{z}}^{\prime},\,\bm{x},\,\bm{z},\,\bm{z}^{\prime}\right)\right.\right.\\
&-\,\Theta_{3}\,S_{qggqg}^{(1)}\left(\bm{\overline{x}},\,\overline{\bm{z}},\,\overline{\bm{z}}^{\prime},\,\bm{y},\,\bm{z}\right)\,+\,\overline{\Theta}_{4}\,\Theta_{3}\,S_{qgqg}\left(\bm{\overline{y}},\,\overline{\bm{z}}^{\prime},\,\bm{y},\,\bm{z}^{\prime}\right)\,+\,\overline{\Theta}_{3}\,\Theta_{4}\,S_{qqgg}^{(1)}\left(\bm{\overline{w}},\,\bm{x},\,\bm{z},\,\bm{z}^{\prime}\right)\\
&\left.\,+\,S_{qqgg}^{(1)}\left(\bm{\overline{x}},\,\bm{w},\,\overline{\bm{z}}^{\prime},\,\overline{\bm{z}}\right)-\,\overline{\Theta}_{4}\,S_{qqg}\left(\bm{\overline{w}},\,\bm{y},\,\bm{z}^{\prime}\right)\,-\,\Theta_{4}\,S_{qqg}\left(\bm{\overline{y}},\,\bm{w},\,\overline{\bm{z}}^{\prime}\right)\,+\,\overline{\Theta}_{4}\,\Theta_{4}\,\mathcal{S}\left(\bm{\overline{w}},\,\bm{w}\right)\right]\\
&+\,K_{qgg}^{2}\left(\bm{\overline{x}},\,\bm{\overline{z}},\,\bm{\overline{z}}^{\prime},\,\bm{x},\,\bm{z},\,\bm{z}^{\prime}\right)\left[\Theta_{3}\,S_{qggqgg}^{(1)}\left(\bm{\overline{x}},\,\overline{\bm{z}},\,\overline{\bm{z}}^{\prime},\,\bm{x},\,\bm{z},\,\bm{z}^{\prime}\right)\,-\,S_{qggqg}^{(1)}\left(\bm{\overline{x}},\,\overline{\bm{z}},\,\overline{\bm{z}}^{\prime},\,\bm{y},\,\bm{z}\right)\right.\\
&\left.\,-\,\Theta_{3}\,S_{qqgg}^{(1)}\left(\bm{\overline{w}},\,\bm{x},\,\bm{z},\,\bm{z}^{\prime}\right)\,+\,\Theta_{4}\,S_{qqgg}^{(1)}\left(\bm{\overline{x}},\,\bm{w},\,\overline{\bm{z}}^{\prime},\,\overline{\bm{z}}\right)\,+\,S_{qqg}\left(\bm{\overline{w}},\,\bm{y},\,\bm{z}^{\prime}\right)\,-\,\Theta_{4}\,\mathcal{S}\left(\bm{\overline{w}},\,\bm{w}\right)\right]\\
&+\,K_{qgg}^{2}\left(\bm{x},\,\bm{z},\,\bm{z}^{\prime},\,\bm{\overline{x}},\,\bm{\overline{z}},\,\bm{\overline{z}}^{\prime}\right)\left[\overline{\Theta}_{3}\,S_{qggqgg}^{(1)}\left(\bm{\overline{x}},\,\overline{\bm{z}},\,\overline{\bm{z}}^{\prime},\,\bm{x},\,\bm{z},\,\bm{z}^{\prime}\right)\,-\,S_{qqggg}^{(1)}\left(\bm{\overline{y}},\,\overline{\bm{z}}^{\prime},\,\bm{x},\,\bm{z},\,\bm{z}^{\prime}\right)\right.\\
&\left.\,+\,\overline{\Theta}_{4}\,S_{qqgg}^{(1)}\left(\bm{\overline{w}},\,\bm{x},\,\bm{z},\,\bm{z}^{\prime}\right)\,-\,\overline{\Theta}_{3}\,S_{qqgg}^{(1)}\left(\bm{\overline{x}},\,\bm{w},\,\overline{\bm{z}}^{\prime},\,\overline{\bm{z}}\right)\,+\,S_{qqg}\left(\bm{\overline{y}},\,\bm{w},\,\overline{\bm{z}}^{\prime}\right)\,-\,\overline{\Theta}_{4}\,\mathcal{S}\left(\bm{\overline{w}},\,\bm{w}\right)\right]\\
&+\,K_{qgg}^{3}\left(\bm{\overline{x}},\,\bm{\overline{z}},\,\bm{\overline{z}}^{\prime},\,\bm{x},\,\bm{z},\,\bm{z}^{\prime}\right)\left[S_{qggqgg}^{(1)}\left(\bm{\overline{x}},\,\overline{\bm{z}},\,\overline{\bm{z}}^{\prime},\,\bm{x},\,\bm{z},\,\bm{z}^{\prime}\right)\,-\,S_{qqgg}^{(1)}\left(\bm{\overline{w}},\,\bm{x},\,\bm{z},\,\bm{z}^{\prime}\right)\right.\\
&\left.\left.-\,S_{qqgg}^{(1)}\left(\bm{\overline{x}},\,\bm{w},\,\overline{\bm{z}}^{\prime},\,\overline{\bm{z}}\right)\,+\,\mathcal{S}\left(\bm{\overline{w}},\,\bm{w}\right)\right]\right\} \,+\,\left(q_{2}^{+}\leftrightarrow q_{3}^{+},\:\bm{q}_{2}\leftrightarrow\bm{q}_{3}\right).
 \end{split}\end{equation}
 where the  additional transverse coordinates (besides those to be integrated over) are defined as:
 \begin{align}
\bm{y}&\,\equiv\,(1-\xi)\bm{x}\,+\,\xi\bm{z},\qquad  \bm{w}\,\equiv \,(1-\xi)(1-\vartheta)\bm{x}\,+\,\xi(1-\vartheta)\bm{z}\,+\,\vartheta\bm{z}^{\prime},\nn
 \overline{\bm{y}}&\,\equiv\,(1-\xi)\overline{\bm{x}}\,+\,\xi\overline{\bm{z}},\qquad
 \overline{\bm{w}}\,\equiv\,(1-\xi)(1-\vartheta)\overline{\bm{x}}\,+\,\xi(1-\vartheta)\overline{\bm{z}}\,+\,\vartheta\overline{\bm{z}}^{\prime},
\end{align} 
with the following values for the longitudinal momentum fractions:
\begin{equation}
 \vartheta\,=\,\frac{q_{2}^{+}}{q^{+}},\qquad\qquad\xi\,=\,\frac{q_{1}^{+}}{q_{1}^{+}+q_{3}^{+}}.
  \end{equation}

Eq.~(\ref{gpar1}) features three new kernels, defined as 
  \begin{equation}\begin{split}
 &K_{qgg}^{1}\left(\bm{\overline{x}},\,\bm{\overline{z}},\,\bm{\overline{z}}^{\prime},\,\bm{x},\,\bm{z},\,\bm{z}^{\prime}\right)\,\equiv\,\frac{2(1-\vartheta)\,\left(1+(1-\vartheta)^{2}\right)\,\left(1+(1-\xi)^{2}\right)\,\bm{\overline{X}}\cdot\bm{X}\,\left(\xi\overline{\bm{X}}-\bm{\overline{X}^{\prime}}\right)\cdot\left(\xi\bm{X}-\bm{X^{\prime}}\right)}{\xi\,\vartheta^{2}\,\overline{\bm{X}}^{2}\,\bm{X}^{2}\left(\xi\overline{\bm{X}}-\bm{\overline{X}^{\prime}}\right)^{2}\,\left(\xi\bm{X}-\bm{X^{\prime}}\right)^{2}},
  \end{split}\end{equation}
  where  $\bm{X}\equiv\bm{x}-\bm{z}$, $\bm{X}^{\prime}\,\equiv\,\bm{x}\,-\,\bm{z}^{\prime}$ and similarly for $\overline{\bm{X}}$ and $\bm{\overline{X}^{\prime}}$. Furthermore,
\begin{equation}\begin{split}
&K_{qgg}^{2}\left(\bm{\overline{x}},\,\bm{\overline{z}},\,\bm{\overline{z}}^{\prime},\,\bm{x},\,\bm{z},\,\bm{z}^{\prime}\right)\,\equiv\,-\frac{(1-\xi)\,\chi_{\lambda}^{\dagger}\left(\delta^{ij}+i\varepsilon^{ij}\sigma^{3}\right)\chi_{\lambda_{2}}\phi_{\lambda_{2}\lambda_{1}}^{jm}(\xi)\,\phi_{\lambda_{1}\lambda}^{il}(\vartheta)\,\left(\xi\bm{X}-\bm{X^{\prime}}\right)^{l}\,\bm{X}^{m}}{\sqrt{\vartheta}\,\left(\vartheta\left(\xi\overline{\bm{X}}-\bm{\overline{X}^{\prime}}\right)^{2}+\xi(1-\xi)\bm{\overline{\bm{X}}}^{2}\right)\,\bm{X}^{2}\,\left(\xi\bm{X}-\bm{X^{\prime}}\right)^{2}},
\end{split}\end{equation}
and respectively
\begin{equation}\begin{split}
&K_{qgg}^{3}\left(\bm{\overline{x}},\,\bm{\overline{z}},\,\bm{\overline{z}}^{\prime},\,\bm{x},\,\bm{z},\,\bm{z}^{\prime}\right)\,\equiv\,\frac{32\xi\vartheta(1-\xi)^{2}}{(1-\vartheta)\,\left(\vartheta\left(\xi\overline{\bm{X}}-\bm{\overline{X}^{\prime}}\right)^{2}+\xi(1-\xi)\bm{\overline{\bm{X}}}^{2}\right)\,\left(\vartheta\left(\xi\bm{X}-\bm{X^{\prime}}\right)^{2}+\xi(1-\xi)\bm{\bm{X}}^{2}\right)}.
\end{split}\end{equation}
The identities (\ref{simptheta}) and (\ref{simpphi}) were useful in deriving the final expressions for these kernels. 

The new $S$-matrices which appear in (\ref{gpar1}) involve both fundamental and adjoint Wilson lines. They are presented, together with their simplified versions at large $N_c$, in Eqs.~(\ref{wils3})--(\ref{wils4}) of  App. \ref{wilsdict}.

 \begin{figure}[!h]
    \includegraphics[scale=0.51]{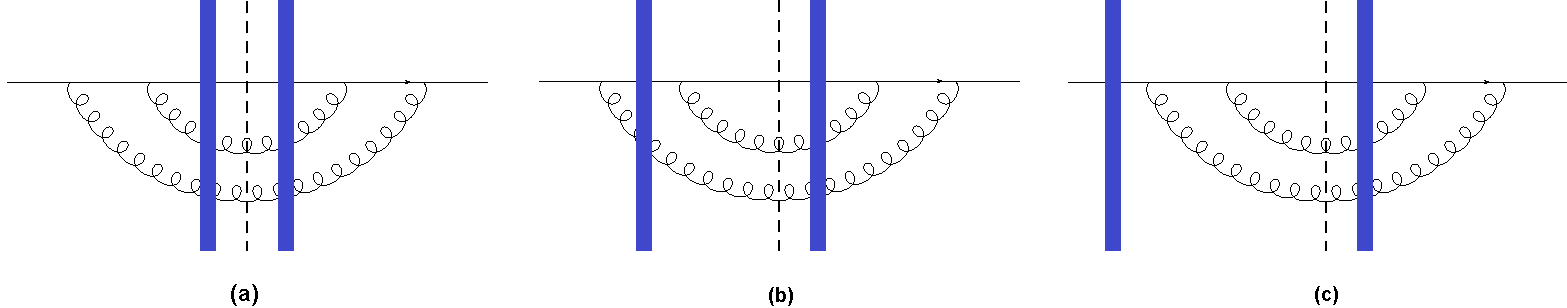}
  \caption{Three among the  nine diagrams of type which contribute to the production of two gluons. Both gluons are emitted by the initial quark and the intermediate quark propagator is not instantaneous. All such diagrams yield contributions proportional with the kernel $K_{qgg}^{1}$. The particular color structures associated with diagrams $(a)$, $(b)$ and $(c)$ are $S_{q\overline{q}gggg}^{(1)}$, $S_{q\overline{q}ggg}^{(1)}$, and $S_{q\overline{q}gg}^{(2)}$, respectively.\label{qgg1fig1}}
\end{figure}

  \begin{figure}[!h]
    \includegraphics[scale=0.51]{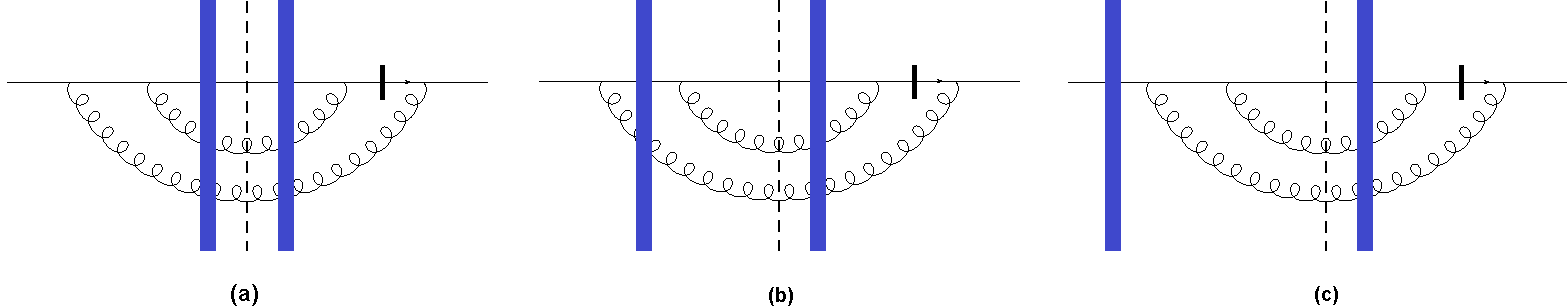}
  \caption{Three examples of interference diagrams between one ``regular'' intermediate quark propagator and one instantaneous one. Such diagrams gives contributions proportional with $K_{qgg}^{2}$.\label{qgg1fig2}}
\end{figure}

 \begin{figure}[!h]\center
    \includegraphics[scale=0.51]{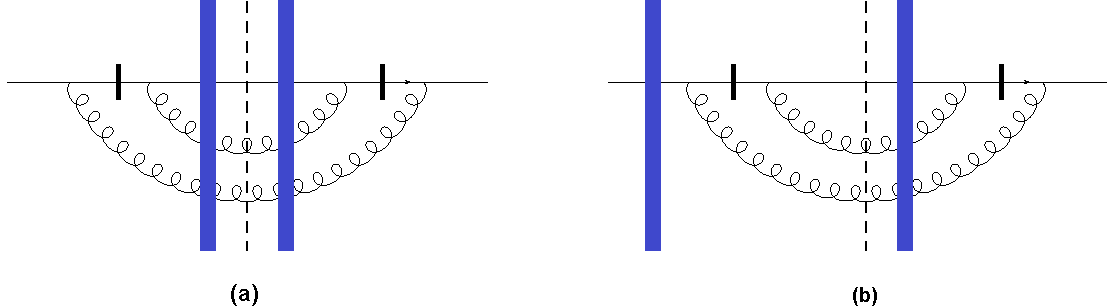}
  \caption{Two among the four diagrams in which the intermediate quark propagator is instantaneous in both the DA and the CCA. The associated contributions involve the kernel $K_{qgg}^{3}$.\label{qgg1fig3}}
\end{figure}

To illustrate the result in Eq.~(\ref{gpar1}), we exhibit in Figs.~\ref{qgg1fig1}, \ref{qgg1fig2}, and \ref{qgg1fig3} some examples of diagrams of the various types.  
 
  \subsubsection{Contributions with an intermediate gluon\label{qggtrijbb}}
 
In this case, the relevant Fock space components are shown in Eqs.~\eqref{fin.qgg3}--\eqref{fin.qgg4}. Using again the matrix element (\ref{matqgg}), one finds after simple algebra
  \begin{equation}\label{gpar2}
  \begin{split}
&\frac{d\sigma_{GG}^{qA\rightarrow qgg+X}}{d^{3}q_{1}\,d^{3}q_{2}\,d^{3}q_{3}}\equiv\,\frac{\alpha_{s}^{2}\,C_{F}\,N_{c}}{(2\pi)^{10}(q^{+})^{2}}\,\delta(q^{+}-q_{1}^{+}-q_{2}^{+}-q_{3}^{+})\,\int_{\bm{\overline{x}},\,\bm{\overline{\bm{z}}},\,\bm{\overline{\bm{z}}}^{\prime},\,\bm{x},\,\bm{\bm{z}},\,\bm{\bm{z}}^{\prime}}\:e^{-i\bm{q}_{1}\cdot(\bm{x}-\bm{\overline{x}})-i\bm{q}_{2}\cdot(\bm{\bm{z}}-\bm{\overline{\bm{z}}})-i\bm{q}_{3}\cdot(\bm{\bm{z}}^{\prime}-\bm{\overline{\bm{z}}}^{\prime})}\\
&\times\left\{ K_{qgg}^{4}\left(\bm{\overline{x}},\,\bm{\overline{z}},\,\bm{\overline{z}}^{\prime},\,\bm{x},\,\bm{z},\,\bm{z}^{\prime}\right)\left[\overline{\Theta}_{1}\,\Theta_{1}\,S_{qggqgg}^{(2)}\left(\overline{\bm{x}},\,\overline{\bm{z}},\,\overline{\bm{z}}^{\prime},\,\bm{x},\,\bm{z},\,\bm{z}^{\prime}\right)\,-\,\overline{\Theta}_{1}\,S_{qggqg}^{(2)}\left(\overline{\bm{x}},\,\overline{\bm{z}},\,\overline{\bm{z}}^{\prime},\,\bm{x},\,\bm{y}\right)\right.\right.\\
&-\,\Theta_{1}\,S_{qgqgg}^{(2)}\left(\overline{\bm{x}},\,\overline{\bm{y}},\,\bm{x},\,\bm{z},\,\bm{z}^{\prime}\right)\,+\,\overline{\Theta}_{2}\,\Theta_{1}\,S_{qqgg}^{(2)}\left(\overline{\bm{w}},\,\bm{x},\,\bm{z},\,\bm{z}^{\prime}\right)\,+\,\overline{\Theta}_{1}\,\Theta_{2}\,S_{qqgg}^{(2)}\left(\bm{\overline{x}},\,\bm{w},\,\bm{\overline{z}},\,\bm{\overline{z}}^{\prime}\right)\\
&\left.\,+\,S_{qgqg}\left(\overline{\bm{x}},\,\overline{\bm{y}},\,\bm{x},\,\bm{y}\right)-\,\Theta_{2}\,S_{qqg}\left(\overline{\bm{x}},\,\bm{w},\,\overline{\bm{y}}\right)\,-\,\overline{\Theta}_{2}\,S_{qqg}\left(\overline{\bm{w}},\,\bm{x},\,\bm{y}\right)\,+\,\overline{\Theta}_{2}\,\Theta_{2}\,\mathcal{S}\left(\overline{\bm{w}},\,\bm{w}\right)\right]\\
&+\,K_{qgg}^{5}\left(\bm{\overline{x}},\,\bm{\overline{z}},\,\bm{\overline{z}}^{\prime},\,\bm{x},\,\bm{z},\,\bm{z}^{\prime}\right)\left[\Theta_{1}\,S_{qggqgg}^{(2)}\left(\overline{\bm{x}},\,\overline{\bm{z}},\,\overline{\bm{z}}^{\prime},\,\bm{x},\,\bm{z},\,\bm{z}^{\prime}\right)\,-\,S_{qggqg}^{(2)}\left(\overline{\bm{x}},\,\overline{\bm{z}},\,\overline{\bm{z}}^{\prime},\,\bm{x},\,\bm{y}\right)\right.\\
&\left.-\,\Theta_{1}\,S_{qqgg}^{(2)}\left(\overline{\bm{w}},\,\bm{x},\,\bm{z},\,\bm{z}^{\prime}\right)\,+\,\Theta_{2}\,S_{qqgg}^{(2)}\left(\bm{\overline{x}},\,\bm{w},\,\bm{\overline{z}},\,\bm{\overline{z}}^{\prime}\right)\,+\,S_{qqg}\left(\overline{\bm{w}},\,\bm{x},\,\bm{y}\right)\,-\,\Theta_{2}\,\mathcal{S}\left(\overline{\bm{w}},\,\bm{w}\right)\right]\\
&+\,K_{qgg}^{5}\left(\bm{x},\,\bm{z},\,\bm{z}^{\prime},\,\bm{\overline{x}},\,\bm{\overline{z}},\,\bm{\overline{z}}^{\prime}\right)\left[\overline{\Theta}_{1}\,S_{qggqgg}^{(2)}\left(\overline{\bm{x}},\,\overline{\bm{z}},\,\overline{\bm{z}}^{\prime},\,\bm{x},\,\bm{z},\,\bm{z}^{\prime}\right)\,-\,S_{qgqgg}^{(2)}\left(\overline{\bm{x}},\,\overline{\bm{y}},\,\bm{x},\,\bm{z},\,\bm{z}^{\prime}\right)\right.\\
&\left.+\,\overline{\Theta}_{2}\,S_{qqgg}^{(2)}\left(\overline{\bm{w}},\,\bm{x},\,\bm{z},\,\bm{z}^{\prime}\right)\,-\,\overline{\Theta}_{1}\,S_{qqgg}^{(2)}\left(\bm{\overline{x}},\,\bm{w},\,\bm{\overline{z}},\,\bm{\overline{z}}^{\prime}\right)\,+\,S_{qqg}\left(\overline{\bm{x}},\,\bm{w},\,\overline{\bm{y}}\right)\,-\,\overline{\Theta}_{2}\,\mathcal{S}\left(\overline{\bm{w}},\,\bm{w}\right)\right]\\
&+\,K_{qgg}^{6}\left(\bm{\overline{x}},\,\bm{\overline{z}},\,\bm{\overline{z}}^{\prime},\,\bm{x},\,\bm{z},\,\bm{z}^{\prime}\right)\left[S_{qggqgg}^{(2)}\left(\bm{\overline{x}},\,\overline{\bm{z}},\,\overline{\bm{z}}^{\prime},\,\bm{x},\,\bm{z},\,\bm{z}^{\prime}\right)\,-\,S_{qqgg}^{(2)}\left(\overline{\bm{w}},\,\bm{x},\,\bm{z},\,\bm{z}^{\prime}\right)\right.\\
&\left.\left.-\,S_{qqgg}^{(2)}\left(\bm{\overline{x}},\,\bm{w},\,\bm{\overline{z}},\,\bm{\overline{z}}^{\prime}\right)\,+\,\mathcal{S}\left(\bm{\overline{w}},\,\bm{w}\right)\right]\right\} \,+\,\left(q_{2}^{+}\leftrightarrow q_{3}^{+},\:\bm{q}_{2}\leftrightarrow\bm{q}_{3}\right).
  \end{split}\end{equation}
  In this equation, $\bm{x},\,\bm{\bm{z}},\,\bm{\bm{z}}^{\prime}$ denote the transverse coordinates of the final quark and of the two final gluons in the DA,  while $\bm{y}$ and $\bm{w}$ refer to the initial quark and the intermediate gluon, respectively. The corresponding coordinates in the CCA are indicated with a bar. We have
   \begin{align}
\bm{y}&\,\equiv\,\xi\bm{z}+(1-\xi)\bm{z}^{\prime},\qquad  \bm{w}\,\equiv \,(1-\vartheta)\bm{x}\,+\,\xi\vartheta\bm{z}^{\prime}\,+\,(1-\xi)\vartheta\bm{z},\nn
 \overline{\bm{y}}&\,\equiv\,\xi\overline{\bm{z}}+(1-\xi)\overline{\bm{z}}^{\prime},\qquad
 \overline{\bm{w}}\,\equiv\,(1-\vartheta)\overline{\bm{x}}\,+\,\xi\vartheta\overline{\bm{z}}^{\prime}\,+\,(1-\xi)\vartheta\overline{\bm{z}}.
\end{align}

   \begin{figure}[!h]
    \includegraphics[scale=1]{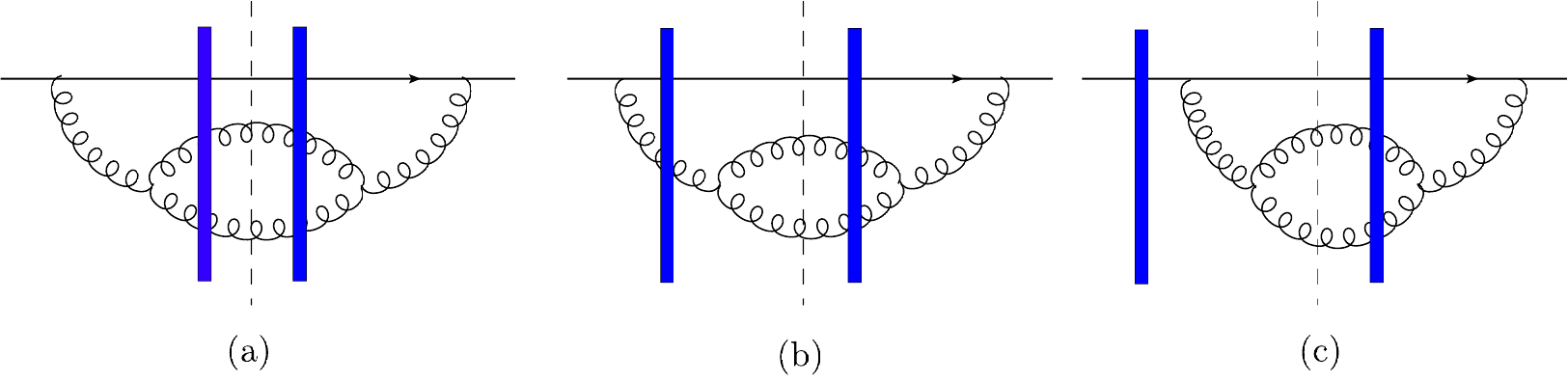}
  \caption{Three examples of diagrams for 2-gluon production which involve a propagating intermediate gluon in both the DA and the CCA. There are nine such graphs, each of them proportional with the kernel $K_{qgg}^{4}$. The particular diagrams shown here are associated with the following $S$--matrices:  $S_{q\overline{q}q\overline{q}q\overline{q}}$ for $(a)$, $S_{q\overline{q}q\overline{q}g}^{(2)}$ for $(b)$, and $S_{q\overline{q}q\overline{q}}^{(1)}$ for $(c)$.\label{qgg2fig1}}
\end{figure}

  \begin{figure}[!h]
    \includegraphics[scale=1]{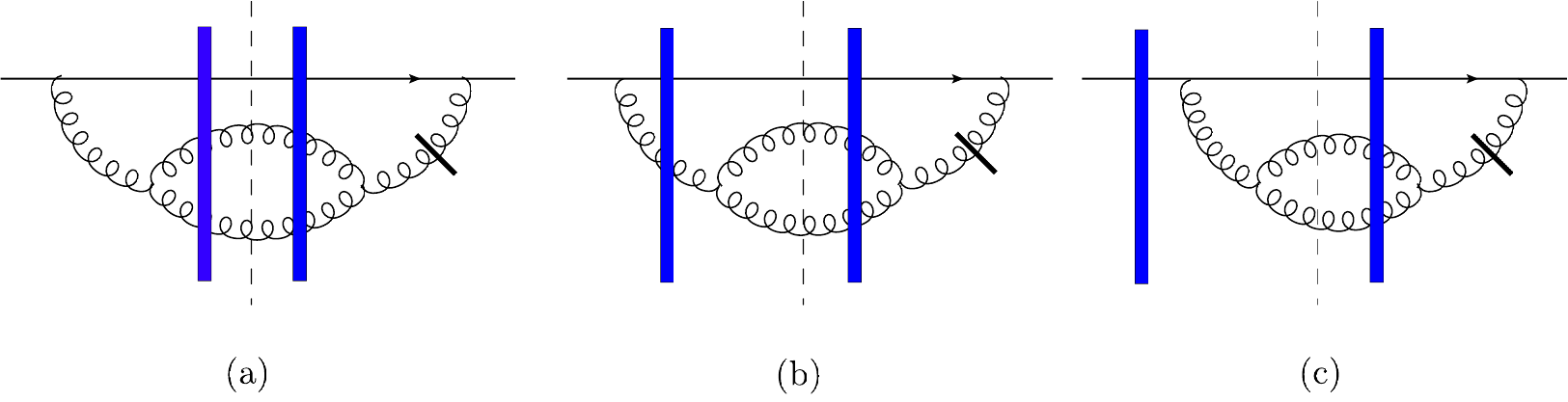}
  \caption{Three examples of diagrams for 2-gluon production which involve a propagating intermediate gluon in the DA and an instantaneous one in the CCA. There are 12 such diagrams altogether (after permuting DA and CCA), all multiplying the kernel $K_{qgg}^{5}$.\label{qgg2fig2}}
\end{figure}

\begin{figure}[!h]\center
    \includegraphics[scale=0.75]{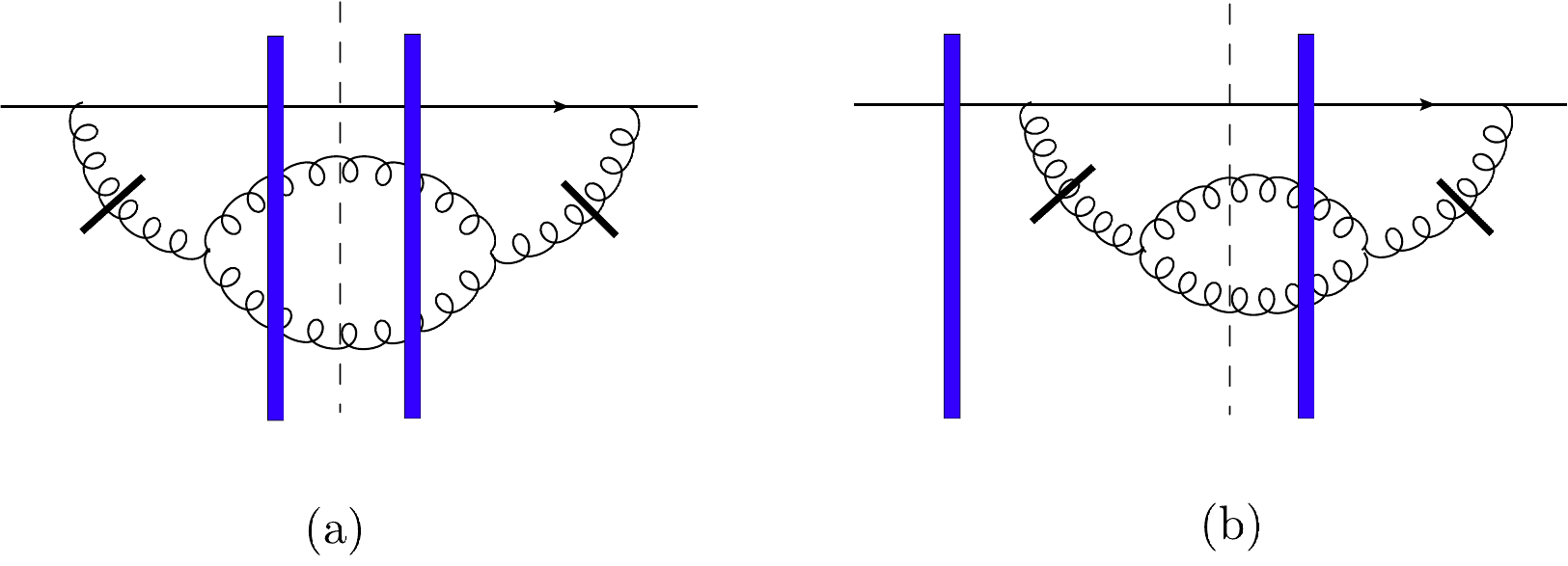}
  \caption{Two diagrams 2-gluon production which involve an instantaneous gluon exchange in both the DA and the CCA. There are 4 such graphs, all proportional with the kernel $K_{qgg}^{6}$.\label{qgg2fig3}}
\end{figure}

The longitudinal momentum fractions take the following values
\begin{equation}
 \vartheta\,=\,\frac{q_{2}^{+}+q_{3}^{+}}{q^{+}},\qquad\qquad\xi\,=\,\frac{q_{2}^{+}}{q_{2}^{+}+q_{3}^{+}}.
  \end{equation}
The three new kernels which appear in  Eq.~(\ref{gpar2}) are defined as (with $\bm{Z}\equiv\bm{z}-\bm{z}^{\prime}$, $\bm{X}\equiv\bm{x}-\bm{z}$ etc.)
     \begin{align}
&K_{qgg}^{4}\left(\bm{\overline{x}},\,\bm{\overline{z}},\,\bm{\overline{z}}^{\prime},\,\bm{x},\,\bm{z},\,\bm{z}^{\prime}\right)\,\equiv\,\frac{\left(\overline{\bm{X}}^{m}+\xi\overline{\bm{Z}}^{m}\right)\left(\bm{X}^{s}+\xi\bm{Z}^{s}\right)}{\overline{\bm{Z}}^{2}\,\left(\overline{\bm{X}}+\xi\overline{\bm{Z}}\right)^{2}\,\bm{Z}^{2}\,\left(\bm{X}+\xi\bm{Z}\right)^{2}}\nn
&\times\left[2\xi(1-\xi)\left((\vartheta-2)^{2}\bm{\overline{Z}}^{m}\bm{Z}^{s}-\bm{\overline{Z}}^{s}\bm{Z}^{m}+\delta^{ms}\,\bm{\overline{Z}}\cdot\bm{Z}\right)\,+\,\left((\vartheta-2)^{2}+1\right)\,\left(\frac{1-\xi}{\xi}+\frac{\xi}{1-\xi}\right)\,\delta^{ms}\,\bm{\overline{Z}}\cdot\bm{Z}\right],
\end{align}
   \begin{equation}\begin{split}
&K_{qgg}^{5}\left(\bm{\overline{x}},\,\bm{\overline{z}},\,\bm{\overline{z}}^{\prime},\,\bm{x},\,\bm{z},\,\bm{z}^{\prime}\right)\equiv\,-\frac{(2-\vartheta)(2\xi(1-\xi)-1)(1-\vartheta)(1-2\xi\vartheta)\,\left(\bm{X}\cdot\bm{Z}\,+\,\xi\bm{Z}^{2}\right)}{\left(\xi(1-\xi)\overline{\bm{Z}}^{2}\,+\,(1-\vartheta)\left(\overline{\bm{X}}+\xi\overline{\bm{Z}}\right)^{2}\right)\,\bm{Z}^{2}\,\left(\bm{X}+\xi\bm{Z}\right)^{2}},
\end{split}\end{equation}
  \begin{equation}\begin{split}
&K_{qgg}^{6}\left(\bm{\overline{x}},\,\bm{\overline{z}},\,\bm{\overline{z}}^{\prime},\,\bm{x},\,\bm{z},\,\bm{z}^{\prime}\right)\\
&\equiv\,\frac{(1-2\xi\vartheta)^{2}\,(1-\vartheta)^{2}\,\xi(1-\xi)}{\left(\xi(1-\xi)\overline{\bm{Z}}^{2}\,+\,(1-\vartheta)\left(\overline{\bm{X}}+\xi\overline{\bm{Z}}\right)^{2}\right)\,\left(\xi(1-\xi)\bm{Z}^{2}\,+\,(1-\vartheta)\left(\bm{X}+\xi\bm{Z}\right)^{2}\right)}.
\end{split}\end{equation}
The explicit expressions for the $S$-matrices in terms of Wilson lines can be found in Eqs.~(\ref{wils5})--(\ref{wils6}), together with their simplified versions at large $N_c$.

In Figs.~\ref{qgg2fig1}, \ref{qgg2fig2} and \ref{qgg2fig3}, we exhibit examples of Feynman graphs which illustrate the various types of terms appearing in  Eq.~(\ref{gpar2}).

 \subsubsection{Interference contributions\label{qggtrijab}}
  
\begin{figure}[!h]
    \includegraphics[scale=0.55]{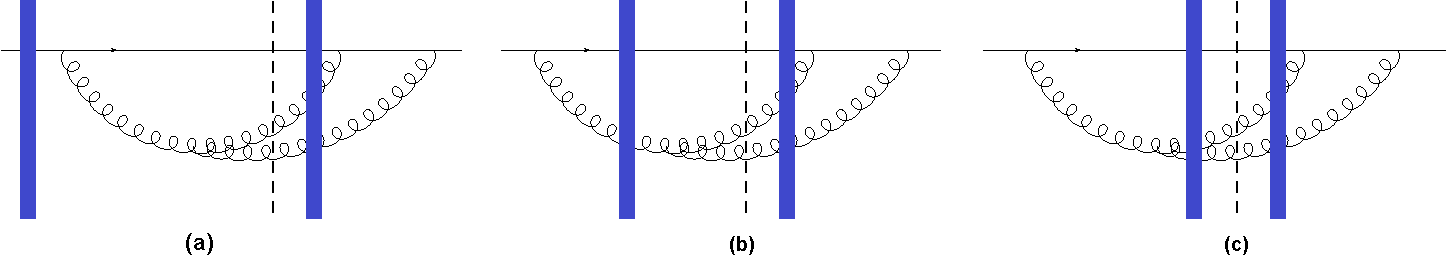}
  \caption{Three examples of interference diagrams in which both the intermediate gluon and the intermediate quark are propagating fields. There are nine such diagrams and they yield contributions proportional with the kernel $K_{qgg}^{7}$.\label{interfig1}}
\end{figure}

\begin{figure}[!h]\center
    \includegraphics[scale=0.55]{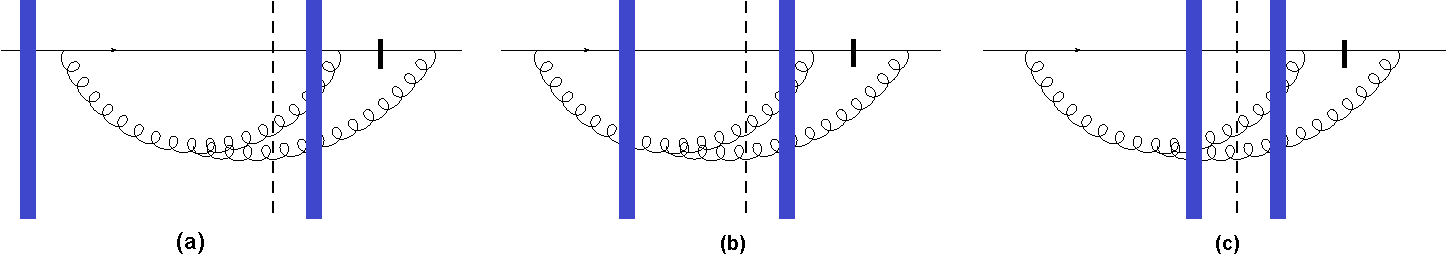}
  \caption{Three examples of interference diagrams in which the intermediate gluon is a propagating field, whereas the  intermediate quark is an instantaneous exchange. There are six such contributions, all proportional to the kernel $K_{qgg}^{8}$.\label{interfig3}}
\end{figure}

\begin{figure}[!h]\center
    \includegraphics[scale=0.57]{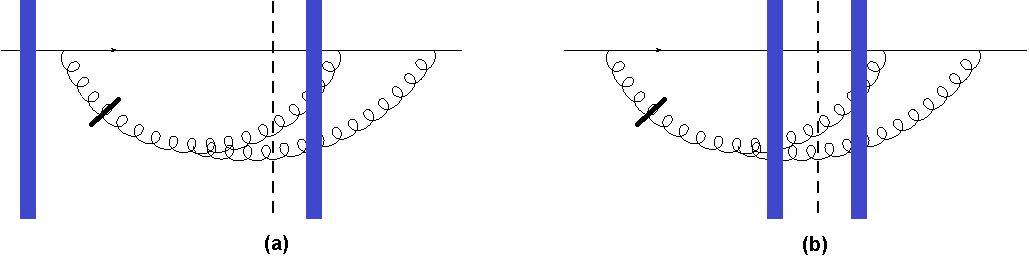}
  \caption{Interference diagrams in which the intermediate gluon is an instantaneous exchange, whereas the  intermediate quark is is a propagating field. There are six such contributions, all proportional to the kernel $K_{qgg}^{9}$.\label{interfig2}}
\end{figure}

\begin{figure}[!h]\center
    \includegraphics[scale=0.57]{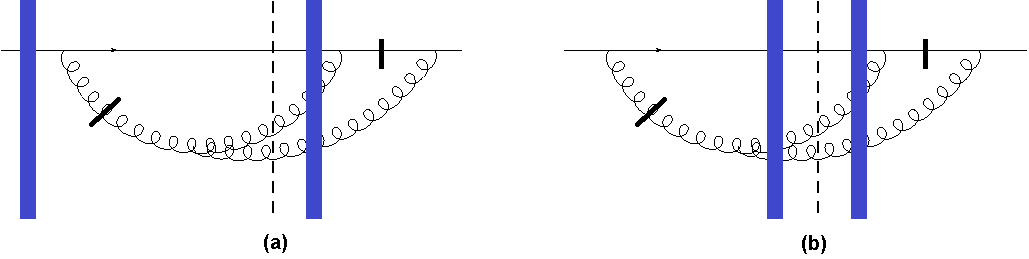}
  \caption{Two among the four graphs in which both the intermediate gluon and the intermediate quark are 
  instantaneous exchanges. All such contributions are proportional to the kernel $K_{qgg}^{10}$.\label{interfig4}}
\end{figure}

 The final contributions we have to consider are those which express the interference between graphs with an intermediate gluon and those with an intermediate quark. Let us first some examples of such production diagrams. In Fig.~\ref{interfig1}, we show diagrams in which both the intermediate gluon and the intermediate quark are propagating fields.  Figs.~\ref{interfig2} and \ref{interfig3} show examples where one of the exchange is instantaneous; finally,  in the examples included in Fig.~\ref{interfig4}, both exchanges are instantaneous.

 All the ingredients that are needed to evaluate such graphs have already been mentioned in relation with the previous calculations; so, here we only show the final result for the interference contribution to the cross-section for $qgg$ production; this reads
  \begin{equation}\begin{split}\label{res.ab}
&\frac{d\sigma_{QG}^{qA\rightarrow qgg+X}}{d^{3}q_{1}\,d^{3}q_{2}\,d^{3}q_{3}}\,\equiv\,\frac{\alpha_{s}^{2}\,C_{F}\,N_{c}}{(2\pi)^{10}(q^{+})^{2}}\,\delta(q^{+}-q_{1}^{+}-q_{2}^{+}-q_{3}^{+})\,\mathrm{Re}\int_{\bm{\overline{x}},\,\bm{\overline{\bm{z}}},\,\bm{\overline{\bm{z}}}^{\prime},\,\bm{x},\,\bm{\bm{z}},\,\bm{\bm{z}}^{\prime}}\:e^{-i\bm{q}_{1}\cdot(\bm{x}-\bm{\overline{x}})-i\bm{q}_{2}\cdot(\bm{\bm{z}}-\bm{\overline{\bm{z}}})-i\bm{q}_{3}\cdot(\bm{\bm{z}}^{\prime}-\bm{\overline{\bm{z}}}^{\prime})}\\
&\times\left\{ K_{qgg}^{7}\left(\bm{\overline{x}},\,\bm{\overline{z}},\,\bm{\overline{z}}^{\prime},\,\bm{x},\,\bm{z},\,\bm{z}^{\prime}\right)\left[\overline{\Theta}_{3}\,\Theta_{1}\,S_{qggqgg}^{(3)}\left(\overline{\bm{x}},\,\overline{\bm{z}},\,\overline{\bm{z}}^{\prime},\,\bm{x},\,\bm{z},\,\bm{z}^{\prime}\right)\,-\,\overline{\Theta}_{3}\,S_{qggqg}^{(3)}\left(\overline{\bm{x}},\,\overline{\bm{z}},\,\overline{\bm{z}}^{\prime},\,\bm{x},\,\bm{y}\right)\right.\right.\\
&-\,\Theta_{1}\,S_{qgqgg}^{(3)}\left(\overline{\bm{y}},\,\overline{\bm{z}}^{\prime},\,\bm{x},\,\bm{z},\,\bm{z}^{\prime}\right)\,+\,\overline{\Theta}_{3}\,\Theta_{2}\,S_{qggq}\left(\overline{\bm{x}},\,\overline{\bm{z}},\,\overline{\bm{z}}^{\prime},\,\bm{w}\right)\,+\,\overline{\Theta}_{4}\,\Theta_{1}\,S_{qqgg}^{(3)}\left(\overline{\bm{w}},\,\bm{x},\,\bm{z},\,\bm{z}^{\prime}\right)\\
&\left.+\,S_{qggq}\left(\overline{\bm{y}},\,\overline{\bm{z}}^{\prime},\,\bm{y},\,\bm{x}\right)\,-\,\Theta_{2}\,S_{qqg}\left(\overline{\bm{y}},\,\bm{w},\,\overline{\bm{z}}^{\prime}\right)\,-\,\overline{\Theta}_{4}\,S_{qqg}\left(\overline{\bm{w}},\,\bm{y},\,\bm{x}\right)\,+\,\overline{\Theta}_{4}\,\Theta_{2}\,\mathcal{S}\left(\overline{\bm{w}},\,\bm{w}\right)\right]\\
&+\,K_{qgg}^{8}\left(\bm{\overline{x}},\,\bm{\overline{z}},\,\bm{\overline{z}}^{\prime},\,\bm{x},\,\bm{z},\,\bm{z}^{\prime}\right)\left[\Theta_{1}\,S_{qggqgg}^{(3)}\left(\overline{\bm{x}},\,\overline{\bm{z}},\,\overline{\bm{z}}^{\prime},\,\bm{x},\,\bm{z},\,\bm{z}^{\prime}\right)\,-\,S_{qggqg}^{(3)}\left(\overline{\bm{x}},\,\overline{\bm{z}},\,\overline{\bm{z}}^{\prime},\,\bm{x},\,\bm{y}\right)\right.\\
&\left.+\,\Theta_{2}\,S_{qggq}\left(\overline{\bm{x}},\,\overline{\bm{z}},\,\overline{\bm{z}}^{\prime},\,\bm{w}\right)\,-\,\Theta_{1}\,S_{qqgg}^{(3)}\left(\overline{\bm{w}},\,\bm{x},\,\bm{z},\,\bm{z}^{\prime}\right)\,+\,S_{qqg}\left(\overline{\bm{w}},\,\bm{y},\,\bm{x}\right)\,-\,\Theta_{2}\,\mathcal{S}\left(\overline{\bm{w}},\,\bm{w}\right)\right]\\
&+\,K_{qgg}^{9}\left(\bm{\overline{x}},\,\bm{\overline{z}},\,\bm{\overline{z}}^{\prime},\,\bm{x},\,\bm{z},\,\bm{z}^{\prime}\right)\left[\overline{\Theta}_{3}\,S_{qggqgg}^{(3)}\left(\overline{\bm{x}},\,\overline{\bm{z}},\,\overline{\bm{z}}^{\prime},\,\bm{x},\,\bm{z},\,\bm{z}^{\prime}\right)\,-\,S_{qgqgg}^{(3)}\left(\overline{\bm{y}},\,\overline{\bm{z}}^{\prime},\,\bm{x},\,\bm{z},\,\bm{z}^{\prime}\right)\right.\\
&\left.-\,\overline{\Theta}_{3}\,S_{qggq}\left(\overline{\bm{x}},\,\overline{\bm{z}},\,\overline{\bm{z}}^{\prime},\,\bm{w}\right)\,+\,\overline{\Theta}_{4}\,S_{qqgg}^{(3)}\left(\overline{\bm{w}},\,\bm{x},\,\bm{z},\,\bm{z}^{\prime}\right)\,+\,S_{qqg}\left(\overline{\bm{y}},\,\bm{w},\,\overline{\bm{z}}^{\prime}\right)\,-\,\overline{\Theta}_{4}\,\mathcal{S}\left(\overline{\bm{w}},\,\bm{w}\right)\right]\\
&+\,K_{qgg}^{10}\left(\bm{\overline{x}},\,\bm{\overline{z}},\,\bm{\overline{z}}^{\prime},\,\bm{x},\,\bm{z},\,\bm{z}^{\prime}\right)\left[S_{qggqgg}^{(3)}\left(\bm{\overline{x}},\,\overline{\bm{z}},\,\overline{\bm{z}}^{\prime},\,\bm{x},\,\bm{z},\,\bm{z}^{\prime}\right)\,-\,S_{qqgg}^{(3)}\left(\bm{\overline{w}},\,\bm{x},\,\bm{z},\,\bm{z}^{\prime}\right)\right.\\
&\left.\left.-\,S_{qggq}\left(\bm{\overline{x}},\,\overline{\bm{z}},\,\overline{\bm{z}}^{\prime},\,\bm{w}\right)\,+\,\mathcal{S}\left(\bm{\overline{w}},\,\bm{w}\right)\right]\right\} \,+\,\left(q_{2}^{+}\leftrightarrow q_{3}^{+},\:\bm{q}_{2}\leftrightarrow\bm{q}_{3}\right).
 \end{split}\end{equation}
 Notice the ``real part'' (Re) sign in front of the integral in  \eqn{res.ab}: by taking twice the real part of diagrams like those illustrated in Figs.~\ref{interfig1},  \ref{interfig2}, \ref{interfig3}  and \ref{interfig4}, we have also taken into account the corresponding diagrams in which the positions of the intermediate gluon and of the intermediate quark are interchanged.

 In  \eqn{res.ab} we introduced the following new kernels:
    \begin{equation}\label{Kqgg7}\begin{split}
&K_{qgg}^{7}\left(\bm{\overline{x}},\,\bm{\overline{z}},\,\bm{\overline{z}}^{\prime},\,\bm{x},\,\bm{z},\,\bm{z}^{\prime}\right)\,\equiv\,-\,\frac{\sqrt{(1-\overline{\vartheta})\xi(1-\xi)}\,\phi_{\lambda_{2}\lambda}^{in\dagger}(\overline{\vartheta})\,\varphi_{\lambda_{1}\lambda_{2}}^{jm\dagger}(\overline{\xi})\,\phi_{\lambda_{1}\lambda}^{sr}(\vartheta)}{2\overline{\vartheta}\,\sqrt{\overline{\xi}}\,\bm{\overline{X}}^{2}\,\left(\xi\overline{\bm{X}}-\bm{\overline{X}^{\prime}}\right)^{2}\,\bm{Z}^{2}\,\left(\bm{X}+\xi\bm{Z}\right)^{2}}\\
&\times\left(\overline{\xi}\,\overline{\bm{X}}-\bm{\overline{X}^{\prime}}\right)^{n}\,\overline{\bm{X}}^{m}\,\left(\bm{X}^{r}+\xi\,\bm{Z}^{r}\right)\left(\bm{Z}^{s}\delta_{ji}\,-\,\frac{1}{\xi}\bm{Z}^{j}\delta_{si}\,-\,\frac{1}{1-\xi}\bm{Z}^{i}\delta_{sj}\right),
  \end{split}\end{equation}
  \begin{equation}\begin{split}
&K_{qgg}^{8}\left(\bm{\overline{x}},\,\bm{\overline{z}},\,\bm{\overline{z}}^{\prime},\,\bm{x},\,\bm{z},\,\bm{z}^{\prime}\right)\\
&\equiv\,-\frac{\sqrt{(1-\overline{\vartheta})\xi(1-\xi)}\,(1-2\xi\vartheta)(1-\vartheta)\,\phi_{\lambda\lambda_{1}}^{im\dagger}(\overline{\xi})\,\phi_{\lambda_{1}\lambda}^{il\dagger}(\overline{\vartheta})\,\left(\overline{\xi}\bm{\overline{X}}-\bm{\overline{X}^{\prime}}\right)^{l}\,\bm{\overline{X}}^{m}}{2\overline{\vartheta}\,\sqrt{\overline{\xi}}\,\overline{\bm{X}}^{2}\,\left(\xi\overline{\bm{X}}-\bm{\overline{X}^{\prime}}\right)^{2}\,\left(\xi(1-\xi)\bm{Z}^{2}\,+\,(1-\vartheta)\left(\bm{X}+\xi\bm{Z}\right)^{2}\right)},
 \end{split}\end{equation}
  \begin{equation}\begin{split}
&K_{qgg}^{9}\left(\bm{\overline{x}},\,\bm{\overline{z}},\,\bm{\overline{z}}^{\prime},\,\bm{x},\,\bm{z},\,\bm{z}^{\prime}\right)\,\equiv\,\frac{2\sqrt{\overline{\xi}\,\overline{\vartheta}\xi(1-\xi)}(1-\overline{\xi})\,\chi_{\lambda}^{\dagger}\left(\delta^{lj}+i\varepsilon^{lj}\sigma^{3}\right)\chi_{\lambda_{1}}\phi_{\lambda_{1}\lambda}^{im}(\vartheta)}{\sqrt{1-\overline{\vartheta}}\,\left(\overline{\vartheta}\left(\xi\overline{\bm{X}}-\bm{\overline{X}^{\prime}}\right)^{2}+\overline{\xi}(1-\overline{\xi})\bm{\overline{\bm{X}}}^{2}\right)\,\bm{Z}^{2}\,\left(\bm{X}+\xi\bm{Z}\right)^{2}}\\
&\times\left(\bm{X}^{m}+\xi\bm{Z}^{m}\right)\,\left(\bm{Z}^{i}\delta_{jl}\,-\,\frac{1}{\xi}\bm{Z}^{j}\delta_{il}\,-\,\frac{1}{1-\xi}\bm{Z}^{l}\delta_{ij}\right),\\
 \end{split}\end{equation}
 \begin{equation}\begin{split}
&K_{qgg}^{10}\left(\bm{\overline{x}},\,\bm{\overline{z}},\,\bm{\overline{z}}^{\prime},\,\bm{x},\,\bm{z},\,\bm{z}^{\prime}\right)\\
&\equiv\,\frac{8\sqrt{\overline{\xi}\,\overline{\vartheta}\xi(1-\xi)}(1-\overline{\xi})(1-2\xi\vartheta)(1-\vartheta)}{\sqrt{1-\overline{\vartheta}}\,\left(\vartheta\left(\xi\overline{\bm{X}}-\bm{\overline{X}^{\prime}}\right)^{2}+\xi(1-\xi)\bm{\overline{\bm{X}}}^{2}\right)\,\left(\xi(1-\xi)\bm{Z}^{2}\,+\,(1-\vartheta)\left(\bm{X}+\xi\bm{Z}\right)^{2}\right)}.
 \end{split}\end{equation}
The sums over polarizations and helicities in \eqn{Kqgg7} can be explicitly performed as follows,
  \begin{equation}\begin{split}
&\phi_{\lambda_{2}\lambda}^{in\dagger}(\overline{\vartheta})\,\varphi_{\lambda_{1}\lambda_{2}}^{jm\dagger}(\overline{\xi})\,\phi_{\lambda_{1}\lambda}^{sr}(\vartheta)\\
&=2\left((2-\overline{\vartheta})(2-\vartheta)(2\xi-1)\delta^{in}\delta^{jm}\delta^{sr}-\vartheta(2-\overline{\vartheta})\delta^{in}\varepsilon^{jm}\varepsilon^{sr}+\vartheta\overline{\vartheta}(2\overline{\xi}-1)\delta^{jm}\varepsilon^{in}\varepsilon^{sr}+(2-\vartheta)\overline{\vartheta}\delta^{sr}\varepsilon^{in}\varepsilon^{jm}\right),
  \end{split}\end{equation}
but the ensuing result is not specially illuminating.

  The longitudinal momentum fractions take the following values:
 \begin{equation}
\overline{\vartheta}\,=\,\frac{q_{2}^{+}+q_{3}^{+}}{q^{+}},\qquad\vartheta\,=\,\frac{q_{3}^{+}}{q^{+}},\qquad\overline{\xi}\,=\,\frac{q_{2}^{+}}{q_{1}^{+}+q_{2}^{+}},\qquad\xi\,=\,\frac{q_{2}^{+}}{q_{2}^{+}+q_{3}^{+}}.
 \end{equation}
 Finally, the new scattering operators appearing in \eqn{res.ab} are defined in terms of Wilson lines in Eqs.~(\ref{wils7})--(\ref{wils8}). Once again, they drastically simplify in the large $N_c$ limit, where they all reduce to products of dipoles and quadrupoles, as expected.

\section{Conclusions and perspectives}
\label{Conc}

In this paper we have presented the first calculation of the cross-section for the production of three colored partons (``jets'')  at forward rapidities in proton-nucleus collisions, to leading order in perturbative QCD. We have thus generalized recent calculations in the literature which focused on three-particle finale states as well, but in somewhat simpler contexts  \cite{Ayala:2016lhd,Boussarie:2016ogo,Altinoluk:2018uax},  where there are considerably less topologies for the Feynman graphs contributing to the final state.

For simplicity, we have considered the quark channel alone, that is, we have only included the processes initiated by a quark that was originally collinear with the proton. The main ingredient of our calculation is the light-cone wave function of this quark and, more precisely, its three-parton Fock space components. These components are generated via two successive (initial-state and/or final-state) parton branchings followed --- for the incoming quark and also for the partons produced via initial-state evolution --- by multiple scattering off the gluon distribution in the target, as computed in the eikonal approximation.

The relevant Fock space components of the LCWF have been explicitly constructed in transverse coordinate space: they are built with Weisz{\"a}cker-Williams kernels for the gluon emissions and Wilson lines for the partons which participate in the scattering. Correspondingly, our final results for the cross-sections for three-parton production are expressed as Fourier transforms from coordinate space to transverse momentum space. But albeit the physical picture is fully explicit (in particular, the structure of the microscopic process can be directly read from the formulae), it is fair to say that our current results are still far away from direct applications to the phenomenology.

First, our cross-sections are written as multiple convolutions in coordinate space (six convolutions, more precisely; see e.g. \eqn{qqcr}), which, besides the emission kernels, also involve $S$-matrices for the elastic scattering of  colorless systems built with up to 6 partons. In principle, all these $S$-matrices can be computed (including their high-energy evolution) by numerically solving the JIMWLK equation \cite{Blaizot:2002xy,Rummukainen:2003ns,Lappi:2011ju}, but this is tedious in practice for the multi-point correlations \cite{Dumitru:2011vk}. As already stressed, the calculations simplify considerably in the multi-color limit $N_c\to\infty$, where all the $S$-matrices reduce to products of dipoles and quadrupoles (see e.g. \eqn{qqqNc}). Still, the evolution equation obeyed by the quadrupole remains complicated even at large $N_c$ \cite{Dominguez:2011gc} and has not yet been solved numerically. A drastic simplification occurs only within the Gaussian approximation to the CGC weight function, which allows for explicit, analytic, results for both the dipole and the quadrupole \cite{Blaizot:2004wv,Dominguez:2011wm,Iancu:2011ns,Iancu:2011nj,Lappi:2012nh}. In that case, the only remaining challenge is the numerical evaluation of the multiple transverse integrations.

Yet, the recent experience with multi-particle production in the CGC formalism demonstrates that there are particular correlations and kinematical limits which are easier to compute, while also being interesting for the phenomenology. In particular, following previous studies in \cite{Dominguez:2011wm,Iancu:2013dta,vanHameren:2016ftb,Ayala:2016lhd,Altinoluk:2018uax}, it would  be interesting to consider the ``correlation limit'', in which the sum of the transverse momenta of the produced particles is much smaller than their individual momenta. This is necessarily the case whenever the individual momenta are large compared to the saturation momentum in the nuclear target. In that case the broadening of the three-particle distribution in the total transverse momentum ${\bm Q}\equiv \bq_1+\bq_2+\bq_3$ around the collinear limit ${\bm Q}=0$ gives us a measure of the nuclear effects, that can be studied via the gradient expansion of the original $S$-matrices. Such an expansion also permits to make contact with Transverse Momentum-dependent Distributions (TMD) characterizing the gluon distribution in the nucleus
 \cite{Dominguez:2011wm,Iancu:2013dta,vanHameren:2016ftb,Ayala:2016lhd,Altinoluk:2018uax}. The calculations are expected to become more tractable for symmetric configurations like the ``Mercedes star'' \cite{Ayala:2016lhd,Altinoluk:2018uax}. We plan to perform similar investigations in the future, on the basis of our current results.
 
Still in view of applications to the phenomenology, one should keep in mind that there are other processes which contribute to the three-parton final states, already at leading order in pQCD --- albeit one can argue that the quark channel that was considered here gives the dominant contribution in the forward kinematics of interest. Clearly, there is also a gluon channel, in which all the processes are initiated by a gluon originally collinear with the proton. We have briefly considered this channel in Sect.~\ref{locros}, but only in relation with the two-parton final states.
The corresponding calculations for a three-parton final state are similar to those for the quark channel, but the results should look even more cumbersome, due to wave function graphs which involve two three-gluon ($g\to gg$) vertices.

There are furthermore contributions associated with the independent scattering of several (two or three) partons that were originally collinear with the incoming proton (see e.g. \cite{Kovner:2017ssr} for a recent analysis). For instance, a $qgg$ final state can also be produced if three collinear partons --- a quark and two gluons --- scatter off the nucleus and thus emerge in the final state (with transverse momenta which are necessarily comparable with the nuclear saturation momentum); the respective contribution to the cross-section is proportional to a triple parton collinear distribution in the proton. Alternatively, the same final state can be achieved if two collinear partons, a quark and a gluon, scatter off the nucleus and one of them radiates a gluon in the final state;  this channel involves a two-parton (quark-gluon) collinear distribution. Such processes are formally of lower order in pQCD, as they involve less explicit emission vertices, but on the other hand they are proportional to multi-parton collinear distributions, which are suppressed when $x_p$ (the longitudinal fraction taken from the proton) is not that small. Interestingly, such independent-scattering processes can also give access to final states, like $qqq$ or $qqg$, that cannot be reached if one starts with a single parton.

At this point, we should recall that yet another main motivation for computing three-parton production is the calculation of the di-jet cross-section at NLO: by integrating out one (any) of the three produced partons, one deduces one-loop, `real', corrections to the impact factor for the production of two partons (here, in the quark channel). Of course, the complete respective calculation must also include `virtual' corrections to the impact factor, i.e. one loop corrections to the quark LCWF itself. This is another direction of research that we plan to pursue.

\bigskip
\noindent
\textbf{Acknowledgments}

We would like to thank M. Lublinsky, C. Marquet, and A. H. Mueller for useful discussions. The work of Y.M. was in part supported by the 2016-2017 Chateaubriand fellowship of the French embassy in Israel. This work is supported in part by the Agence Nationale de la Recherche project  ANR-16-CE31-0019-01.
\appendix

\section{The light-cone QCD Hamiltonian\label{hamilt}}
Our starting point is  QCD Hamiltonian in  light-cone gauge \cite{Venugopalan:1998zd}. In  light-cone coordinates,  four-vectors are $x^{\mu}=\left(x^{+},\, x^{-},\, \bm{x}\right)$, where $x^{+}\,\equiv\,\frac{1}{\sqrt{2}}\left(x^{0}\,+\,x^{3}\right)$ and $x^{-}\,\equiv\,\frac{1}{\sqrt{2}}\left(x^{0}\,-\,x^{3}\right)$ stand for longitudinal, while $\bm{x}=\left(x_{1},\, x_{2}\right)$ for transverse components.  In the light-cone gauge:
\begin{equation}
A^{+}_a\equiv\frac{1}{\sqrt{2}}\big(
A^{0}_a+A^{3}_a\big)=0.
\end{equation}
The QCD Hamiltonian is given by the following expression:

\begin{equation}\begin{split}\label{qcdhamii}
&H_{LC\; QCD}\\
&\quad =\int dx^{-}d^{2}\bm{x}\left(\frac{1}{2}\Pi^{a}(x^{-},\, \bm{x})\,\Pi^{a}(x^{-},\, \bm{x})\,+\,\frac{1}{4}F_{ij}^{a}(x^{-},\, \bm{x})\, F_{ij}^{a}(x^{-},\, \bm{x})\,+\, i\bar{\psi}\gamma^{+}D_{+}\psi\right),\\
  \end{split}\end{equation}
 where  the electric and magnetic pieces have the form:
 \begin{equation}\begin{split}
&\Pi^{a}(x^{-},\,\bm{x})\,\equiv\,-\,\frac{1}{\partial^{+}}(D_{i}^{ab}\partial^{+}A_{i}^{b}\,-\,2g\psi_{+}^{\dagger}t^{a}\psi_{+}),\\
&F_{ij}^{a}(x^{-},\,\bm{x})\,\equiv\,\partial_{i}A_{j}^{a}\,-\,\partial_{i}A_{j}^{a}\,-\, gf^{abc}A_{i}^{b}A_{j}^{c}.\\
\end{split}\end{equation}
$\psi$ denotes  Dirac's 4-component quark spinor\footnote{The flavour index is suppressed in this section.}, while
$\psi_+$ is a corresponding 2-component spinor build from the two non vanishing components of the projected vector $\frac{1}{2}\gamma^{0}\gamma_{+}\psi$. After substitution of $\Pi^{a}(x^{-},\, \bm{x})$ and $F_{ij}^{a}(x^{-},\, \bm{x})$ in $H_{LC\; QCD}$, the result can be written as  $H_{LC\; QCD}=H_{0}+H_{\rm int}$, with free Hamiltonian $H_{0}$ given by:
\begin{equation}\begin{split}\label{hzero}
&H_{0}\,\equiv\,\int dx^{-}\, d^{2}\bm{x}\,\left(\frac{1}{2}(\partial_{i}A_{j}^{a})^{2}\,+\, i\psi_{+}^{\dagger}\frac{\partial_{i}\partial_{i}}{\partial^{+}}\psi_{+}\right).\\
  \end{split}\end{equation}
The interaction Hamiltonian $H_{\rm int}$ reads
\begin{eqnarray}\label{hint}
H_{\rm int}&&\equiv\,\int dx^{-}\, d^{2}\bm{x}\,\left(-gf^{abc}A_{i}^{b}A_{j}^{c}\partial_{i}A_{j}^{a}\,+\,\frac{g^{2}}{4}f^{abc}f^{ade}A_{i}^{b}A_{j}^{c}A_{i}^{d}A_{j}^{e}\,\right.\nonumber\\
&&-\, gf^{abc}(\partial_{i}A_{i}^{a})\frac{1}{\partial^{+}}(A_{j}^{b}\partial^{+}A_{j}^{c})\,+\,\frac{g^{2}}{2}f^{abc}f^{ade}\frac{1}{\partial^{+}}(A_{i}^{b}\partial^{+}A_{i}^{c})\frac{1}{\partial^{+}}(A_{j}^{d}\partial^{+}A_{j}^{e})\\
&&\,+\,2g^{2}f^{abc}\frac{1}{\partial^{+}}(A_{i}^{b}\partial^{+}A_{i}^{c})\frac{1}{\partial^{+}}(\psi_{+}^{\dagger}t^{a}\psi_{+})+\,2g^{2}\frac{1}{\partial^{+}}(\psi_{+}^{\dagger}t^{a}\psi_{+})\frac{1}{\partial^{+}}(\psi_{+}^{\dagger}t^{a}\psi_{+})\nonumber\\
&&\,-\,2g(\partial_{i}A_{i}^{a})\frac{1}{\partial^{+}}(\psi_{+}^{\dagger}t^{a}\psi_{+})\,-\, g\psi_{+}^{\dagger}t^{a}(\sigma_{i}\partial_{i})\frac{1}{\partial^{+}}(\sigma_{j}A_{j}^{a}\psi_{+})-\, g\psi_{+}^{\dagger}t^{a}\sigma_{i}A_{i}^{a}\frac{1}{\partial^{+}}(\sigma_{j}\partial_{j}\psi_{+})\nonumber\\
&&\left.\,-\, ig^{2}\psi_{+}^{\dagger}t^{a}t^{b}\sigma_{i}A_{i}^{a}\frac{1}{\partial^{+}}(\sigma_{j}A_{j}^{b}\psi_{+})\right).\nonumber\end{eqnarray}
$\sigma^{i}$ are  Pauli matrices, the matrices $t^{a}$ denote the $SU(N_{c})$ gauge group generators in  fundamental representation (the gauge group generators in  adjoint representation will be denoted by $T^{a}$), which obey the algebra $\left[t^{a},t^{b}\right]=if^{abc}t^{c}$, where $f^{abc}$ are 
structure constants of the gauge group.

\section{Field quantization \label{fieldef}}

Quantisation of the fields is performed in  usual manner introducing creation/annihilation operators and imposing commutation (anti-commutation) relations among them.
For the gauge fields we can use the following expansion:
\begin{equation}\label{glufield}
A_{i}^{a}(x)=\int_{0}^{\infty}\frac{dk^{+}}{2\pi}\int\frac{d^{2}\bm{k}}{(2\pi)^{2}}\frac{1}{\sqrt{2k^{+}}}\left(a_{i}^{a}(k^{+},\bm{k})e^{-ik\cdot x}+a_{i}^{a\dagger}(k^{+},\bm{k})e^{ik\cdot x}\right).
  \end{equation}
The creation and annihilation operators obey the bosonic algebra:
\begin{equation}
\left[a_{i}^{a}(k^{+},\bm{k}),\, a_{j}^{b\dagger}(p^{+},\bm{p})\right]=(2\pi)^{3}\delta^{ab}\delta_{ij}\delta(k^{+}-p^{+})\delta^{(2)}(\bm{k}-\bm{p}).
 \end{equation}
Transforming to coordinate space,
     \begin{equation}\begin{split}\label{tran}
a_{i}^{a}(k^{+},\,\bm{k})\,=\,\int d^2\bm{z}\,\rme^{-i\bm{k}\cdot\bm{z}}\,a_{i}^{a}(k^{+},\,\bm{z})\;,
 \end{split}\end{equation}
the commutation relation becomes:
 \begin{equation}\label{composw}
 \left[a_{i}^{a}(k^{+},\bm{x}),\, a_{j}^{b\dagger}(p^{+},\bm{y})\right]=2\pi\delta^{ab}\delta_{ij}\delta(k^{+}-p^{+})\delta^{(2)}(\bm{x}-\bm{y}).
  \end{equation}
The quark fields can be expanded by the following expression:
\begin{equation}\label{quarfield}
\psi_{+}^{\alpha}(x)\,=\,\chi_{\lambda}\int_{0}^{\infty}\frac{dk^{+}}{2\pi}\int\frac{d^{2}\bm{k}}{(2\pi)^{2}}\frac{1}{\sqrt{2}}\left(b_{\lambda}^{\alpha}(k^{+},\bm{k})e^{-ik\cdot x}\,+\,d_{\lambda}^{\alpha\dagger}(k^{+},\bm{k})e^{ik\cdot x}\right).
 \end{equation}
The polarisation vectors are:
\begin{equation}\label{chis}
\chi_{+\frac{1}{2}}\,=\,\left(\begin{array}{c}
1\\
0
\end{array}\right),\qquad\quad\chi_{-\frac{1}{2}}\,=\,\left(\begin{array}{c}
0\\
1
\end{array}\right),
 \end{equation}
 \begin{equation}\label{lamiden}
\chi_{\lambda}\chi_{\lambda}^{\dagger}\,=\,I,\qquad\chi_{\lambda}^{\dagger}\chi_{\lambda}\,=\,2,\qquad\chi_{\lambda_{1}}^{\dagger}\,I\,\chi_{\lambda_{2}}\,=\,\delta_{\lambda_{1}\lambda_{2}},\qquad\chi_{\lambda_{1}}^{\dagger}\,\sigma^{3}\,\chi_{\lambda_{2}}\,=\,2\lambda_{1}\delta_{\lambda_{1}\lambda_{2}}.
 \end{equation}
The anti-commutation relations:
 \begin{equation}\begin{split}
\left\{ b_{\lambda_{1}}^{\alpha}(k^{+},\,\bm{k}),\, b_{\lambda_{2}}^{\beta\dagger}(p^{+},\,\bm{p})\right\} &=\,\left\{ d_{\lambda_{1}}^{\alpha}(k^{+},\,\bm{k}),\, d_{\lambda_{2}}^{\beta\dagger}(p^{+},\,\bm{p})\right\} \\
&=\,(2\pi)^{3}\,\delta_{\lambda_{1}\lambda_{2}}\,\delta^{\alpha\beta}\,\delta^{(2)}(\bm{k}-\bm{p})\,\delta(k^{+}-p^{+}).\\
 \end{split}\end{equation}
 Transforming the fields to coordinate space a la (\ref{tran}):
 \begin{equation}\begin{split}
\left\{ b_{\lambda_{1}}^{\alpha}(k^{+},\,\bm{x}),\, b_{\lambda_{2}}^{\beta\dagger}(p^{+},\,\bm{y})\right\} &=\,\left\{ d_{\lambda_{1}}^{\alpha}(k^{+},\,\bm{x}),\, d_{\lambda_{2}}^{\beta\dagger}(p^{+},\,\bm{y})\right\} \,\\
&=\,2\pi\,\delta_{\lambda_{1}\lambda_{2}}\,\delta^{\alpha\beta}\,\delta^{(2)}(\bm{x}-\bm{y})\,\delta(k^{+}-p^{+}).
 \end{split}\end{equation}
Inserting the field expansions (\ref{glufield}) and (\ref{quarfield}) into (\ref{hzero}), the free  Hamiltonian  becomes:
  \begin{equation}\begin{split}
 H_{0}=\int_{0}^{\infty}\frac{dk^{+}}{2\pi}&\int\frac{d^{2}\bm{k}}{(2\pi)^{2}}\frac{\bm{k}^{2}}{2k^{+}}\bigg(a_{i}^{a\dagger}(k^{+},\bm{k})\, a_{i}^{a}(k^{+},\bm{k})\\
&\left.+\,b_{\lambda}^{\alpha\dagger}(k^{+},\,\bm{k})\,b_{\lambda}^{\alpha}(k^{+},\,\bm{k})\,-\,d_{\lambda}^{\alpha}(k^{+},\,\bm{k})\,d_{\lambda}^{\alpha\dagger}(k^{+},\,\bm{k})\right),\\
  \end{split}\end{equation}
from which the dispersion relation for free quarks and gluons is $E_{k}=\frac{\bm{k}^{2}}{2k^{+}}$. The multi-parton bare Fock states are obtained by acting with the relevant creation operators on the bare vacuum state. In this paper, we use both the 3-momentum representation  $k=(k^+,\bm{k})$ and  the mixed representation $(k^+,\bm{x})$, as obtained via the Fourier transform from transverse momenta to transverse coordinates. Let us present here a few representative examples.

\smallskip
$\bullet$ \textit{\textbf{The bare vacuum state  $\left|0\right\rangle $.}} This state obeys the following conditions:
  \begin{equation}\label{vacuum}
a_{i}^{a}(q^{+},\,\bm{q})\left|0\right\rangle \,=\,b_{\lambda}^{\alpha}(q^{+},\,\bm{q})\left|0\right\rangle \,=\,d_{\lambda}^{\alpha}(q^{+},\,\bm{q})\left|0\right\rangle \,=\,0.
  \end{equation}
  
  \smallskip
$\bullet$ \textit{\textbf{The bare quark state.}} In momentum space, this state is constructed as
  \begin{equation}\label{gsta}
\left|q_{\lambda}^{\alpha}(k^{+},\,\bm{k})\right\rangle \,\equiv\,{b_{\lambda}^{\alpha\dagger}(k^{+},\,\bm{k})}\left|0\right\rangle.
  \end{equation}
It has a LC energy  $E_{q}(k)=\frac{\bm{k}^{2}}{2k^{+}}$ and its scalar product is normalized as follows:
\begin{equation}\label{norcon1}
\left\langle q_{\lambda_{2}}^{\beta}(p^{+},\,\bm{p})\left|q_{\lambda_{1}}^{\alpha}(k^{+},\,\bm{k})\right.\right\rangle \,=\,(2\pi)^{3}\delta^{\alpha\beta}\,\delta_{\lambda_{1}\lambda_{2}}\,\delta^{(2)}(\bm{k}-\bm{p})\,\delta(k^{+}-p^{+}).
\end{equation}
The mixed representation of the bare quark state is obtained as 
\begin{align}\label{chanrep2}
\left|q_{\lambda}^{\alpha}(k^{+},\,\bm{x})\right\rangle \equiv\int 
\frac{d^{2}\bm{k}}{(2\pi)^{2}}\,\rme^{-i\bm{x}\cdot\bm{k}}\left|q_{\lambda}^{\alpha}(k^{+},\,\bm{k})\right\rangle=
 {b_{\lambda}^{\alpha\dagger}(k^{+},\,\bm{x})}\left|0\right\rangle,
\end{align}
and the dot product reads
\begin{equation}\label{norcon2}
\left\langle q_{\lambda_{2}}^{\beta}(p^{+},\,\bm{y})\left|q_{\lambda_{1}}^{\alpha}(k^{+},\,\bm{x})\right.\right\rangle \,=\,2\pi \,
\delta^{\alpha\beta}\,\delta_{\lambda_{1}\lambda_{2}}\,\delta^{(2)}(\bm{x}-\bm{y})\,\delta(k^{+}-p^{+}).
\end{equation}

\smallskip
$\bullet$ \textit{\textbf{The bare quark-gluon state.}} In momentum space, this state reads
  \begin{equation}\begin{split}\label{ggst}
&\left|q_{\lambda}^{\alpha}(p^{+},\,\bm{p})\,g_{i}^{a}(k^{+},\,\bm{k})\right\rangle \,\equiv\,{b_{\lambda}^{\alpha\dagger}(p^{+},\,\bm{p})\,a_{i}^{a\dagger}(k^{+},\,\bm{k})}\left|0\right\rangle\,,
\end{split}\end{equation}
and has an energy $E_{qg}(k,\, p)=\frac{\bm{k}^{2}}{2k^{+}}+\frac{\bm{p}^{2}}{2p^{+}}$.  The mixed representation of this state reads
\begin{align}\label{qgchanrep}
\left|q_{\lambda}^{\alpha}(p^{+},\,\bm{x})\,g_{i}^{a}(k^{+},\,\bm{y})\right\rangle & \equiv\int \frac{d^{2}\bm{p}}{(2\pi)^{2}}
\frac{d^{2}\bm{k}}{(2\pi)^{2}}\,\rme^{-i\bm{x}\cdot\bm{p}-i\bm{y}\cdot\bm{k}}
\left|q_{\lambda}^{\alpha}(p^{+},\,\bm{p})\,g_{i}^{a}(k^{+},\,\bm{k})\right\rangle=\nonumber\\*[.2cm]
&= {b_{\lambda}^{\alpha\dagger}(p^{+},\,\bm{x})\,a_{i}^{a\dagger}(k^{+},\,\bm{y})}\left|0\right\rangle.
\end{align}

\smallskip
$\bullet$ \textit{\textbf{A bare quark wavepacket.}} This is constructed as a linear superposition  of the single bare quark Fock states which is normalized to unity. For instance, in momentum space with $p=(p^{+},\,\bm{p})$ etc, the following wavepacket
\beq\label{bqwp}
\left|\phi_{\lambda}^{\alpha}(p)\right\rangle \,\equiv\, \int \frac{d^{3}k}{(2\pi)^{3}}\,\phi_p(k)\, \left|q_{\lambda}^{\alpha}(k)\right\rangle,
\eeq
where the (generally complex) function $\phi_p(k)$ is peaked around the central value $p$, is properly normalized provided
\beq 
2N_c\int \frac{d^{3}k}{(2\pi)^{3}}\,|\phi_p(k)|^2=1\qquad\Longrightarrow\qquad  \left\langle \phi_{\lambda}^{\alpha}(p)\left| \phi_{\lambda}^{\alpha}(p)\right.\right\rangle=1.
\eeq
This wavepacket represents a  bare quark state with definite quantum numbers for color ($\alpha$) and spin ($\lambda$), but whose 3-momentum is specified only on the average (this is peaked around the 3-vector $p$) and hence
 is localized in space (unlike the momentum eigenstate \eqref{gsta}). It is now easy to check that by using the definition \eqref{Nq} for the bare quark number density operator, one obtains
 $ \left\langle \phi_{\lambda}^{\alpha}(p) \right|\hat{\mathcal{N}}_{q}(k) \left| \phi_{\lambda}^{\alpha}(p)\right\rangle = 2N_c |\phi_p(k)|^2/(2\pi)^3$. This in turn implies $ \left\langle \phi_{\lambda}^{\alpha}(p) \right|  \hat {n}_q\left| \phi_{\lambda}^{\alpha}(p)\right\rangle = 1$, with $\hat {n}_q\equiv \int d^3k \,\hat{\mathcal{N}}_{q}(k)$ the bare quark number operator. This is indeed the expected result for a state containing exactly one bare quark.

 \section{Matrix elements for the trijet calculation\label{mateleapp}}
Not all the interaction terms in the Hamiltonian (\ref{hint}) enter to our calculation here. Let us list here the interaction terms of the QCD Hamiltonian which are relevant for our calculation:

  \begin{equation}\begin{split}\label{gqq} 
&H_{gqq}\,\equiv\,-g\int dx^{-}\,d^{2}\bm{x}\,\left(2(\partial_{i}A_{i}^{a})\frac{1}{\partial^{+}}(\psi_{+}^{\dagger}t^{a}\psi_{+})+\psi_{+}^{\dagger}t^{a}(\sigma_{i}\partial_{i})\frac{1}{\partial^{+}}(\sigma_{j}A_{j}^{a}\psi_{+})\right.\\
&\left.+\psi_{+}^{\dagger}t^{a}(\sigma_{i}A_{i}^{a})\frac{1}{\partial^{+}}(\sigma_{j}\partial_{j}\psi_{+})\right),\\
 \end{split}\end{equation}
 
\begin{equation}\label{hggg}
H_{ggg}\,\equiv\,-gf^{abc}\int dx^{-}\,d^{2}\bm{x}\,\left((\partial_{i}A_{j}^{a})A_{i}^{b}A_{j}^{c}+(\partial_{i}A_{i}^{a})\frac{1}{\partial^{+}}(A_{j}^{b}\partial^{+}A_{j}^{c})\right),
 \end{equation}
 
   \begin{equation}
H_{qq-inst}\,=\,2g^{2}\int dx^{-}\,d^{2}\bm{x}\,\frac{1}{\partial^{+}}(\psi_{+}^{\dagger}t^{a}\psi_{+})\frac{1}{\partial^{+}}(\psi_{+}^{\dagger}t^{a}\psi_{+}).
\end{equation}

 \begin{equation}\label{hqg}
H_{gg-inst}^{q}\,\equiv\,-ig^{2}\int dx^{-}\,d^{2}\bm{x}\,\psi_{+}^{\dagger}t^{a}t^{b}\sigma_{i}\sigma_{j}A_{i}^{a}\frac{1}{\partial^{+}}(A_{j}^{b}\psi_{+}),
 \end{equation}

  \begin{equation}\begin{split}\label{ggi}
&H_{gg-inst}^{g}\,\equiv\,2g^{2}f^{abc}\int dx^{-}\,d^{2}\bm{x}\,\frac{1}{\partial^{+}}(A_{i}^{b}\partial^{+}A_{i}^{c})\frac{1}{\partial^{+}}(\psi_{+}^{\dagger}t^{a}\psi_{+}),
 \end{split}\end{equation}
 
 By inserting the field expansions (\ref{glufield}) and (\ref{quarfield}) to the interaction terms defined by (\ref{gqq}) - (\ref{hqg}), we can recast the last expressions and write them in terms of the creation and annihilation operators:
 
   \begin{equation}\begin{split}\label{hami_start}
&\mathsf{H}_{q\rightarrow qg}\equiv\int_{0}^{\infty}\frac{dk^{+}}{2\pi}\frac{dp^{+}}{2\pi}dq^{+}\,\int\frac{d^{2}\bm{k}}{(2\pi)^{2}}\frac{d^{2}\bm{p}}{(2\pi)^{2}}d^{2}\bm{q}\,\frac{gt_{\alpha\beta}^{a}}{2\sqrt{2k^{+}}}\,\Gamma_{\lambda_{1}\lambda_{2}}^{i}\\
&\times\,\left[a_{i}^{a\dagger}(k^{+},\bm{k})\,\delta^{(3)}(k+p-q)\,+\,a_{i}^{a}(k^{+},\bm{k})\,\delta^{(3)}(k-p+q)\right]\,b_{\lambda_{1}}^{\alpha\dagger}(p^{+},\bm{p})\,b_{\lambda_{2}}^{\beta}(q^{+},\bm{q}),
 \end{split}\end{equation}
 
   \begin{eqnarray}
&&\mathsf{H}_{g\rightarrow q\bar{q}}\equiv\int_{0}^{\infty}\frac{dk^{+}}{2\pi}\frac{dp^{+}}{2\pi}dq^{+}\,\int\frac{d^{2}\bm{k}}{(2\pi)^{2}}\frac{d^{2}\bm{p}}{(2\pi)^{2}}d^{2}\bm{q}\,\frac{gt_{\alpha\beta}^{a}}{2\sqrt{2k^{+}}}\,\Gamma_{\lambda_{1}\lambda_{2}}^{i}\nonumber\\
&&\times\,\delta^{(3)}(k-p-q)\,(b_{\lambda_{1}}^{\alpha\dagger}(p^{+},\,\bm{p})\,d_{\lambda_{2}}^{\beta\dagger}(q^{+},\,\bm{q})\,a_{i}^{a}(k^{+},\,\bm{k})\,+\,h.c.),\end{eqnarray}

 with
  \begin{equation}
\Gamma_{\lambda_{1}\lambda_{2}}^{i}\,\equiv\,\chi_{\lambda_{1}}^{\dagger}\left[\frac{2\bm{k}^{i}}{k^{+}}-\frac{\sigma\cdot\bm{p}}{p^{+}}\sigma^{i}-\sigma^{i}\frac{\sigma\cdot\bm{q}}{q^{+}}\right]\chi_{\lambda_{2}},
 \end{equation}
  
    \begin{equation}\begin{split}
\mathsf{H}_{g\rightarrow gg}=&\int_{\Lambda}^{e^{\delta\mathsf{Y}}\Lambda}\frac{dk^{+}}{2\pi}\,\frac{dp^{+}}{2\pi}\, dq^{+}\,\int\frac{d^{2}\bm{k}}{(2\pi)^{2}}\,\frac{d^{2}\bm{p}}{(2\pi)^{2}}\, d^{2}\bm{q}\,\frac{igf^{abc}}{2\sqrt{2k^{+}p^{+}q^{+}}}\\
&\times\left[\left(\bm{p}^{i}+\frac{q^{+}}{p^{+}+q^{+}}\bm{k}^{i}\right)a_{i}^{a}(k)a_{j}^{b\dagger}(p)a_{j}^{c\dagger}(q)\delta^{(3)}(-k+p+q)\right.\\
&+\left.\left(\bm{q}^{i}+\bm{p}^{i}-\frac{p^{+}+q^{+}}{q^{+}-p^{+}}\bm{k}^{i}\right)a_{i}^{a\dagger}(k)a_{j}^{b\dagger}(p)a_{j}^{c}(q)\delta^{(3)}(k+p-q)\,+\,h.c.\right]\\
 \end{split}\end{equation}
 In additional, for the instantaneous parts:
 \begin{equation}\begin{split}
&\mathsf{H}_{q\rightarrow qq\bar{q}}\,=\,g^{2}t_{\beta\alpha}^{a}t_{\delta\gamma}^{a}\,\int_{0}^{\infty}\frac{ds^{+}}{2\pi}\,\frac{dq^{+}}{2\pi}\,\frac{dk^{+}}{2\pi}\,dp^{+}\int\frac{d^{2}\bm{s}}{(2\pi)^{2}}\,\frac{d^{2}\bm{q}}{(2\pi)^{2}}\,\frac{d^{2}\bm{k}}{(2\pi)^{2}}\,d^{2}\bm{p}\,\frac{\delta(k-p+q+s)}{(q^{+}+s^{+})^{2}}\\
&\times\chi_{\lambda_{4}}^{\dagger}\,\chi_{\lambda_{3}}\,\chi_{\lambda_{2}}^{\dagger}\,\chi_{\lambda_{1}}b_{\lambda_{4}}^{\delta\dagger}(s^{+},\bm{s})\,d_{\lambda_{3}}^{\gamma\dagger}(q^{+},\bm{q})\,b_{\lambda_{2}}^{\beta\dagger}(k^{+},\bm{k})\,b_{\lambda_{1}}^{\alpha}(p^{+},\bm{p}).
\end{split}\end{equation}

  \begin{equation}\begin{split}
&\mathsf{H}_{q\rightarrow qgg}^{q}\,\equiv\,ig^{2}f^{abc}t_{\alpha\beta}^{a}\,\int_{0}^{\infty}\frac{ds^{+}}{2\pi}\,\frac{dt^{+}}{2\pi}\,\frac{dk^{+}}{2\pi}\,dp^{+}\int\frac{d^{2}\bm{s}}{(2\pi)^{2}}\,\frac{d^{2}\bm{t}}{(2\pi)^{2}}\,\frac{d^{2}\bm{k}}{(2\pi)^{2}}\,d^{2}\bm{p}\,\frac{t^{+}}{2\sqrt{t^{+}s^{+}}(s^{+}+t^{+})^{2}}\\
&\times\chi_{\lambda_{1}}^{\dagger}\,\chi_{\lambda_{2}}\,a_{i}^{b\dagger}(s^{+},\bm{s})\,a_{i}^{c\dagger}(t^{+},\bm{t})\,b_{\lambda_{1}}^{\alpha\dagger}(k^{+},\bm{k})\,b_{\lambda_{2}}^{\beta}(p^{+},\bm{p})\,\delta(s+t+k-p),
\end{split}\end{equation}

\begin{equation}\begin{split}\label{hami_end}
&\mathsf{H}_{q\rightarrow qgg}^{g}\,\equiv\,g^{2}\chi_{\lambda_{1}}^{\dagger}\sigma_{i}\sigma_{j}\chi_{\lambda_{2}}\int_{0}^{\infty}\frac{ds^{+}}{2\pi}\,\frac{dk^{+}}{2\pi}\,\frac{dp^{+}}{2\pi}\,dq^{+}\int\frac{d^{2}\bm{s}}{(2\pi)^{2}}\,\frac{d^{2}\bm{k}}{(2\pi)^{2}}\,\frac{d^{2}\bm{p}}{(2\pi)^{2}}\,d^{2}\bm{q}\\
&\times\frac{t_{\alpha\beta}^{a}t_{\beta\gamma}^{b}}{4\sqrt{q^{+}k^{+}}(q^{+}+s^{+})}\,b_{\lambda_{1}}^{\alpha\dagger}(s^{+},\bm{s})\,b_{\lambda_{2}}^{\gamma}(p^{+},\bm{p})\,a_{i}^{a\dagger}(q^{+},\bm{q})\,a_{j}^{b\dagger}(k^{+},\bm{k})\,\delta(q+s+k-p).
\end{split}\end{equation}

Based on the above expressions, (\ref{hami_start}) $-$ (\ref{hami_end}), we write down all the matrix elements that are relevant for the forward NLO trijet cross section calculation. \\

$\bullet$ \quad\textit{\textbf{Emission of a gluon from the quark state}}
\begin{equation}\begin{split}\label{qqg}
&\left\langle q_{\lambda_{2}}^{\beta}(p)\,g_{i}^{a}(k)\left|\mathsf{H}_{q\rightarrow qg}\right|q_{\lambda_{1}}^{\alpha}(q)\right\rangle =\frac{gt_{\alpha\beta}^{a}}{2\sqrt{2k^{+}}}\chi_{\lambda_{1}}^{\dagger}\left[\frac{2\bm{k}^{i}}{k^{+}}-\frac{\sigma\cdot\bm{p}}{p^{+}}\sigma^{i}-\sigma^{i}\frac{\sigma\cdot\bm{q}}{q^{+}}\right]\chi_{\lambda_{2}}(2\pi)^{3}\delta^{(3)}(k-p-q).
 \end{split}\end{equation}
  
 $\bullet$ \quad\textit{\textbf{Emission of a second gluon from the quark state}}
 \begin{equation}\begin{split}\label{qggqg}
&\left\langle q_{\lambda_{2}}^{\beta}(u)\,g_{j}^{b}(t)\,g_{l}^{c}(p)\left|\mathsf{H}_{q\rightarrow qg}\right|q_{\lambda_{1}}^{\alpha}(s)\,g_{i}^{a}(k)\right\rangle \\
&=\,\frac{gt_{\beta\alpha}^{c}\delta^{ab}\delta_{ij}}{2\sqrt{2p^{+}}}\chi_{\lambda_{2}}^{\dagger}\left[\frac{2\bm{p}^{l}}{p^{+}}-\frac{\sigma\cdot\bm{u}}{u^{+}}\sigma^{l}-\sigma^{l}\frac{\sigma\cdot\bm{s}}{s^{+}}\right]\chi_{\lambda_{1}}(2\pi)^{6}\delta^{(3)}(k-t)\delta^{(3)}(s-u-p)\\
&+\,\frac{gt_{\beta\alpha}^{b}\delta^{ac}\delta_{il}}{2\sqrt{2t^{+}}}\chi_{\lambda_{2}}^{\dagger}\left[\frac{2\bm{t}^{j}}{t^{+}}-\frac{\sigma\cdot\bm{u}}{u^{+}}\sigma^{j}-\sigma^{j}\frac{\sigma\cdot\bm{s}}{s^{+}}\right]\chi_{\lambda_{1}}(2\pi)^{6}\delta^{(3)}(k-p)\delta^{(3)}(s-u-t).
 \end{split}\end{equation}\\

  $\bullet$ \quad\textit{\textbf{Gluon splits into quark and antiquark pair}}
 \begin{equation}\begin{split}\label{qqqqg}
&\left\langle \bar{q}_{\lambda_{4}}^{\epsilon}(u)\,q_{\lambda_{3}}^{\delta}(t)\,q_{\lambda_{2}}^{\gamma}(p)\left|\mathsf{H}_{g\rightarrow qq}\right|q_{\lambda_{1}}^{\beta}(s)\,g_{i}^{a}(k)\right\rangle \\
&=\frac{gt_{\gamma\epsilon}^{a}\delta^{\beta\delta}\delta_{\lambda_{1}\lambda_{3}}}{2\sqrt{2k^{+}}}\chi_{\lambda_{2}}^{\dagger}\left[\frac{2\bm{k}^{i}}{k^{+}}-\frac{\sigma\cdot\bm{p}}{p^{+}}\sigma^{i}-\sigma^{i}\frac{\sigma\cdot\bm{u}}{u^{+}}\right]\chi_{\lambda_{4}}(2\pi)^{6}\delta^{(3)}(s-t)\delta^{(3)}(k-p-u)\\
&+\frac{gt_{\delta\epsilon}^{a}\delta^{\beta\gamma}\delta_{\lambda_{1}\lambda_{2}}}{2\sqrt{2k^{+}}}\chi_{\lambda_{3}}^{\dagger}\left[\frac{2\bm{k}^{i}}{k^{+}}-\frac{\sigma\cdot\bm{t}}{t^{+}}\sigma^{i}-\sigma^{i}\frac{\sigma\cdot\bm{u}}{u^{+}}\right]\chi_{\lambda_{4}}(2\pi)^{6}\delta^{(3)}(s-p)\delta^{(3)}(k-t-u).
 \end{split}\end{equation}\\ 
 
  $\bullet$ \quad\textit{\textbf{Triple gluon interaction in presence of a quark}}
\begin{eqnarray}\label{qgqgg_split}
&&\left\langle q_{\lambda_{2}}^{\beta}(t)\,g_{l}^{c}(q)\,g_{n}^{b}(p)\left|\mathsf{H}_{g\rightarrow gg}\right|q_{\lambda_{1}}^{\alpha}(s)\,g_{m}^{a}(k)\right\rangle =\frac{igf^{abc}\delta^{\alpha\beta}\delta_{\lambda_{1}\lambda_{2}}}{2\sqrt{2k^{+}p^{+}q^{+}}}(2\pi)^{6}\delta^{(3)}(s-t)\delta^{(3)}(k-p-q)\nonumber\\
&&\times\left[\left(\bm{p}^{m}-\bm{q}^{m}+\frac{q^{+}-p^{+}}{k^{+}}\bm{k}^{m}\right)\delta_{nl}+\left(\bm{k}^{n}+\bm{q}^{n}-\frac{k^{+}+q^{+}}{p^{+}}\bm{p}^{n}\right)\delta_{ml}\right.\\
&&\left.+\left(\frac{k^{+}+p^{+}}{q^{+}}\bm{q}^{l}-\bm{p}^{l}-\bm{k}^{l}\right)\delta_{mn}\right].\nonumber\end{eqnarray}\\

      $\bullet$ \quad\textit{\textbf{Quark produces quark and antiquark pair instantaneously}}
 \begin{equation}\begin{split}\label{instqq}
\left\langle q_{\lambda_{2}}^{\beta}(k)\,\bar{q}_{\lambda_{3}}^{\gamma}(q)\,q_{\lambda_{4}}^{\delta}(s)\left|\mathsf{H}_{q\rightarrow qqq}\right|q_{\lambda_{1}}^{\alpha}(p)\right\rangle \,=\,\frac{g^{2}t_{\beta\alpha}^{a}t_{\delta\gamma}^{a}}{(q^{+}+s^{+})^{2}}\chi_{\lambda_{4}}^{\dagger}\,\chi_{\lambda_{3}}\,\chi_{\lambda_{2}}^{\dagger}\,\chi_{\lambda_{1}}(2\pi)^{3}\delta(k-p+q+s).
  \end{split}\end{equation}
  
   $\bullet$ \quad\textit{\textbf{Instantaneous emission of a two gluons from the quark state (gluon channel)}}
  \begin{equation}\begin{split}\label{instq}
&\left\langle q_{\lambda_{2}}^{\gamma}(u)\,g_{j}^{b}(t)\,g_{l}^{c}(p)\left|\mathsf{H}_{q\rightarrow qgg}^{q}\right|q_{\lambda_{1}}^{\alpha}(s)\right\rangle \\
&=\,\frac{g^{2}}{\sqrt{t^{+}p^{+}}}\left(\frac{t_{\gamma\beta}^{b}t_{\beta\alpha}^{c}}{s^{+}-p^{+}}\chi_{\lambda_{2}}^{\dagger}\sigma_{j}\sigma_{l}\chi_{\lambda_{1}}+\frac{t_{\gamma\beta}^{c}t_{\beta\alpha}^{b}}{s^{+}-t^{+}}\chi_{\lambda_{2}}^{\dagger}\sigma_{l}\sigma_{j}\chi_{\lambda_{1}}\right)(2\pi)^{3}\delta(t+p+u-s).
  \end{split}\end{equation}

 $\bullet$ \quad\textit{\textbf{Instantaneous emission of two gluons from the quark state (quark channel)}}
 \begin{equation}\begin{split}\label{instg}
\left\langle q_{\lambda_{2}}^{\beta}(s)\,g_{j}^{c}(p)\,g_{i}^{b}(k)\left|\mathsf{H}_{q\rightarrow qgg}^{g}\right|q_{\lambda_{1}}^{\alpha}(q)\right\rangle \,=\,\frac{ig^{2}f^{abc}t_{\beta\alpha}^{a}(p^{+}-k^{+})}{2\sqrt{k^{+}p^{+}}(k^{+}+p^{+})^{2}}\chi_{\lambda_{2}}^{\dagger}\chi_{\lambda_{1}}(2\pi)^{3}\delta_{ij}\delta(k+p+s-q).
  \end{split}\end{equation}
    
 $\bullet$ \quad\textit{\textbf{Gluon splits into quark and antiquark pair}}
\begin{equation}\begin{split}\label{gqq1}
 \left\langle \overline{q}_{\lambda_{2}}^{\beta}(q)\,q_{\lambda_{1}}^{\alpha}(p)\left|\mathsf{H}_{g\rightarrow qq}\right|g_{i}^{a}(k)\right\rangle =\frac{gt_{\alpha\beta}^{a}}{2\sqrt{2k^{+}}}\chi_{\lambda_{1}}^{\dagger}\left[\frac{2\bm{k}^{i}}{k^{+}}-\frac{\sigma\cdot\bm{p}}{p^{+}}\sigma^{i}-\sigma^{i}\frac{\sigma\cdot\bm{q}}{q^{+}}\right]\chi_{\lambda_{2}}(2\pi)^{3}\delta^{(3)}(k-p-q).
 \end{split}\end{equation}

 $\bullet$ \quad\textit{\textbf{Triple gluon interaction}}
\begin{eqnarray}\label{ggg_split}
&&\left\langle g_{l}^{c}(q)\,g_{n}^{b}(p)\left|\mathsf{H}_{g\rightarrow gg}\right|g_{m}^{a}(k)\right\rangle =\frac{igf^{abc}}{2\sqrt{2k^{+}p^{+}q^{+}}}\left[\left(\bm{p}^{m}-\bm{q}^{m}+\frac{q^{+}-p^{+}}{k^{+}}\bm{k}^{m}\right)\delta_{nl}\right.\nonumber\\
&&\left.+\left(\bm{k}^{n}+\bm{q}^{n}-\frac{k^{+}+q^{+}}{p^{+}}\bm{p}^{n}\right)\delta_{ml}+\left(\frac{k^{+}+p^{+}}{q^{+}}\bm{q}^{l}-\bm{p}^{l}-\bm{k}^{l}\right)\delta_{mn}\right](2\pi)^{3}\delta^{(3)}(k-p-q).\end{eqnarray}
  In addition, we present here two additional matrix elements that will be needed for the computation in this paper:
  
  $\bullet$ \quad\textit{\textbf{Measuring a quark and a gluon}}
  \begin{equation}\begin{split}\label{matqg}
&\left\langle q_{\lambda_{2}}^{\beta}((1-\bar{\vartheta})q^{+},\,\bar{\bm{x}})\,g_{j}^{b}(\bar{\vartheta}q^{+},\,\bar{\bm{z}})\right|\,\hat{\mathcal{N}}_{q}(p)\,\hat{\mathcal{N}}_{g}(k)\,\left|q_{\lambda_{1}}^{\alpha}((1-\vartheta)q^{+},\,\bm{x})\,g_{i}^{a}(\vartheta q^{+},\,\bm{z})\right\rangle \,=\,\frac{e^{i\bm{k}\cdot(\bar{\bm{x}}-\bm{x})+i\bm{p}\cdot(\bar{\bm{z}}-\bm{z})}}{(2\pi)^{2}}\\
&\times\,\delta_{\lambda_{1}\lambda_{2}}\,\delta^{\alpha\beta}\,\delta_{ij}\,\delta^{ab}\,\delta(k^{+}-(1-\bar{\vartheta})q^{+})\,\delta(k^{+}-(1-\vartheta)q^{+})\,\delta(p^{+}-\bar{\vartheta}q^{+})\,\delta(p^{+}-\vartheta q^{+}).
\end{split}\end{equation}

  $\bullet$ \quad\textit{\textbf{Measuring a quark and an antiquark}}
  \begin{equation}\begin{split}\label{matqq}
&\left\langle \overline{q}_{\lambda_{4}}^{\delta}((1-\overline{\vartheta})q^{+},\,\overline{\bm{z}}^{\prime})\,q_{\lambda_{3}}^{\gamma}(\overline{\vartheta}q^{+},\,\overline{\bm{z}})\right|\,\hat{\mathcal{N}}_{q}(p)\,\hat{\mathcal{N}}_{\bar{q}}(k)\,\left|\overline{q}_{\lambda_{2}}^{\beta}((1-\vartheta)q^{+},\,\bm{z}^{\prime})\,q_{\lambda_{1}}^{\alpha}(\vartheta q^{+},\,\bm{z})\right\rangle \,=\,\frac{e^{i\bm{k}\cdot(\overline{\bm{z}}^{\prime}-\bm{z}^{\prime})+i\bm{p}\cdot(\bar{\bm{z}}-\bm{z})}}{(2\pi)^{2}}\\
&\times\,\delta_{\lambda_{2}\lambda_{4}}\,\delta^{\beta\delta}\,\delta_{\lambda_{1}\lambda_{3}}\,\delta^{\alpha\gamma}\,\delta(k^{+}-(1-\bar{\vartheta})q^{+})\,\delta(k^{+}-(1-\vartheta)q^{+})\,\delta(p^{+}-\bar{\vartheta}q^{+})\,\delta(p^{+}-\vartheta q^{+}).
\end{split}\end{equation}

  $\bullet$ \quad\textit{\textbf{Measuring two gluons}}
  \begin{equation}\begin{split}\label{matgg}
&\left\langle g_{n}^{d}((1-\bar{\vartheta})q^{+},\,\bar{\bm{z}}^{\prime})\,g_{m}^{c}(\bar{\vartheta}q^{+},\,\bar{\bm{z}})\right|\,\hat{\mathcal{N}}_{g}(p)\,\hat{\mathcal{N}}_{g}(k)\,\left|g_{j}^{b}((1-\vartheta)q^{+},\,\bm{z}^{\prime})\,g_{i}^{a}(\vartheta q^{+},\,\bm{z})\right\rangle \\
&=\frac{e^{i\bm{k}\cdot\bar{\bm{z}}^{\prime}+i\bm{p}\cdot\bar{\bm{z}}}}{(2\pi)^{2}}\,\delta(k^{+}-(1-\bar{\vartheta})q^{+})\,\delta(p^{+}-\bar{\vartheta}q^{+})\,\left(e^{-i\bm{k}\cdot\bm{z}^{\prime}-i\bm{p}\cdot\bm{z}}\,\delta_{mi}\,\delta^{ac}\,\delta_{nj}\,\delta^{bd}\,\delta(k^{+}-(1-\vartheta)q^{+})\right.\\
&\left.\times\,\delta(p^{+}-\vartheta q^{+})+\,\rme^{-i\bm{k}\cdot\bm{z}-i\bm{p}\cdot\bm{z}^{\prime}}\,\delta_{in}\,\delta^{ad}\,\delta_{jm}\,\delta^{bc}\,\delta(p^{+}-(1-\vartheta)q^{+})\,\delta(k^{+}-\vartheta q^{+})\right)\,+\,\left(k^{+}\leftrightarrow p^{+};\,\bm{k}\leftrightarrow\bm{p}\right).
\end{split}\end{equation}

  $\bullet$ \quad\textit{\textbf{Measuring two quarks and an antiquark}}
  \begin{equation}\begin{split}\label{matqqq}
&\left\langle \overline{q}_{\lambda_{4}}^{\delta}\left(Dq^{+},\,\bm{y}\right)\,q_{\lambda_{5}}^{\zeta}\left(Eq^{+},\,\overline{\bm{z}}\right)\,q_{\lambda_{6}}^{\eta}\left(Fq^{+},\,\bm{\overline{\bm{z}}}^{\prime}\right)\right|\,\hat{\mathcal{N}}_{q}(s)\,\hat{\mathcal{N}}_{q}(t)\,\hat{\mathcal{N}}_{\overline{q}}(u)\,\left|\overline{q}_{\lambda_{1}}^{\alpha}\left(Aq^{+},\,\bm{x}\right)\,q_{\lambda_{2}}^{\beta}\left(Bq^{+},\,\bm{z}\right)\right.\\
&\left.\,q_{\lambda_{3}}^{\gamma}\left(Cq^{+},\,\bm{z}^{\prime}\right)\right\rangle \,=\,\frac{e^{-i\bm{u}\cdot(\bm{x}-\bm{y})}}{(2\pi)^{3}}\,\delta^{\delta\alpha}\,\delta_{\lambda_{4}\lambda_{1}}\,\delta(u^{+}-Aq^{+})\,\delta(u^{+}-Dq^{+})\\
&\times\,\left(\delta^{\zeta\beta}\,\delta^{\eta\gamma}\,\delta_{\lambda_{5}\lambda_{2}}\,\delta_{\lambda_{6}\lambda_{3}}\,\delta(s^{+}-Cq^{+})\,\delta(s^{+}-Fq^{+})\,\delta(t^{+}-Bq^{+})\,\delta(t^{+}-Eq^{+})\,\rme^{-i\bm{s}\cdot(\bm{z}^{\prime}-\bm{\bar{z}}^{\prime})-i\bm{t}\cdot(\bm{z}-\bm{\bar{z}})}\right.\\
&\left.+\,\delta^{\eta\beta}\,\delta^{\zeta\gamma}\,\delta_{\lambda_{6}\lambda_{2}}\,\delta_{\lambda_{5}\lambda_{3}}\,\delta(s^{+}-Cq^{+})\,\delta(s^{+}-Eq^{+})\,\delta(t^{+}-Bq^{+})\,\delta(t^{+}-Fq^{+})\,\rme^{-i\bm{s}\cdot(\bm{z}^{\prime}-\bm{\bar{z}})-i\bm{t}\cdot(\bm{z}-\bm{\overline{\bm{z}}}^{\prime})}\right)\\
&+\,\left(s^{+}\leftrightarrow t^{+};\,\bm{s}\leftrightarrow\bm{t}\right).
\end{split}\end{equation}

 $\bullet$ \quad\textit{\textbf{Measuring quark and two gluons}}
  \begin{equation}\begin{split}\label{matqgg}
&\left\langle q_{\lambda_{2}}^{\beta}\left(Dq^{+},\,\bm{y}\right)\,g_{k}^{d}\left(Eq^{+},\,\overline{\bm{z}}\right)\,g_{l}^{e}\left(Fq^{+},\,\bm{\overline{\bm{z}}}^{\prime}\right)\right|\,\hat{\mathcal{N}}_{g}(s)\,\hat{\mathcal{N}}_{g}(t)\,\hat{\mathcal{N}}_{q}(u)\,\left|q_{\lambda_{1}}^{\alpha}\left(Aq^{+},\,\bm{x}\right)\,g_{i}^{b}\left(Bq^{+},\,\bm{z}\right)\right.\\
&\left.\,g_{j}^{c}\left(Cq^{+},\,\bm{z}^{\prime}\right)\right\rangle \,=\,\frac{e^{-i\bm{u}\cdot(\bm{x}-\bm{y})}}{(2\pi)^{3}}\,\delta^{\beta\alpha}\,\delta_{\lambda_{2}\lambda_{1}}\,\delta(u^{+}-Aq^{+})\,\delta(u^{+}-Dq^{+})\\
&\times\,\left(\delta^{db}\,\delta_{ki}\,\delta^{ec}\,\delta_{lj}\,\delta(s^{+}-Eq^{+})\,\delta(s^{+}-Bq^{+})\,\delta(t^{+}-Fq^{+})\,\delta(t^{+}-Cq^{+})\,\rme^{-i\bm{s}\cdot(\bm{z}^{\prime}-\overline{\bm{z}}^{\prime})-i\bm{t}\cdot(\bm{z}-\bm{\bar{z}})}\right.\\
&\left.+\,\delta^{dc}\,\delta_{kj}\,\delta^{eb}\,\delta_{li}\,\delta(s^{+}-Eq^{+})\,\delta(s^{+}-Cq^{+})\,\delta(t^{+}-Fq^{+})\,\delta(t^{+}-Bq^{+})\,\rme^{-i\bm{s}\cdot(\bm{z}^{\prime}-\bm{\bar{z}})-i\bm{t}\cdot(\bm{z}-\bm{\bar{z}}^{\prime})}\right)\\
&+\,\left(s^{+}\leftrightarrow t^{+};\,\bm{s}\leftrightarrow\bm{t}\right).
\end{split}\end{equation}
 
\section{Fourier transforms \label{fouriers}}

Here we present a list of the integrals necessary to perform the Fourier transformations during the computations.
  \begin{equation}
\frac{1}{(2\pi)^{3}}\int dx^{-}\,d^{2}\bm{x}\,\rme^{i(k-p)\cdot x}\,=\,\delta^{(3)}(k-p).
\end{equation}
  \begin{equation}\label{delta}
\frac{1}{(2\pi)^{2}}\int d^{2}\bm{p}\,\rme^{i\bm{p}\cdot(\bm{x}-\bm{y})}\,=\,\delta^{(2)}(\bm{x}-\bm{y}).
\end{equation}
 \begin{equation}\begin{split}\label{fourier.1}
\int\frac{d^{2}\bm{k}}{2\pi}\,\frac{\bm{k}^{i}}{\bm{k}^{2}}e^{i\bm{k}\cdot\bm{X}}=\frac{i\bm{X}^{i}}{\bm{X}^{2}}.
 \end{split}\end{equation}
    \begin{equation}\begin{split} \label{fourier.3}
   \int\frac{d^{2}\bm{k}}{2\pi}\frac{d^{2}\bm{p}}{2\pi}\,\frac{1}{a\bm{k}^{2}+\bm{p}^{2}}e^{i\bm{k}\cdot\bm{X}+i\bm{p}\cdot\bm{Z}}\,=\,\frac{1}{\bm{X}^{2}+a\bm{Z}^{2}}.
 \end{split}\end{equation}
  \begin{equation}\begin{split} \label{fourier.4}
  \int\frac{d^{2}\bm{k}}{2\pi}\frac{d^{2}\bm{p}}{2\pi}\,\frac{\bm{k}^{i}\bm{p}^{j}}{\bm{k}^{2}\left(a\bm{k}^{2}+b\bm{p}^{2}\right)}e^{i\bm{k}\cdot\bm{X}+i\bm{p}\cdot\bm{Z}}\,=\,-\frac{\bm{X}^{i}\,\bm{Z}^{j}}{\bm{Z}^{2}\,\left(b\bm{X}^{2}+a\bm{Z}^{2}\right)}.
 \end{split}\end{equation}
 
  \section{The gluon wave function at leading order \label{glucha}}
\begin{figure}[!h]
\center \includegraphics[scale=0.7]{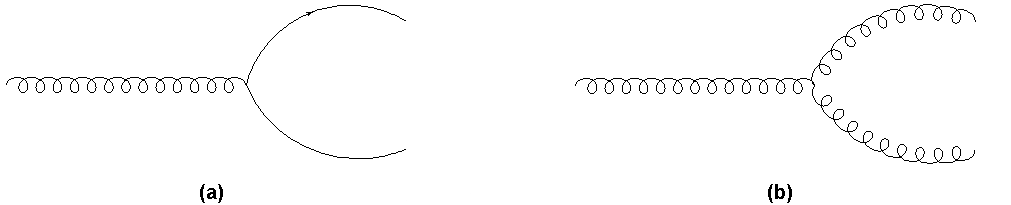}
\caption{The two-parton contributions to the LO gluon LCWF: (a) gluon splitting into a  quark-antiquark pair, cf. Eq.~(\ref{qqofg}); (b) gluon splitting into two gluons, cf. Eq.~(\ref{gg_state}).\label{gLOWF}}
\end{figure}

  At leading order the gluon LCWF has following form (see also Fig.~\ref{gLOWF}):
\begin{equation}\begin{split}
&\left|g_{i}^{a}(q^{+},\,\bm{q})\right\rangle _{\lo}\,\equiv\,U_{I}(0,\,-\infty)_{\lo}\,\left|g_{i}^{a}(q^{+},\,\bm{q})\right\rangle \,=\,\mathcal{Z}_{\lo}\left|g_{i}^{a}(q^{+},\,\bm{q})\right\rangle \,+\,\left|g_{i}^{a}(q^{+},\,\bm{q})\right\rangle _{q\overline{q}}\,+\,\left|g_{i}^{a}(q^{+},\,\bm{q})\right\rangle _{gg}\,,
\end{split}\end{equation}
where $\mathcal{Z}_{\lo}$ takes into account the contribution to the normalization of the wave-function (not shown explicitly in this paper), and the following definitions are used:
 \begin{equation}\begin{split}\label{qqofg}
\left|g_{i}^{a}(q)\right\rangle _{q\bar{q}}\,\equiv\,-\int_{0}^{\infty}\frac{dk^{+}}{2\pi}\,\frac{ds^{+}}{2\pi}\,\int\frac{d^{2}\bm{k}}{(2\pi)^{2}}\,\frac{d^{2}\bm{s}}{(2\pi)^{2}}\:\frac{1}{\Delta(q;\,k,\,s)}\left|\bar{q}_{\lambda_{2}}^{\beta}(s)\,q_{\lambda_{1}}^{\alpha}(k)\right\rangle \left\langle \bar{q}_{\lambda_{2}}^{\beta}(s)\,q_{\lambda_{1}}^{\alpha}(k)\left|H_{gqq}\right|g_{i}^{a}(q)\right\rangle ,
 \end{split}\end{equation}
 
 \begin{equation}\label{gg_state}
\left|g_{i}^{a}(q)\right\rangle _{gg}\,\equiv\,-\frac{1}{2}\int_{0}^{\infty}\frac{dk^{+}}{2\pi}\,\frac{ds^{+}}{2\pi}\,\int\frac{d^{2}\bm{k}}{(2\pi)^{2}}\,\frac{d^{2}\bm{s}}{(2\pi)^{2}}\,\frac{1}{\Delta(q;\,k,\,s)}\,\left|g_{l}^{c}(s)\,g_{j}^{b}(k)\right\rangle \left\langle g_{l}^{c}(s)\,g_{j}^{b}(k)\left|H_{ggg}\right|g_{i}^{a}(q)\right\rangle .
\end{equation}

The various definitions of the states involved in the above expressions are shown in App. \ref{fieldef}. After inserting the matrix element (\ref{gqq1}):
\begin{equation}\begin{split}
&\left|g_{i}^{a}(q)\right\rangle _{q\bar{q}}\,=\,-\int_{0}^{\infty}\frac{dk^{+}}{2\pi}\,\frac{ds^{+}}{2\pi}\,\int\frac{d^{2}\bm{k}}{(2\pi)^{2}}\,\frac{d^{2}\bm{s}}{(2\pi)^{2}}\:\frac{1}{\Delta(q;\,k,\,s)}\\
&\times\frac{gt_{\alpha\beta}^{a}}{2\sqrt{2q^{+}}}\chi_{\lambda_{1}}^{\dagger}\left[\frac{2\bm{q}^{i}}{q^{+}}-\frac{\sigma\cdot\bm{k}}{k^{+}}\sigma^{i}-\sigma^{i}\frac{\sigma\cdot\bm{s}}{s^{+}}\right]\chi_{\lambda_{2}}(2\pi)^{3}\delta^{(3)}(q-k-s)\left|\bar{q}_{\lambda_{2}}^{\beta}(s)\,q_{\lambda_{1}}^{\alpha}(k)\right\rangle .
\end{split}\end{equation}
After adopting the variables in (\ref{newvar}), the last result becomes:
\begin{equation}\begin{split}
&\left|g_{i}^{a}(q^{+},\,\bm{q})\right\rangle _{q\bar{q}}\\
&=\,-\int_{0}^{1}d\vartheta\,\int d^{2}\bm{k}\:\frac{gt_{\alpha\beta}^{a}\varphi_{\lambda_{2}\lambda_{1}}^{ij}(\vartheta)\sqrt{q^{+}}\,\bm{\widetilde{k}}^{j}}{8\sqrt{2}\pi^{3}\bm{\widetilde{k}}^{2}}\left|\bar{q}_{\lambda_{2}}^{\beta}((1-\vartheta)q^{+},\,(1-\vartheta)\bm{q}-\bm{\widetilde{k}})\,q_{\lambda_{1}}^{\alpha}(\vartheta q^{+},\,\vartheta\bm{q}+\bm{\widetilde{k}})\right\rangle .
\end{split}\end{equation}
After Fourier transformation with the aid of (\ref{fourier.1}), one finds
\begin{equation}\begin{split}\label{qqcompoofg}
&\left|g_{i}^{a}(q^{+},\,\bm{w})\right\rangle _{q\bar{q}}\,=\,\int_{\bm{z},\,\bm{z}^{\prime}}\,\int_{0}^{1}d\vartheta\,\frac{igt_{\alpha\beta}^{a}\varphi_{\lambda_{2}\lambda_{1}}^{ij}(\vartheta)\sqrt{q^{+}}\,\bm{Z}^{j}}{4\sqrt{2}\pi^{2}\bm{Z}^{2}}\,\delta^{(2)}(\bm{w}-(1-\vartheta)\bm{z}^{\prime}-\vartheta\bm{z})\\
&\times\left|\bar{q}_{\lambda_{2}}^{\beta}((1-\vartheta)q^{+},\,\bm{z}^{\prime})\,q_{\lambda_{1}}^{\alpha}(\vartheta q^{+},\,\bm{z})\right\rangle .
\end{split}\end{equation}
Consider now the state (\ref{gg_state}): by inserting the matrix element (\ref{ggg_split}), one can deduce 
\begin{eqnarray}
&&\left|g_{i}^{a}(q)\right\rangle _{gg}\,=\,-\frac{1}{2}\int_{0}^{\infty}\frac{ds^{+}}{2\pi}\,\frac{dk^{+}}{2\pi}\,\int\frac{d^{2}\bm{s}}{(2\pi)^{2}}\,\frac{d^{2}\bm{k}}{(2\pi)^{2}}\,\frac{igf^{abc}(2\pi)^{3}\delta(q-k-s)}{2\sqrt{2q^{+}s^{+}k^{+}}\,\Delta(q;\,s,\,k)}\left[\left(\bm{k}^{i}-\bm{s}^{i}+\frac{s^{+}-k^{+}}{q^{+}}\bm{q}^{i}\right)\delta_{jl}\right.\nonumber\\
&&\left.+\left(\bm{q}^{j}+\bm{s}^{j}-\frac{q^{+}+s^{+}}{k^{+}}\bm{k}^{j}\right)\delta_{il}+\left(\frac{q^{+}+k^{+}}{s^{+}}\bm{s}^{l}-\bm{k}^{l}-\bm{q}^{l}\right)\delta_{ij}\right]\left|g_{l}^{c}(s)\,g_{j}^{b}(k)\right\rangle ,
\end{eqnarray}
which takes a simpler form in terms of the variables introduced in Eq.~(\ref{newvar}):
\begin{equation}\begin{split}
&\left|g_{i}^{a}(q^{+},\,\bm{q})\right\rangle _{gg}=\,-\int_{0}^{1}d\vartheta\,\int d^{2}\bm{\widetilde{k}}\,\frac{igf^{abc}\,\sqrt{\vartheta(1-\vartheta)q^{+}}}{8\sqrt{2}\pi^{3}\bm{\widetilde{k}}^{2}}\left(\widetilde{\bm{k}}^{i}\delta_{jl}\,-\,\frac{1}{1-\vartheta}\widetilde{\bm{k}}^{j}\delta_{il}\,-\,\frac{1}{\vartheta}\widetilde{\bm{k}}^{l}\delta_{ij}\right)\\
&\times\left|g_{l}^{c}\left((1-\vartheta)q^{+},\,(1-\vartheta)\bm{q}-\bm{\widetilde{k}}\right)\,g_{j}^{b}\left(\vartheta q^{+},\,\vartheta\bm{q}+\bm{\widetilde{k}}\right)\right\rangle .
 \end{split}\end{equation}
After Fourier transformation with the aid of (\ref{fourier.1}), this gives
\begin{equation}\begin{split}\label{ggcompoofg}
&\left|g_{i}^{a}(q^{+},\,\bm{w})\right\rangle _{gg}=\,\int_{\bm{z},\,\bm{z}^{\prime}}\,\int_{0}^{1}d\vartheta\,\frac{igf^{abc}\,\sqrt{\vartheta(1-\vartheta)q^{+}}}{4\sqrt{2}\pi^{2}\bm{Z}^{2}}\left(\bm{Z}^{i}\delta_{jl}\,-\,\frac{1}{1-\vartheta}\bm{Z}^{j}\delta_{il}\,-\,\frac{1}{\vartheta}\bm{Z}^{l}\delta_{ij}\right)\\
&\times\,\delta^{(2)}(\bm{w}-(1-\vartheta)\bm{z}^{\prime}-\vartheta\bm{z})\left|g_{l}^{c}\left((1-\vartheta)q^{+},\,\bm{z}^{\prime}\right)\,g_{j}^{b}\left(\vartheta q^{+},\,\bm{z}\right)\right\rangle .
 \end{split}\end{equation}

  \section{The three-parton components of the quark LCWF}\label{details}
  
In this Appendix we present an explicit construction of the three-parton Fock components of the quark LCWF at time $0_-$ (i.e. just before to scattering). The respective two-parton (quark-gluon) component has been already constructed in Sect.~\ref{looutgoing} (see notably Eqs.~\eqref{lobeg} and \eqref{lobegon} there). We start with the general formula (cf. the first line in \eqn{UinWV})
\begin{equation}\label{LCWF}
\left|q\right\rangle _{\nlo}\,\equiv\,U_{I}(0,\,-\infty)|_{\nlo}\,\left|q_{\lambda}^{\alpha}(q^{+},\,\bm{q})\right\rangle \,=\,\mathcal{Z}_{\nlo}\left|q\right\rangle \,-\,\left|i\right\rangle \,\frac{\left\langle i\left|H_{\rm int}\right|q\right\rangle }{E_{i}-E_{q}}\,+\,\left|j\right\rangle \,\frac{\left\langle j\left|H_{\rm int}\right|i\right\rangle \,\left\langle i\left|H_{\rm int}\right|q\right\rangle }{(E_{j}-E_{q})\,(E_{i}-E_{q})}.
\end{equation}
As announced, we shall display only those terms which count for the calculation of three parton production in the final state.  After introducing the subscripts $qq\bar{q}$, $qgg$ and $qg$ to denote the particle content of the various free eigenstates, it is possible to represent (\ref{LCWF}) by the following expression\footnote{Note a slight abuse of notations: in Sect.~\ref{sec:NLOprod} and notably in Eqs.~\eqref{nloasi}--\eqref{out.qgg4} we have used exactly the same notations, e.g. $\left|q\right\rangle _{qq\bar{q}}^{reg}$, to denote the three-parton Fock-space components of the {\em outgoing} quark state, including the scattering with the shockwave; in the present Appendix though, they systematically refer to the LCWF before the scattering.}:
\begin{equation}
\label{qNLO}
\left|q\right\rangle _{\nlo}\,=\,\mathcal{Z}_{\nlo}\left|q\right\rangle \,+\,\left|q\right\rangle _{qg}\,+\,\left|q\right\rangle _{qq\bar{q}}^{reg}\,+\,\left|q\right\rangle _{qq\bar{q}}^{inst}\,+\,\sum_{i=1,2}\left|q\right\rangle _{qgg}^{reg,\,i}\,+\,\sum_{i=1,2}\left|q\right\rangle _{qgg}^{inst,\,i},
  \end{equation}
where $\left|q\right\rangle _{qg}$ is defined in eq. (\ref{lobeg}), and the three-parton Fock space components are defined as follows:
\begin{equation}\label{qq1}
\left|q\right\rangle _{qq\bar{q}}^{reg}\,\equiv\,\frac{\left|q\,q\,\bar{q}\right\rangle \left\langle q\,q\,\bar{q}\left|H_{gqq}\right|q\,g\right\rangle \left\langle q\,g\left|H_{gqq}\right|q\right\rangle }{2(E_{qq\bar{q}}-E_{q})\,(E_{qg}-E_{q})},
  \end{equation}
  \begin{equation}\label{qq2}
\left|q\right\rangle _{qq\bar{q}}^{inst}\,\equiv\,-\frac{\left|q\,q\,\bar{q}\right\rangle \left\langle q\,\bar{q}\,q\left|H_{qq-inst}\right|q\right\rangle }{2(E_{qq\bar{q}}-E_{q})},
\end{equation} 
   \begin{equation}\label{agg1}
\left|q\right\rangle _{qgg}^{reg,\,1}\,\equiv\,\frac{\left|q\,g\,g\right\rangle \left\langle q\,g\,g\left|H_{gqq}\right|q\,g\right\rangle \left\langle q\,g\left|H_{gqq}\right|q\right\rangle }{2(E_{qgg}-E_{q})\,(E_{qg}-E_{q})},
  \end{equation}
    \begin{equation}\label{agg2}
\left|q\right\rangle _{qgg}^{inst,\,1}\,\equiv\,-\frac{\left|q\,g\,g\right\rangle \left\langle q\,g\,g\left|H_{gg-inst}^{q}\right|q\right\rangle }{2(E_{qgg}-E_{q})},
 \end{equation}
   \begin{equation}\label{agg3}
\left|q\right\rangle _{qgg}^{reg,\,2}\,\equiv\,\frac{\left|q\,g\,g\right\rangle \left\langle q\,g\,g\left|H_{ggg}\right|q\,g\right\rangle \left\langle q\,g\left|H_{gqq}\right|q\right\rangle }{2(E_{qgg}-E_{q})\,(E_{qg}-E_{q})},
\end{equation}
  \begin{equation}\label{agg4}
\left|q\right\rangle _{qgg}^{inst,\,2}\,\equiv\,-\frac{\left|q\,g\,g\right\rangle \left\langle q\,g\,g\left|H_{gg-inst}^{g}\right|q\right\rangle }{2(E_{qgg}-E_{q})}.
  \end{equation}
  The additional $\frac{1}{2}$ in each of the states defined above is required from the rules of perturbation theory to divide by $n!$ whenever $n$ particles are involved in the intermediate or final state.
   
   In this Appendix we will compute explicitly the states (\ref{qq1})--(\ref{agg4}), by using the results for the matrix elements from appendix \ref{mateleapp}. A similar calculation has been done in \cite{Lappi:2016oup}, but the results presented there do not include all the states appearing in  Eqs.~(\ref{qq1})--(\ref{agg4}). (The states involving the instantaneous interaction vertices are not explicitly shown in \cite{Lappi:2016oup}.) The calculations will be performed in transverse momentum space but the results will be Fourier transformed to the transverse coordinate space with the help of mathematical identities collected in appendix \ref{fouriers}. 
 
 \subsection{The $qq\overline{q}$ component}\label{qqqsubs}

We start our calculation with the Fock states involving 3 quarks, as defined in Eqs.~(\ref{qq1})  and (\ref{qq2}). The corresponding graphs are exhibited in Fig.~\ref{quarfig}.

\subsubsection*{$\bullet$ Computation of $\left|q_{\lambda}^{\alpha}\right\rangle _{qq\bar{q}}^{reg}$}

\begin{figure}[!t]
\center
\includegraphics[scale=0.8]{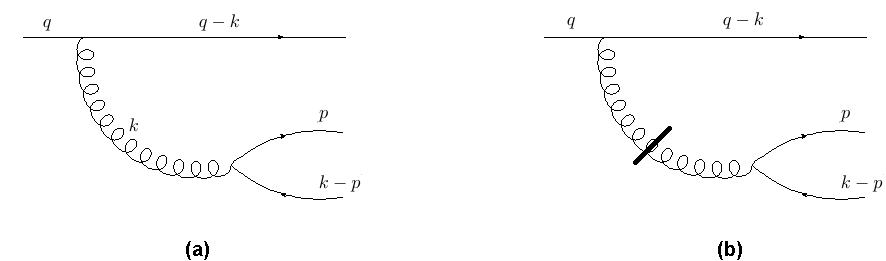}
  \caption{Feynman graphs for the emission of a quark-antiquark pair.  Fig.~(a) involves a genuine (propagating) 
 intermediate gluon and corresponds to 
  $\left|q_{\lambda}^{\alpha}\right\rangle _{qq\bar{q}}^{reg}$, cf. \eqn{qq1}. 
  Fig.~(b) involves an instantaneous gluon exchange, graphically represented as a ``cut''  gluon line; it corresponds to
    $\left|q_{\lambda}^{\alpha}\right\rangle _{qq\bar{q}}^{inst}$, cf. \eqn{qq2}.  \label{quarfig}}
\end{figure}

Using the transverse-momentum representation, the contribution of the diagram in Fig.~\ref{quarfig}.a can be explicitly written as

\begin{equation}\begin{split}
&\left|q_{\lambda}^{\alpha}(q^{+},\,\bm{q})\right\rangle _{qq\bar{q}}^{reg}\\
&=\,\frac{1}{2}\int_{0}^{\infty}\frac{ds^{+}}{2\pi}\,\frac{dk^{+}}{2\pi}\,\frac{du^{+}}{2\pi}\,\frac{dt^{+}}{2\pi}\,\frac{dp^{+}}{2\pi}\,\int\frac{d^{2}\bm{s}}{(2\pi)^{2}}\,\frac{d^{2}\bm{k}}{(2\pi)^{2}}\,\frac{d^{2}\bm{u}}{(2\pi)^{2}}\,\frac{d^{2}\bm{t}}{(2\pi)^{2}}\,\frac{d^{2}\bm{p}}{(2\pi)^{2}}\:\left|\bar{q}_{\lambda_{4}}^{\epsilon}(u)\,q_{\lambda_{3}}^{\delta}(t)\,q_{\lambda_{2}}^{\gamma}(p)\right\rangle \\
&\times\frac{1}{\Delta(q;\,s,\,k)\,\Delta(q;\,u,\,t,\,p)}\left\langle \bar{q}_{\lambda_{4}}^{\epsilon}(u)\,q_{\lambda_{3}}^{\delta}(t)\,q_{\lambda_{2}}^{\gamma}(p)\left|\mathsf{H}_{g\rightarrow qq}\right|q_{\lambda_{1}}^{\beta}(s)\,g_{i}^{a}(k)\right\rangle \left\langle q_{\lambda_{1}}^{\beta}(s)\,g_{i}^{a}(k)\left|\mathsf{H}_{q\rightarrow qg}\right|q_{\lambda}^{\alpha}(q)\right\rangle .
\end{split}\end{equation}
where the energy denominator is defined by:
\begin{equation}\label{deltatwo}
\Delta(q;\,u,\,p,\,t)\,\equiv\,\frac{\bm{u}^{2}}{2u^{+}}\,+\,\frac{\bm{p}^{2}}{2p^{+}}\,+\,\frac{\bm{t}^{2}}{2t^{+}}\,-\,\frac{\bm{q}^{2}}{2q^{+}}\,.
\end{equation}
After inserting the relevant matrix elements, cf.~(\ref{gqq1}), one obtains
\begin{equation}\begin{split}
&\left|q_{\lambda}^{\alpha}(q^{+},\,\bm{q})\right\rangle _{qq\bar{q}}^{reg}\\
&=\,\frac{1}{2}\int_{0}^{\infty}\frac{ds^{+}}{2\pi}\,\frac{dk^{+}}{2\pi}\,\frac{du^{+}}{2\pi}\,\frac{dt^{+}}{2\pi}\,\frac{dp^{+}}{2\pi}\,\int\frac{d^{2}\bm{s}}{(2\pi)^{2}}\,\frac{d^{2}\bm{k}}{(2\pi)^{2}}\,\frac{d^{2}\bm{u}}{(2\pi)^{2}}\,\frac{d^{2}\bm{t}}{(2\pi)^{2}}\,\frac{d^{2}\bm{p}}{(2\pi)^{2}}\:\frac{1}{\Delta(q;\,s,\,k)\,\Delta(q;\,u,\,t,\,p)}\\
&\times\left(\frac{gt_{\gamma\epsilon}^{a}\delta^{\beta\delta}\delta_{\lambda_{1}\lambda_{3}}}{2\sqrt{2k^{+}}}\chi_{\lambda_{2}}^{\dagger}\left[\frac{2\bm{k}^{i}}{k^{+}}-\frac{\sigma\cdot\bm{p}}{p^{+}}\sigma^{i}-\sigma^{i}\frac{\sigma\cdot\bm{u}}{u^{+}}\right]\chi_{\lambda_{4}}(2\pi)^{6}\delta^{(3)}(s-t)\delta^{(3)}(k-p-u)\right.\\
&\left.+\frac{gt_{\delta\epsilon}^{a}\delta^{\beta\gamma}\delta_{\lambda_{1}\lambda_{2}}}{2\sqrt{2k^{+}}}\chi_{\lambda_{3}}^{\dagger}\left[\frac{2\bm{k}^{i}}{k^{+}}-\frac{\sigma\cdot\bm{t}}{t^{+}}\sigma^{i}-\sigma^{i}\frac{\sigma\cdot\bm{u}}{u^{+}}\right]\chi_{\lambda_{4}}(2\pi)^{6}\delta^{(3)}(s-p)\delta^{(3)}(k-t-u)\right)\\
&\times\frac{gt_{\beta\alpha}^{a}}{2\sqrt{2k^{+}}}\chi_{\lambda_{1}}^{\dagger}\left[\frac{2\bm{k}^{i}}{k^{+}}-\frac{\sigma\cdot\bm{s}}{s^{+}}\sigma^{i}-\sigma^{i}\frac{\sigma\cdot\bm{q}}{q^{+}}\right]\chi_{\lambda}(2\pi)^{3}\delta^{(3)}(k+s-q)\left|\bar{q}_{\lambda_{4}}^{\epsilon}(u)\,q_{\lambda_{3}}^{\delta}(t)\,q_{\lambda_{2}}^{\gamma}(p)\right\rangle .
\end{split}\end{equation} 
After using the 3 $\delta$-functions to perform the integrations over $s$, $t$, and $u$, one deduces
\begin{equation}\begin{split}\label{qqsta}
&\left|q_{\lambda}^{\alpha}(q^{+},\,\bm{q})\right\rangle _{qq\bar{q}}^{reg}\,=\,\int_{0}^{q^{+}}dk^{+}\,dp^{+}\,\int d^{2}\bm{k}\,d^{2}\bm{p}\:\frac{g^{2}t_{\delta\epsilon}^{a}t_{\beta\alpha}^{a}}{8(2\pi)^{6}k^{+}\Delta(q;\,k,\,q-k)\,\Delta(q;\,k-p,\,p,\,q-k)}\\
&\times\left(\chi_{\lambda_{2}}^{\dagger}\left[\frac{2\bm{k}^{i}}{k^{+}}-\frac{\sigma\cdot\bm{p}}{p^{+}}\sigma^{i}-\sigma^{i}\frac{\sigma\cdot(\bm{k}-\bm{p})}{k^{+}-p^{+}}\right]\chi_{\lambda_{3}}\right)\left(\chi_{\lambda_{1}}^{\dagger}\left[\frac{2\bm{k}^{i}}{k^{+}}-\frac{\sigma\cdot(\bm{q}-\bm{k})}{q^{+}-k^{+}}\sigma^{i}-\sigma^{i}\frac{\sigma\cdot\bm{q}}{q^{+}}\right]\chi_{\lambda}\right)\\
&\times\left|\bar{q}_{\lambda_{3}}^{\epsilon}(k-p)\,q_{\lambda_{2}}^{\delta}(p)\,q_{\lambda_{1}}^{\beta}(q-k)\right\rangle .
\end{split}\end{equation} 
By introducing new variables according to
\begin{equation}\label{chanvar2}
\vartheta\equiv\frac{k^{+}}{q^{+}},\qquad\xi\equiv\frac{p^{+}}{k^{+}},\qquad\bm{k}\,=\,\vartheta\bm{q}+\bm{\widetilde{k}},\qquad\bm{p}=\xi\bm{k}+\bm{\widetilde{p}}=\xi\vartheta\bm{q}+\xi\bm{\widetilde{k}}+\bm{\widetilde{p}},
\end{equation}
one can easily check that
\begin{equation}
\frac{2\bm{k}^{i}}{k^{+}}-\frac{\sigma\cdot\bm{p}}{p^{+}}\sigma^{i}-\sigma^{i}\frac{\sigma\cdot(\bm{k}-\bm{p})}{k^{+}-p^{+}}\,=\,\frac{(2\xi-1)\delta^{ij}+i\varepsilon^{ij}\sigma^{3}}{\xi(1-\xi)\vartheta q^{+}}\bm{\widetilde{p}}^{j},
\end{equation}
whereas the energy denominator (\ref{deltatwo}) simplifies to
\begin{equation}\begin{split}
&\Delta(q;\,k-p,\,p,\,q-k)\,=\,\frac{(\bm{k}-\bm{p})^{2}}{2(k^{+}-p^{+})}\,+\,\frac{\bm{p}^{2}}{2p^{+}}\,+\,\frac{(\bm{q}-\bm{k})^{2}}{2(q^{+}-k^{+})}\,-\,\frac{\bm{q}^{2}}{2q^{+}}\,=\,\frac{\xi(1-\xi)\bm{\widetilde{k}}^{2}+(1-\vartheta)\bm{\widetilde{p}}^{2}}{2\xi(1-\xi)\vartheta(1-\vartheta)q^{+}}.
\end{split}\end{equation}
Putting together the previous results, Eq.~(\ref{qqsta}) takes the form:
\begin{equation}\begin{split}\label{qqqsta}
&\left|q_{\lambda}^{\alpha}(q^{+},\,\bm{q})\right\rangle _{qq\bar{q}}^{reg}\,\equiv\,\int_{0}^{1}d\vartheta\,d\xi\,\int d^{2}\bm{\widetilde{k}}\,d^{2}\bm{\widetilde{p}}\:\frac{g^{2}t_{\delta\epsilon}^{a}t_{\beta\alpha}^{a}\varphi_{\lambda_{2}\lambda_{3}}^{il}(\xi)\,\phi_{\lambda_{1}\lambda}^{ij}(\vartheta)\,\bm{\widetilde{p}}^{l}\,\bm{\widetilde{k}}^{j}\,(1-\vartheta)q^{+}}{2(2\pi)^{6}\bm{\widetilde{k}}^{2}\left(\xi(1-\xi)\bm{\widetilde{k}}^{2}+(1-\vartheta)\bm{\widetilde{p}}^{2}\right)}\\
&\times\left|\bar{q}_{\lambda_{3}}^{\epsilon}((1-\xi)\vartheta q^{+},\,\vartheta(1-\xi)\bm{q}+(1-\xi)\bm{\widetilde{k}}-\bm{\widetilde{p}})\,q_{\lambda_{2}}^{\delta}(\xi\vartheta q^{+},\,\xi\vartheta\bm{q}+\xi\bm{\widetilde{k}}+\bm{\widetilde{p}})\right.\\
&\left.\;\quad q_{\lambda_{1}}^{\beta}((1-\vartheta)q^{+},(1-\vartheta)\bm{q}-\bm{\widetilde{k}})\right\rangle,
\end{split}\end{equation}
where we introduced the new notation
\begin{equation}
\varphi_{\lambda_{2}\lambda_{1}}^{ij}(\xi)\,\equiv\,\chi_{\lambda_{2}}^{\dagger}\left[(2\xi-1)\delta^{ij}+i\varepsilon^{ij}\sigma^{3}\right]\chi_{\lambda_{1}}\,=\,\delta_{\lambda_{1}\lambda_{2}}\left[(2\xi-1)\delta^{ij}+2i\varepsilon^{ij}\lambda_{1}\right].
\end{equation}
The following identity will also be useful:
 \begin{equation}\begin{split}\label{simpphi}
&\varphi_{\lambda_{1}\lambda}^{ij\dagger}(\xi)\,\varphi_{\lambda_{1}\lambda}^{ik}(\xi)\,=\,\chi_{\lambda}^{\dagger}\left[(2\xi-1)\delta^{ij}-i\varepsilon^{ij}\sigma^{3}\right]\,I\,\left[(2\xi-1)\delta^{ik}+i\varepsilon^{ik}\sigma^{3}\right]\chi_{\lambda}\\
&=\,2\delta^{jk}\,\left[1+(1-2\xi)^{2}\right].
\end{split}\end{equation}
The Fourier transform to coordinate space, $\bm{q}\to \bm{w}$, can now be easily computed with the help of Eq.~(\ref{fourier.4}). One finds
\begin{equation}\begin{split}\label{qqqreg}
&\left|q_{\lambda}^{\alpha}(q^{+},\,\bm{w})\right\rangle _{qq\bar{q}}^{reg}\,=\,-\int_{0}^{1}d\vartheta\,d\xi\,\int_{\bm{x},\,\bm{z},\,\bm{z}^{\prime}}\:\frac{g^{2}t_{\delta\epsilon}^{a}t_{\beta\alpha}^{a}\varphi_{\lambda_{2}\lambda_{3}}^{il}(\xi)\,\phi_{\lambda_{1}\lambda}^{ij}(\vartheta)\,\bm{Z}^{l}\,\left(\bm{X}^{j}+\xi\bm{Z}^{j}\right)\,(1-\vartheta)q^{+}}{2(2\pi)^{4}\bm{Z}^{2}\left(\xi(1-\xi)\bm{Z}^{2}\,+\,(1-\vartheta)\left(\bm{X}+\xi\bm{Z}\right)^{2}\right)}\\
&\times\delta^{(2)}\left(\bm{w}-\bm{C}\right)\left|\bar{q}_{\lambda_{3}}^{\epsilon}((1-\xi)\vartheta q^{+},\,\bm{z})\,q_{\lambda_{2}}^{\delta}(\xi\vartheta q^{+},\,\bm{z}^{\prime})\,q_{\lambda_{1}}^{\beta}((1-\vartheta)q^{+},\bm{x})\right\rangle .
\end{split}\end{equation}
with $\bm{Z}\equiv\bm{z}-\bm{z}^{\prime}$ and the 2-dimensional $\delta$-function ensures that transverse coordinate $\bm{w}$ of the incoming quark coincides, as it should, with the centre of energy of the three-fermion system in the final state:
 \begin{equation}\label{conditionc}
\bm{C}\,\equiv\,(1-\vartheta)\bm{x}\,+\,\xi\vartheta\bm{z}^{\prime}\,+\,(1-\xi)\vartheta\bm{z}.
 \end{equation}

  \subsubsection*{$\bullet$ Computation of $\left|q_{\lambda}^{\alpha}\right\rangle _{qq\bar{q}}^{inst}$}
  
The contribution of the diagram with an instantaneous gluon exchange to the three-quark Fock component can be evaluated as (see Fig.~\ref{quarfig}.b)
\begin{equation}\begin{split}
&\left|q_{\lambda}^{\alpha}\right\rangle _{qq\bar{q}}^{inst}\,=-\frac{1}{2}\int_{0}^{\infty}\frac{dp^{+}}{2\pi}\,\frac{du^{+}}{2\pi}\,\frac{dt^{+}}{2\pi}\,\int\frac{d^{2}\bm{p}}{(2\pi)^{2}}\,\frac{d^{2}\bm{u}}{(2\pi)^{2}}\,\frac{d^{2}\bm{t}}{(2\pi)^{2}}\\
&\times\frac{1}{\Delta(q;\,t,\,p,\,u)}\left|\overline{q}_{\lambda_{3}}^{\epsilon}(t)\,q_{\lambda_{2}}^{\delta}(p)\,q_{\lambda_{1}}^{\beta}(u)\right\rangle \left\langle \overline{q}_{\lambda_{3}}^{\epsilon}(t)\,q_{\lambda_{2}}^{\delta}(p)\,q_{\lambda_{1}}^{\beta}(u)\left|\mathsf{H}_{qq-inst}\right|q_{\lambda}^{\alpha}(q)\right\rangle .
\end{split}\end{equation} 
After inserting the matrix element (\ref{instqq}) and changing variables according to (\ref{chanvar2}), one finds
\begin{equation}\begin{split}\label{qq_cont2}
&\left|q_{\lambda}^{\alpha}\right\rangle _{qq\bar{q}}^{inst}\,=\,-\int_{0}^{1}d\vartheta\,d\xi\,\int d^{2}\bm{\widetilde{k}}\,d^{2}\bm{\widetilde{p}}\:\frac{g^{2}t_{\delta\epsilon}^{a}t_{\beta\alpha}^{a}\,(1-\vartheta)\xi(1-\xi)q^{+}}{(2\pi)^{6}\left(\xi(1-\xi)\bm{\widetilde{k}}^{2}+(1-\vartheta)\bm{\widetilde{p}}^{2}\right)}\,\delta_{\lambda_{3}\lambda_{2}}\,\delta_{\lambda_{1}\lambda}\\
&\times\left|\bar{q}_{\lambda_{3}}^{\epsilon}((1-\xi)\vartheta q^{+},\,\vartheta(1-\xi)\bm{q}+(1-\xi)\bm{\widetilde{k}}-\bm{\widetilde{p}})\,q_{\lambda_{2}}^{\delta}(\xi\vartheta q^{+},\,\xi\vartheta\bm{q}+\xi\bm{\widetilde{k}}+\bm{\widetilde{p}})\right.\\
&\left.\;\quad q_{\lambda_{1}}^{\beta}((1-\vartheta)q^{+},(1-\vartheta)\bm{q}-\bm{\widetilde{k}})\right\rangle 
\end{split}\end{equation}
One can now perform the Fourier transform to coordinate space using (\ref{fourier.3}), to find
\begin{equation}\begin{split}\label{qqqinst}
&\left|q_{\lambda}^{\alpha}\right\rangle _{qq\bar{q}}^{inst}\,\equiv\,-\int_{0}^{1}d\vartheta\,d\xi\,\int_{\bm{x},\,\bm{z},\,\bm{z}^{\prime}}\:\frac{g^{2}t_{\delta\epsilon}^{a}t_{\beta\alpha}^{a}\,(1-\vartheta)\xi(1-\xi)q^{+}}{(2\pi)^{4}\left(\xi(1-\xi)\bm{Z}^{2}\,+\,(1-\vartheta)\left(\bm{X}+\xi\bm{Z}\right)^{2}\right)}\\
&\times\delta^{(2)}\left(\bm{w}-\bm{C}\right)\left|\bar{q}_{\lambda_{1}}^{\epsilon}((1-\xi)\vartheta q^{+},\,\bm{z})\,q_{\lambda_{1}}^{\delta}(\xi\vartheta q^{+},\,\bm{z}^{\prime})\,q_{\lambda}^{\beta}((1-\vartheta)q^{+},\bm{x})\right\rangle ,
\end{split}\end{equation}
with $\bm{C}$ as defined in \eqn{conditionc}.

\subsection{The $qgg$ component} \label{qggsubs}
We now turn to the four Fock space components which involve a gluon pair in the final state, together with the incoming quark.
\subsubsection*{$\bullet$ Computation of $\left|q_{\lambda}^{\alpha}\right\rangle _{qgg}^{reg,\,1}$}
    \begin{figure}[!t]
\center \includegraphics[scale=0.8]{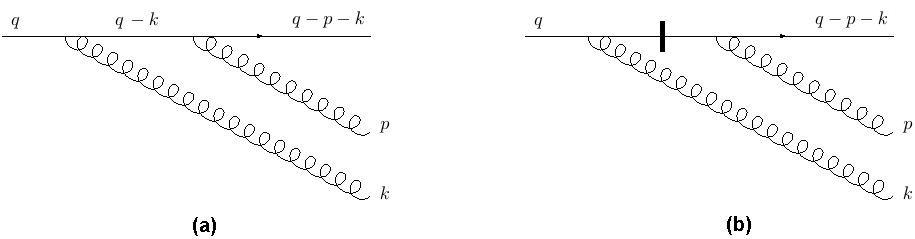}
\caption{The diagrams for two gluons emissions by the initial quark. The fermion propagator with a cut represents an instantaneous quark exchange. Fig.~(a) corresponds to the state  $\left|q_{\lambda}^{\alpha}\right\rangle _{qgg}^{reg,\,1}$, cf. Eq.~(\ref{agg1}), while Fig.~(b) to  $\left|q_{\lambda}^{\alpha}\right\rangle _{qgg}^{inst,\,1}$, cf. Eq.~(\ref{agg2}). \label{figugg}}
\end{figure}

This state is defined in (\ref{agg1}) and represented by the diagram in Fig.~\ref{figugg}.a, which is
computed as
\begin{equation}\begin{split}\label{ggone}
&\left|q_{\lambda}^{\alpha}\right\rangle _{qgg}^{reg,\,1}\equiv\,\frac{1}{2}\int_{0}^{\infty}\frac{dk^{+}}{2\pi}\,\frac{ds^{+}}{2\pi}\,\frac{dp^{+}}{2\pi}\,\frac{du^{+}}{2\pi}\,\frac{dt^{+}}{2\pi}\,\int\frac{d^{2}\bm{k}}{(2\pi)^{2}}\,\frac{d^{2}\bm{s}}{(2\pi)^{2}}\,\frac{d^{2}\bm{p}}{(2\pi)^{2}}\,\frac{d^{2}\bm{u}}{(2\pi)^{2}}\,\frac{d^{2}\bm{t}}{(2\pi)^{2}}\,\left|q_{\lambda_{2}}^{\gamma}(u)\,g_{j}^{b}(t)\,g_{l}^{c}(p)\right\rangle  \\
&\times\frac{1}{\Delta(q;\,k,\,s)\,\Delta(q;\,p,\,t,\,u)}\left\langle q_{\lambda_{2}}^{\gamma}(u)\,g_{j}^{b}(t)\,g_{l}^{c}(p)\left|\mathsf{H}_{q\rightarrow qg}\right|q_{\lambda_{1}}^{\beta}(s)\,g_{i}^{a}(k)\right\rangle \left\langle q_{\lambda_{1}}^{\beta}(s)\,g_{i}^{a}(k)\left|\mathsf{H}_{q\rightarrow qg}\right|q_{\lambda}^{\alpha}(q)\right\rangle ,
\end{split}\end{equation}
where the overall factor $\frac{1}{2}$ is to avoid the double-counting of the two-gluon states. After inserting the matrix elements in Eqs.~(\ref{qqg}) and (\ref{qggqg}), one obtains
\begin{eqnarray}
&&\left|q_{\lambda}^{\alpha}\right\rangle _{qgg}^{reg,\,1}=\,\frac{1}{2}\int_{0}^{\infty}\frac{dk^{+}}{2\pi}\,\frac{ds^{+}}{2\pi}\,\frac{dp^{+}}{2\pi}\,\frac{du^{+}}{2\pi}\,\frac{dt^{+}}{2\pi}\,\int\frac{d^{2}\bm{k}}{(2\pi)^{2}}\,\frac{d^{2}\bm{s}}{(2\pi)^{2}}\,\frac{d^{2}\bm{p}}{(2\pi)^{2}}\,\frac{d^{2}\bm{u}}{(2\pi)^{2}}\,\frac{d^{2}\bm{t}}{(2\pi)^{2}}\,\frac{1}{\Delta(q;\,k,\,s)\,\Delta(q;\,t,\,p,\,u)}\nonumber\\
&&\times\left(\frac{gt_{\gamma\beta}^{b}}{2\sqrt{2t^{+}}}\chi_{\lambda_{2}}^{\dagger}\left[\frac{2\bm{t}^{j}}{t^{+}}-\frac{\sigma\cdot\bm{u}}{u^{+}}\sigma^{j}-\sigma^{j}\frac{\sigma\cdot\bm{s}}{s^{+}}\right]\chi_{\lambda_{1}}(2\pi)^{6}\delta^{(3)}(u+t-s)\delta^{ac}\delta_{il}\delta^{(3)}(k-p)\right.\\
&&\left.+\frac{gt_{\gamma\beta}^{c}}{2\sqrt{2p^{+}}}\chi_{\lambda_{2}}^{\dagger}\left[\frac{2\bm{p}^{l}}{p^{+}}-\frac{\sigma\cdot\bm{u}}{u^{+}}\sigma^{l}-\sigma^{l}\frac{\sigma\cdot\bm{s}}{s^{+}}\right]\chi_{\lambda_{1}}(2\pi)^{6}\delta^{(3)}(u+p-s)\delta^{ab}\delta_{ij}\delta^{(3)}(k-t)\right)\nonumber\\
&&\times\left(\frac{gt_{\beta\alpha}^{a}}{2\sqrt{2k^{+}}}\chi_{\lambda_{1}}^{\dagger}\left[\frac{2\bm{k}^{i}}{k^{+}}-\frac{\sigma\cdot\bm{s}}{s^{+}}\sigma^{i}-\sigma^{i}\frac{\sigma\cdot\bm{q}}{q^{+}}\right]\chi_{\lambda}(2\pi)^{3}\delta^{(3)}(k+s-q)\right)\left|q_{\lambda_{2}}^{\gamma}(u)\,g_{j}^{b}(t)\,g_{l}^{c}(p)\right\rangle .\nonumber
\end{eqnarray}
Integrating over $s$, $u$ and $t$ yields
\begin{equation}\begin{split}\label{step1aa}
&\left|q_{\lambda}^{\alpha}\right\rangle _{qgg}^{reg,\,1}=\,\int_{0}^{q^{+}}dk^{+}\,dp^{+}\,\int d^{2}\bm{k}\,d^{2}\bm{p}\,\frac{g^{2}t_{\gamma\beta}^{b}t_{\beta\alpha}^{a}}{8(2\pi)^{6}\sqrt{k^{+}p^{+}}\,\Delta(q;\,k,\,q-k)\,\Delta(q;\,k,\,p,\,q-p-k)}\\
&\times\chi_{\lambda_{2}}^{\dagger}\,\left[\frac{2\bm{p}^{j}}{p^{+}}-\frac{\sigma\cdot(\bm{q}-\bm{k}-\bm{p})}{q^{+}-k^{+}-p^{+}}\sigma^{j}-\sigma^{j}\frac{\sigma\cdot(\bm{q}-\bm{k})}{q^{+}-k^{+}}\right]\chi_{\lambda_{1}}\,\chi_{\lambda_{1}}^{\dagger}\,\left[\frac{2\bm{k}^{i}}{k^{+}}-\frac{\sigma\cdot(\bm{q}-\bm{k})}{q^{+}-k^{+}}\sigma^{i}-\sigma^{i}\frac{\sigma\cdot\bm{q}}{q^{+}}\right]\\
&\times\chi_{\lambda}\,\left|q_{\lambda_{2}}^{\gamma}(q-p-k)\, g_{j}^{b}(p)\, g_{i}^{a}(k)\right\rangle .
\end{split}\end{equation}
At this point we introduce ``better'' variables,
\begin{equation}\label{vari1}
\vartheta\equiv\frac{k^{+}}{q^{+}},\qquad\xi\equiv\frac{p^{+}}{q^{+}-k^{+}},\qquad\bm{k}\,=\,\vartheta\bm{q}+\bm{\widetilde{k}},\qquad\bm{p}=\xi(\bm{q}-\bm{k})+\bm{\widetilde{p}}=\xi(1-\vartheta)\bm{q}-\xi\bm{\widetilde{k}}+\bm{\widetilde{p}},
\end{equation}
in terms of which the energy denominators take relatively simple forms, that is, \eqn{enede1} for
$\Delta(q;\,k,\,q-k)$ and respectively
\begin{equation}\begin{split}
&\Delta(q;\,k,\,p,\,q-k-p)\,=\,\frac{\bm{k}^{2}}{2k^{+}}\,+\,\frac{\bm{p}^{2}}{2p^{+}}\,+\,\frac{(\bm{q}-\bm{k}-\bm{p})^{2}}{2(q^{+}-k^{+}-p^{+})}\,-\,\frac{\bm{q}^{2}}{2q^{+}}\,=\,\frac{\vartheta\bm{\widetilde{p}}^{2}+\xi(1-\xi)\bm{\widetilde{k}}^{2}}{2\xi(1-\xi)\vartheta(1-\vartheta)q^{+}}.
\end{split}\end{equation}
Then, the r.h.s. of Eq.~(\ref{step1aa}) becomes:
\begin{equation}\begin{split}
&\left|q_{\lambda}^{\alpha}\right\rangle _{qgg}^{reg,\,1}=\,\int_{0}^{1}d\xi\,d\vartheta\,\int d^{2}\bm{\widetilde{k}}\,d^{2}\bm{\widetilde{p}}\,\frac{g^{2}t_{\gamma\beta}^{b}t_{\beta\alpha}^{a}\,\phi_{\lambda_{2}\lambda_{1}}^{jm}(\xi)\,\phi_{\lambda_{1}\lambda}^{il}(\vartheta)\,\bm{\widetilde{k}}^{l}\,\bm{\widetilde{p}}^{m}\,\sqrt{1-\vartheta}\,q^{+}}{2(2\pi)^{6}\sqrt{\xi}\,\bm{\widetilde{k}}^{2}\,\left(\vartheta\bm{\widetilde{p}}^{2}+\xi(1-\xi)\bm{\widetilde{k}}^{2}\right)}\\
&\times\left|q_{\lambda_{2}}^{\gamma}\left((1-\xi)(1-\vartheta)q^{+},\,(1-\xi)(1-\vartheta)\bm{q}-(1-\xi)\bm{\widetilde{k}}-\bm{\widetilde{p}}\right)\right.\\
&\left. \qquad g_{j}^{b}\left(\xi(1-\vartheta)q^{+},\,\xi(1-\vartheta)\bm{q}-\xi\bm{\widetilde{k}}+\bm{\widetilde{p}}\right)\, g_{i}^{a}\left(\vartheta q^{+},\,\vartheta\bm{q}+\bm{\widetilde{k}}\right)\right\rangle .
\end{split}\end{equation}
After a Fourier transformation with the help of Eq.~(\ref{fourier.4}), we finally deduce
\begin{equation}\label{qggreg1}
\begin{split}
&\left|q_{\lambda}^{\alpha}\right\rangle _{qgg}^{reg,\,1}\,\equiv\,\int_{\bm{x},\,\bm{z},\,\bm{z}^{\prime}}\int_{0}^{1}d\xi\,d\vartheta\,\frac{g^{2}t_{\gamma\beta}^{b}t_{\beta\alpha}^{a}\,\phi_{\lambda_{2}\lambda_{1}}^{jm}(\xi)\,\phi_{\lambda_{1}\lambda}^{il}(\vartheta)\,\left(\xi\bm{X}-\bm{X^{\prime}}\right)^{l}\,\bm{X}^{m}\,\sqrt{(1-\vartheta)}q^{+}}{2(2\pi)^{4}\sqrt{\xi}\,\bm{X}^{2}\,\left(\vartheta\left(\xi\bm{X}-\bm{X^{\prime}}\right)^{2}+\xi(1-\xi)\bm{\bm{X}}^{2}\right)}\\
&\times\delta^{(2)}\left(\bm{w}-\bm{D}\right)\left|q_{\lambda_{2}}^{\gamma}((1-\xi)(1-\vartheta)q^{+},\,\bm{x})\,g_{j}^{b}(\xi(1-\vartheta)q^{+},\,\bm{z})\,g_{i}^{a}(\vartheta q^{+},\,\bm{z}^{\prime})\right\rangle .
 \end{split}\end{equation}
where $\bm{D}$ is the centre of energy of the three-parton final state:
\begin{equation}\label{conditiond}
\bm{D}\,\equiv\,(1-\xi)(1-\vartheta)\bm{x}\,+\,\xi(1-\vartheta)\bm{z}\,+\,\vartheta\bm{z}^{\prime}.
\end{equation}

\subsubsection*{$\bullet$  Computation of $\left|q_{\lambda}^{\alpha}\right\rangle _{qgg}^{inst,\,1}$}

This state, which is defined in Eq.~(\ref{agg2}), involves the instantaneous quark interaction, cf.
Fig.~\ref{figugg}.b, and reads:
\begin{equation}\begin{split}
&\left|q_{\lambda}^{\alpha}\right\rangle _{qgg}^{inst,\,1}\,\equiv\,-\frac{1}{2}\int_{0}^{\infty}\frac{dp^{+}}{2\pi}\,\frac{dt^{+}}{2\pi}\,\frac{du^{+}}{2\pi}\,\int\frac{d^{2}\bm{p}}{(2\pi)^{2}}\,\frac{d^{2}\bm{t}}{(2\pi)^{2}}\,\frac{d^{2}\bm{u}}{(2\pi)^{2}}\,\frac{1}{\Delta(q;\,p,\,t,\,u)}\\
&\times\left|q_{\lambda_{1}}^{\gamma}(u)\,g_{j}^{b}(t)\,g_{i}^{c}(p)\right\rangle \left\langle q_{\lambda_{1}}^{\gamma}(u)\,g_{j}^{b}(t)\,g_{i}^{c}(p)\left|\mathsf{H}_{q\rightarrow qgg}^{inst\:q}\right|q_{\lambda}^{\alpha}(q)\right\rangle .
\end{split}\end{equation}
After inserting the matrix element (\ref{instq}) and using the symmetry between the 2 variables $t$ and $p$, one finds
\begin{equation}\begin{split}
&\left|q_{\lambda}^{\alpha}\right\rangle _{qgg}^{inst,\,1}\,=\,-\,\int_{0}^{q^{+}}\frac{dp^{+}}{2\pi}\,\frac{dt^{+}}{2\pi}\,\frac{du^{+}}{2\pi}\,\int\frac{d^{2}\bm{p}}{(2\pi)^{2}}\,\frac{d^{2}\bm{t}}{(2\pi)^{2}}\,\frac{d^{2}\bm{u}}{(2\pi)^{2}}\,\frac{g^{2}(2\pi)^{3}\delta(p+t+u-q)}{\sqrt{\xi\vartheta(1-\vartheta)}\Delta(p,t,u)}\\
&\times\frac{t_{\gamma\beta}^{b}t_{\beta\alpha}^{a}}{1-\vartheta}\,\chi_{\lambda_{1}}^{\dagger}\,\sigma_{j}\,\sigma_{i}\,\chi_{\lambda}\left|q_{\lambda_{1}}^{\gamma}(u)\,g_{j}^{b}(t)\,g_{i}^{a}(p)\right\rangle .
\end{split}\end{equation}
In terms of the variables introduced in Eq.~(\ref{vari1}), one can write
\begin{equation}\begin{split}
&\left|q_{\lambda}^{\alpha}\right\rangle _{qgg}^{inst,\,1}\,=\,-\int_{0}^{1}d\xi\,d\vartheta\,\int d^{2}\bm{\widetilde{k}}\,d^{2}\bm{\widetilde{p}}\,\frac{2g^{2}t_{\gamma\beta}^{b}t_{\beta\alpha}^{a}\,\sqrt{\xi\vartheta}(1-\xi)\chi_{\lambda_{1}}^{\dagger}\left(\delta^{ij}-i\varepsilon^{ij}\sigma^{3}\right)\chi_{\lambda}q^{+}}{(2\pi)^{6}\sqrt{(1-\vartheta)}\,\left(\vartheta\bm{\widetilde{p}}^{2}+\xi(1-\xi)\bm{\widetilde{k}}^{2}\right)}\\
&\times\left|q_{\lambda_{1}}^{\gamma}((1-\xi)(1-\vartheta)q^{+},\,(1-\xi)(1-\vartheta)\bm{q}+\xi\bm{\widetilde{k}}-\bm{\widetilde{p}}-\bm{\widetilde{k}})\right.\\
&\left.\qquad g_{j}^{b}(\xi(1-\vartheta)q^{+},\,\xi(1-\vartheta)\bm{q}-\xi\bm{\widetilde{k}}+\bm{\widetilde{p}})\,g_{i}^{a}(\vartheta q^{+},\,\vartheta\bm{q}+\bm{\widetilde{k}})\right\rangle.
\end{split}\end{equation}
The Fourier transform to coordinate space finally yields (cf. Eqs.~(\ref{fourier.3}) and \eqref{conditiond})
\begin{equation}\label{qgginst1}
\begin{split}
&\left|q_{\lambda}^{\alpha}\right\rangle _{qgg}^{inst,\,1}\,=\,-\int_{\bm{x},\,\bm{z},\,\bm{z}^{\prime}}\int_{0}^{1}d\xi\,d\vartheta\,\frac{2g^{2}t_{\gamma\beta}^{b}t_{\beta\alpha}^{a}\,\sqrt{\xi\vartheta}(1-\xi)\chi_{\lambda_{1}}^{\dagger}\left(\delta^{ij}-i\varepsilon^{ij}\sigma^{3}\right)\chi_{\lambda}\,q^{+}}{(2\pi)^{4}\sqrt{1-\vartheta}\,\left(\vartheta\left(\xi\bm{X}-\bm{X^{\prime}}\right)^{2}+\xi(1-\xi)\bm{\bm{X}}^{2}\right)}\\
&\times\delta^{(2)}\left(\bm{w}-\bm{D}\right)\left|q_{\lambda_{1}}^{\gamma}((1-\xi)(1-\vartheta)q^{+},\,\bm{x})\,g_{j}^{b}(\xi(1-\vartheta)q^{+},\,\bm{z})\,g_{i}^{a}(\vartheta q^{+},\,\bm{z}^{\prime})\right\rangle .
\end{split}\end{equation}

\subsubsection*{$\bullet$ Computation of $\left|q_{\lambda}^{\alpha}\right\rangle _{qgg}^{reg,\,2}$}
  \begin{figure}[!t]
\center \includegraphics[scale=0.8]{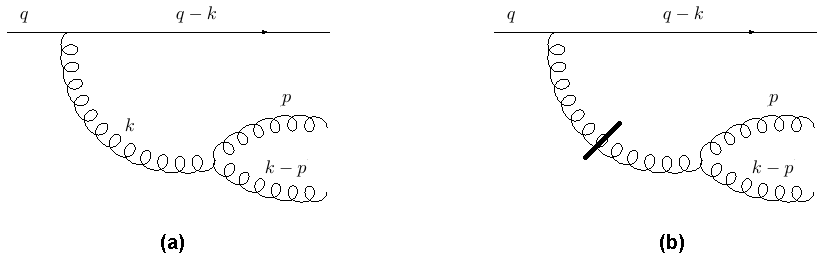}
\caption{Two-gluon production with an intermediate gluon state, corresponding to the two states defined in
 Eq.~(\ref{agg3}) and Eq.~(\ref{agg4}), respectively. \label{ggsplit}}
\end{figure}

This state is defined in (\ref{agg3}) and associated with the diagram in Fig.~\ref{ggsplit}.a. One has
\begin{equation}\begin{split}\label{ggtwo}
&\left|q_{\lambda}^{\alpha}\right\rangle _{qgg}^{reg,\,2}\\
&\equiv\frac{1}{2}\int_{0}^{\infty}\frac{dk^{+}}{2\pi}\,\frac{ds^{+}}{2\pi}\,\frac{dp^{+}}{2\pi}\,\frac{dt^{+}}{2\pi}\,\frac{du^{+}}{2\pi}\,\int\frac{d^{2}\bm{k}}{(2\pi)^{2}}\,\frac{d^{2}\bm{s}}{(2\pi)^{2}}\,\frac{d^{2}\bm{p}}{(2\pi)^{2}}\,\frac{d^{2}\bm{t}}{(2\pi)^{2}}\,\frac{d^{2}\bm{u}}{(2\pi)^{2}}\,\frac{1}{\Delta(q;\,p,\,t,\,u)\,\Delta(q;\,k,s)}\\
&\times\left|q_{\lambda_{2}}^{\gamma}(u)\,g_{j}^{b}(t)\,g_{l}^{c}(p)\right\rangle \left\langle q_{\lambda_{2}}^{\gamma}(u)\,g_{j}^{b}(t)\,g_{l}^{c}(p)\left|\mathsf{H}_{g\rightarrow gg}\right|q_{\lambda_{1}}^{\beta}(s)\,g_{i}^{a}(k)\right\rangle \left\langle q_{\lambda_{1}}^{\beta}(s)\,g_{i}^{a}(k)\left|\mathsf{H}_{q\rightarrow qg}\right|q_{\lambda}^{\alpha}(q)\right\rangle .
\end{split}\end{equation}
By inserting the relevant matrix elements, cf. Eqs.~(\ref{qqg}) and (\ref{qgqgg_split}),  one can deduce
\begin{equation}\begin{split}
&\left|q_{\lambda}^{\alpha}\right\rangle _{qgg}^{reg,\,2}\,=\,\frac{1}{2}\int_{0}^{q^{+}}\frac{dk^{+}}{2\pi}\,\frac{ds^{+}}{2\pi}\,\frac{dp^{+}}{2\pi}\,\frac{du^{+}}{2\pi}\,\frac{dt^{+}}{2\pi}\,\int\frac{d^{2}\bm{k}}{(2\pi)^{2}}\,\frac{d^{2}\bm{s}}{(2\pi)^{2}}\,\frac{d^{2}\bm{p}}{(2\pi)^{2}}\,\frac{d^{2}\bm{u}}{(2\pi)^{2}}\,\frac{d^{2}\bm{t}}{(2\pi)^{2}}\\
&\times\frac{igf^{abc}\delta^{\beta\gamma}\delta_{\lambda_{1}\lambda_{2}}}{2\sqrt{2k^{+}p^{+}t^{+}}\Delta(s,\,k)\,\Delta(u,\,t,\,p)}(2\pi)^{6}\delta^{(3)}(s-u)\delta^{(3)}(k-p-t)\left[\left(\bm{t}^{i}-\bm{p}^{i}+\frac{p^{+}-t^{+}}{k^{+}}\bm{k}^{i}\right)\delta_{jl}\right.\\
&\left.+\left(\bm{k}^{j}+\bm{p}^{j}-\frac{k^{+}+p^{+}}{t^{+}}\bm{t}^{j}\right)\delta_{il}+\left(\frac{k^{+}+t^{+}}{p^{+}}\bm{p}^{l}-\bm{t}^{l}-\bm{k}^{l}\right)\delta_{ij}\right]\\
&\times\left(\frac{gt_{\beta\alpha}^{a}}{2\sqrt{2k^{+}}}\chi_{\lambda_{1}}^{\dagger}\left[\frac{2\bm{k}^{i}}{k^{+}}-\frac{\sigma\cdot\bm{s}}{s^{+}}\sigma^{i}-\sigma^{i}\frac{\sigma\cdot\bm{q}}{q^{+}}\right]\chi_{\lambda}(2\pi)^{3}\delta^{(3)}(k+s-q)\right)\left|q_{\lambda_{2}}^{\gamma}(u)\,g_{j}^{b}(t)\,g_{l}^{c}(p)\right\rangle .\\
\end{split}\end{equation}
The final result takes a simpler form in terms of the variables introduced in Eq.~(\ref{chanvar2}):
\begin{equation}\begin{split}\label{contgg}
&\left|q_{\lambda}^{\alpha}\right\rangle _{qgg}^{reg,\,2}\equiv\,\int_{0}^{1}d\xi\,d\vartheta\,\int d^{2}\bm{\widetilde{k}}\,d^{2}\bm{\widetilde{p}}\,\frac{ig^{2}f^{abc}t_{\beta\alpha}^{a}\phi_{\lambda_{1}\lambda}^{im}(\vartheta)\bm{\widetilde{k}}^{m}\,(1-\vartheta)^{2}\,\sqrt{\xi(1-\xi)}q^{+}}{2(2\pi)^{6}\sqrt{\vartheta}\left(\xi(1-\xi)\bm{\widetilde{k}}^{2}+(1-\vartheta)\bm{\widetilde{p}}^{2}\right)\bm{\widetilde{k}}^{2}}\\
&\times\left(\widetilde{\bm{p}}^{i}\delta_{jl}-\frac{1}{1-\xi}\widetilde{\bm{p}}^{j}\delta_{il}-\frac{1}{\xi}\widetilde{\bm{p}}^{l}\delta_{ij}\right)\left|q_{\lambda_{1}}^{\beta}\left((1-\vartheta)q^{+},\,(1-\vartheta)\bm{q}-\bm{\widetilde{k}}\right)\,\right.\\
&\left.\quad g_{j}^{b}\left(\vartheta(1-\xi)q^{+},\,\vartheta(1-\xi)\bm{q}+(1-\xi)\bm{\widetilde{k}}-\widetilde{\bm{p}}\right)\, g_{l}^{c}\left(\xi\vartheta q^{+},\,\vartheta\xi\bm{q}+\xi\bm{\widetilde{k}}+\widetilde{\bm{p}}\right)\right\rangle .
 \end{split}\end{equation}
 Finally, the Fourier transform to coordinate space involves Eq.~(\ref{fourier.4}):
\begin{equation}\begin{split} \label{qggreg2} 
&\left|q_{\lambda}^{\alpha}\right\rangle _{qgg}^{reg,\,2}\,=\,-\int_{\bm{x},\,\bm{z},\,\bm{z}^{\prime}}\,\int_{0}^{1}d\vartheta\,d\xi\,\frac{ig^{2}t_{\beta\alpha}^{a}f^{abc}\sqrt{\xi(1-\xi)}(1-\vartheta)\phi_{\lambda_{1}\lambda}^{im}(\vartheta)\,q^{+}}{2(2\pi)^{4}\left((1-\vartheta)\left(\bm{X}+\xi\bm{Z}\right)^{2}\,+\,\xi(1-\xi)\bm{Z}^{2}\right)\bm{Z}^{2}}\\
&\times\left(\bm{X}^{m}+\xi\bm{Z}^{m}\right)\left(\bm{Z}^{i}\delta_{jl}-\frac{1}{\xi}\bm{Z}^{j}\delta_{il}-\frac{1}{1-\xi}\bm{Z}^{l}\delta_{ij}\right)\delta^{(2)}\left(\bm{w}-\bm{C}\right)\\
&\times\left|q_{\lambda_{1}}^{\alpha}\left((1-\vartheta)q^{+},\,\bm{x}\right)\,g_{j}^{b}\left(\vartheta\xi q^{+},\,\bm{z}^{\prime}\right)\,g_{l}^{c}\left(\vartheta(1-\xi)q^{+},\,\bm{z}\right)\right\rangle .
 \end{split}\end{equation}
 We recall that the variable $\bm{C}$ was defined in  \eqn{conditionc} and it represents the centre of energy of the final three partons in any of the two diagrams in Fig.~\ref{ggsplit}.

\subsubsection*{$\bullet$ Computation of $\left|q_{\lambda}^{\alpha}\right\rangle _{qgg}^{inst,\,2}$}

This state is defined in (\ref{agg4}) and corresponds to the diagram in Fig.~\ref{ggsplit}.b. One can write
\begin{equation}\begin{split}\label{ggthree}
&\left|q_{\lambda}^{\alpha}\right\rangle _{qgg}^{inst,\,2}=\,-\frac{1}{2}\int_{0}^{\infty}\frac{dp^{+}}{2\pi}\,\frac{du^{+}}{2\pi}\,\frac{dt^{+}}{2\pi}\,\int\frac{d^{2}\bm{p}}{(2\pi)^{2}}\,\frac{d^{2}\bm{u}}{(2\pi)^{2}}\,\frac{d^{2}\bm{t}}{(2\pi)^{2}}\\
&\times\frac{1}{\Delta(q;\,p,\,t,\,u)}\left|q_{\lambda_{1}}^{\gamma}(u)\,g_{j}^{b}(t)\,g_{i}^{a}(p)\right\rangle \left\langle q_{\lambda_{1}}^{\gamma}(u)\,g_{j}^{b}(t)\,g_{i}^{a}(p)\left|\mathsf{H}_{q\rightarrow qgg}^{inst\:g}\right|q_{\lambda}^{\alpha}(q)\right\rangle .
\end{split}\end{equation}
After inserting the relevant matrix element (\ref{instg}), changing variables according to (\ref{chanvar2}) and using symmetry properties under the exchange of the two gluons, we can write the previous result as
\begin{eqnarray}\label{contgg3}
&&\left|q_{\lambda}^{\alpha}\right\rangle _{qgg}^{inst,\,2}\,=\,-\int_{0}^{1}d\xi\,d\vartheta\,\int d^{2}\bm{k}\,d^{2}\widetilde{\bm{p}}\,\frac{ig^{2}t_{\beta\alpha}^{a}f^{abc}(1-2\xi\vartheta)(1-\vartheta)\sqrt{\xi(1-\xi)}q^{+}\delta_{ij}}{2(2\pi)^{6}\left(\xi(1-\xi)\bm{\widetilde{k}}^{2}+(1-\vartheta)\bm{\widetilde{p}}^{2}\right)}\nonumber\\
&&\times\left|q_{\lambda}^{\beta}\left((1-\vartheta)q^{+},\,(1-\vartheta)\bm{q}-\bm{\widetilde{k}}\right)\,g_{i}^{b}\left(\xi\vartheta q^{+},\,\vartheta\xi\bm{q}+\xi\bm{\widetilde{k}}+\widetilde{\bm{p}}\right)\right.\\
&&\quad\left.\,g_{j}^{c}\left(\vartheta(1-\xi)q^{+},\,\vartheta(1-\xi)\bm{q}+(1-\xi)\bm{\widetilde{k}}-\widetilde{\bm{p}}\right)\right\rangle .\nonumber
\end{eqnarray}
The final Fourier transform to transverse coordinate space gives, cf. Eq.~(\ref{fourier.3}),
 \begin{equation}\begin{split}\label{qgginst2} 
&\left|q_{\lambda}^{\alpha}\right\rangle _{qgg}^{inst,\,2}\,=\,-\int_{\bm{x},\,\bm{z},\,\bm{z}^{\prime}}\,\int_{0}^{1}d\vartheta\,d\xi\,\frac{ig^{2}t_{\beta\alpha}^{a}f^{abc}(1-2\xi\vartheta)(1-\vartheta)\sqrt{\xi(1-\xi)}q^{+}\delta_{ij}}{2(2\pi)^{4}\left((1-\vartheta)\left(\bm{X}+\xi\bm{Z}\right)^{2}\,+\,\xi(1-\xi)\bm{Z}^{2}\right)}\\
&\times\delta^{(2)}\left(\bm{w}-\bm{C}\right)\left|q_{\lambda}^{\beta}\left((1-\vartheta)q^{+},\,\bm{x}\right)\,g_{i}^{b}\left(\vartheta\xi q^{+},\,\bm{z}^{\prime}\right)\,g_{j}^{c}\left(\vartheta(1-\xi)q^{+},\,\bm{z}\right)\right\rangle .
\end{split}\end{equation}

 \section{Contracting Wilson lines \label{wilsdict}}
 
In this Appendix we collect our definitions for the various $S$-matrices introduced in Sect.~\ref{trijetfinal} . These $S$-matrices are built as contractions of Wilson lines in either the fundamental or the adjoint representation, corresponding to the partons which exist at the time of scattering ($x^+=0$) in either the direct amplitude, or the complex conjugate amplitude. By using the relation \eqref{Uadj} between the Wilson lines in the two representations, together with the following  Fiertz identities: $t_{ij}^{a}\,t_{kl}^{a}\,=\,\left(\delta_{il}\,\delta_{jk}\,-\,\frac{1}{N_{c}}\delta_{ij}\,\delta_{kl}\right)/\,2$,  $f^{abc}\,t_{\alpha\beta}^{a}\,t_{\gamma\delta}^{b}\,t_{\epsilon\varrho}^{c}\,=\,\frac{i}{4}\left(\delta_{\epsilon\delta}\,\delta_{\beta\gamma}\,\delta_{\alpha\varrho}-\delta_{\alpha\delta}\,\delta_{\varrho\gamma}\,\delta_{\beta\epsilon}\right)$, and $f^{abc}\,t_{\delta\alpha}^{a}\,t_{\gamma\beta}^{b}\,=\,\frac{i}{2}\left(t_{\delta\beta}^{c}\delta_{\alpha\gamma}\,-\,t_{\gamma\alpha}^{c}\,\delta_{\delta\beta}\right)$, it is possible to reexpress all the $S$-matrices as products of  fundamental Wilson lines alone and to also get rid of all the fundamental (Gell-Mann) matrices in between the Wilson lines. For the purpose of compactly presenting the final results, we will also need the hextupole:
 \begin{equation}\label{Hexdef}
\mathcal{H}\left(\overline{\bm{x}},\,\bm{x},\,\overline{\bm{z}},\,\bm{z},\,\overline{\bm{z}}^{\prime},\,\bm{z}^{\prime}\right)\,\equiv\,\frac{1}{N_{c}}\,\mathrm{tr}\left(V^{\dagger}(\overline{\bm{x}})\,V(\bm{x})\,V^{\dagger}(\overline{\bm{z}})\,V(\bm{z})\,V^{\dagger}(\overline{\bm{z}}^{\prime})\,V(\bm{z}^{\prime})\right).
\end{equation}
The average over the color fields in the target is implicitly understood in \eqn{Hexdef} and in all the other equations from this Appendix.

In all the subsequent formulae, we shall also present the limit of the respective $S$-matrix in the limit of a large number of colors  $N_{c}\to \infty$. In fact, for a few of them, we shall exhibit only the original definition and its large $N_c$ limit, but not also the intermediate expression in terms of fundamental Wilson lines alone (since this latter looks rather complicated).

\smallskip
$\bullet$ {\bf Definitions which appear in Sect.~\ref{qqqtrij}}
 \begin{equation}\begin{split}\label{wils1}
&S_{q\overline{q}qq\overline{q}q}\left(\overline{\bm{x}},\,\overline{\bm{z}},\,\overline{\bm{z}}^{\prime},\,\bm{x},\,\bm{z},\,\bm{z}^{\prime}\right)\equiv\,\frac{2}{C_{F}\,N_{c}}\,\mathrm{tr}\left(V^{\dagger}(\bm{\overline{x}})\,V(\bm{x})\,t^{a}\,t^{b}\right)\,\mathrm{tr}\left(V(\bm{\overline{z}})\,t^{b}\,V^{\dagger}(\bm{\overline{z}}^{\prime})\,V(\bm{z}^{\prime})\,t^{a}\,V^{\dagger}(\bm{z})\right)\\
&=\,\frac{1}{2C_{F}N_{c}}\,\left(N_{c}^{2}\,\mathcal{Q}(\overline{\bm{x}},\,\bm{x},\,\bm{z},\,\overline{\bm{z}})\,\mathcal{S}(\overline{\bm{z}}^{\prime},\,\bm{z}^{\prime})\,-\,\mathcal{H}(\overline{\bm{x}},\,\bm{x},\,\bm{z},\,\overline{\bm{z}},\,\overline{\bm{z}}^{\prime},\,\bm{z}^{\prime})-\,\mathcal{H}(\overline{\bm{x}},\,\bm{x},\,\overline{\bm{z}}^{\prime},\,\bm{z}^{\prime},\,\bm{z},\,\overline{\bm{z}})\right.\\
&\left.\,+\,\mathcal{S}(\overline{\bm{x}},\,\bm{x})\,\mathcal{Q}(\overline{\bm{z}}^{\prime},\,\bm{z}^{\prime},\,\bm{z},\,\overline{\bm{z}})\right)\,\simeq\,{\mathcal{Q}(\overline{\bm{x}},\,\bm{x},\,\bm{z},\,\overline{\bm{z}})\,\mathcal{S}(\overline{\bm{z}}^{\prime},\,\bm{z}^{\prime})},
 \end{split}\end{equation}

 \begin{equation}\begin{split}
&S_{q\overline{q}qqg}\left(\overline{\bm{x}},\,\overline{\bm{z}},\,\overline{\bm{z}}^{\prime}\,,\bm{x},\,\bm{y}\right)\,\equiv\,\frac{2}{C_{F}\,N_{c}}\,\mathrm{tr}\left[t^{a}\,V^{\dagger}(\overline{\bm{x}})\,V(\bm{x})\,t^{d}\right]\,\mathrm{tr}\left[t^{a}\,V^{\dagger}(\overline{\bm{z}}^{\prime})\,t^{c}\,V(\overline{\bm{z}})\right]\,U^{cd}(\bm{y})\,=\,\frac{1}{2C_{F}\,N_{c}}\\
&\times\left(N_{c}^{2}\,\mathcal{Q}(\overline{\bm{x}},\,\bm{x},\,\bm{y},\,\overline{\bm{z}})\,\mathcal{S}(\overline{\bm{z}}^{\prime},\,\bm{y})-\mathcal{H}(\overline{\bm{x}},\,\bm{x},\,\bm{y},\,\overline{\bm{z}},\,\overline{\bm{z}}^{\prime},\,\bm{y})-\mathcal{Q}(\bm{\overline{\bm{x}}},\,\bm{x},\,\overline{\bm{z}}^{\prime},\,\overline{\bm{z}})+\mathcal{S}(\overline{\bm{x}},\,\bm{x})\,\mathcal{S}(\overline{\bm{z}}^{\prime},\,\overline{\bm{z}})\right)\\
&\simeq\,\mathcal{Q}(\overline{\bm{x}},\,\bm{x},\,\bm{y},\,\overline{\bm{z}})\,\mathcal{S}(\overline{\bm{z}}^{\prime},\,\bm{y}).
\end{split}\end{equation}

 \begin{equation}\begin{split}
&S_{qgq\overline{q}q}\left(\bm{\overline{x}},\,\bm{\overline{y}},\,\bm{x},\,\bm{z},\,\bm{z}^{\prime}\right)\,\equiv\,\frac{2}{C_{F}\,N_{c}}\,\mathrm{tr}\left[t^{d}V^{\dagger}(\bm{\overline{x}})\,V(\bm{x})\,t^{a}\right]\,\mathrm{tr}\left[t^{c}\,V(\bm{z}^{\prime})\,t^{a}\,V^{\dagger}(\bm{z})\right]\,U^{cd}(\bm{\overline{y}})\,=\,\frac{1}{2C_{F}\,N_{c}}\\
&\times\left(N_{c}^{2}\,\mathcal{Q}(\overline{\bm{x}},\,\bm{x},\,\bm{z},\,\overline{\bm{y}})\,\mathcal{S}(\overline{\bm{y}},\,\bm{z}^{\prime})-\mathcal{H}(\overline{\bm{x}},\,\bm{x},\,\overline{\bm{y}},\,\bm{z}^{\prime},\,\bm{z},\,\overline{\bm{y}})-\mathcal{Q}(\bm{\overline{\bm{x}}},\,\bm{x},\,\bm{z},\,\bm{z}^{\prime})+\mathcal{S}(\overline{\bm{x}},\,\bm{x})\,\mathcal{S}(\bm{z},\,\bm{z}^{\prime})\right)\\
&\simeq\,\mathcal{Q}(\overline{\bm{x}},\,\bm{x},\,\bm{z},\,\overline{\bm{y}})\,\mathcal{S}(\overline{\bm{y}},\,\bm{z}^{\prime}),
\end{split}\end{equation}

\begin{equation}\begin{split}\label{wils2}
&S_{qq\overline{q}q}\left(\overline{\bm{w}},\,\bm{x},\,\bm{z},\,\bm{z}^{\prime}\right)\,=\,\frac{2}{C_{F}\,N_{c}}\,\mathrm{tr}\left[V^{\dagger}(\overline{\bm{w}})\,t^{b}\,V(\bm{x})\,t^{a}\right]\,\mathrm{tr}\left[V(\bm{z}^{\prime})\,t^{a}\,V^{\dagger}(\bm{z})\,t^{b}\right]\,=\,\frac{1}{2C_{F}\,N_{c}}\\
&\times\left(N_{c}^{2}\,\mathcal{S}(\overline{\bm{w}},\,\bm{z}^{\prime})\,\mathcal{S}(\bm{z},\,\bm{x})\,-\,\mathcal{Q}(\overline{\bm{w}},\,\bm{x},\,\bm{z},\,\bm{z}^{\prime})\,-\,\mathcal{Q}(\overline{\bm{w}},\,\bm{z}^{\prime},\,\bm{z},\,\bm{x})\,+\,\mathcal{S}(\overline{\bm{w}},\,\bm{x})\,\mathcal{S}(\bm{z},\,\bm{z}^{\prime})\right)\\
&\simeq\,\mathcal{S}(\overline{\bm{w}},\,\bm{z}^{\prime})\,\mathcal{S}(\bm{z},\,\bm{x}).
 \end{split}\end{equation}

\smallskip
$\bullet$ {\bf Definitions which appear in \ref{qggtrijaa}}

\begin{equation}\begin{split}\label{wils3}
&S_{qggqgg}^{(1)}\left(\bm{\overline{x}},\,\overline{\bm{z}},\,\overline{\bm{z}}^{\prime},\,\bm{x},\,\bm{z},\,\bm{z}^{\prime}\right)\,\equiv\frac{1}{C_{F}^{2}\,N_{c}}\,\mathrm{tr}\left[V^{\dagger}(\bm{\overline{x}})\,V(\bm{x})\,t^{b}\,t^{a}\,t^{f}\,t^{e}\right]\,\left[U^{\dagger}(\overline{\bm{z}}^{\prime})\,U(\bm{z}^{\prime})\right]^{fa}\,\left[U^{\dagger}(\bm{\overline{z}})\,U(\bm{z})\right]^{eb}\\
&\simeq\,\mathcal{Q}\left(\bm{\overline{x}},\,\bm{x},\,\bm{z},\,\overline{\bm{z}}\right)\,\mathcal{Q}\left(\bm{z}^{\prime},\,\overline{\bm{z}}^{\prime},\,\overline{\bm{z}},\,\bm{z}\right)\,\mathcal{S}\left(\overline{\bm{z}}^{\prime},\,\bm{z}^{\prime}\right),
\end{split}\end{equation}

\begin{equation}\begin{split}
&S_{qgqgg}^{(1)}\left(\bm{\overline{y}},\,\overline{\bm{z}}^{\prime},\,\bm{x},\,\bm{z},\,\bm{z}^{\prime}\right)\,\equiv\,\frac{1}{C_{F}^{2}\,N_{c}}\,\mathrm{tr}\left[V^{\dagger}(\bm{\overline{y}})\,t^{d}\,V(\bm{x})\,t^{b}\,t^{a}\,t^{e}\right]\,\left[U^{\dagger}(\overline{\bm{z}}^{\prime})\,U(\bm{z}^{\prime})\right]^{ea}\,U^{db}(\bm{z})\\
&\simeq\,\mathcal{Q}\left(\overline{\bm{y}},\,\overline{\bm{z}}^{\prime},\,\bm{z}^{\prime},\,\bm{z}\right)\,\mathcal{S}\left(\overline{\bm{z}}^{\prime},\,\bm{x}\right)\,\mathcal{S}\left(\bm{z},\,\bm{z}^{\prime}\right),
\end{split}\end{equation}

 \begin{equation}\begin{split}
&S_{qggqg}^{(1)}\left(\bm{\overline{x}},\,\overline{\bm{z}},\,\overline{\bm{z}}^{\prime},\,\bm{y},\,\bm{z}\right)\,\equiv\,\frac{1}{C_{F}^{2}\,N_{c}}\,\mathrm{tr}\left[V^{\dagger}(\bm{\overline{x}})\,t^{d}\,V(\bm{y})\,t^{a}\,t^{f}\,t^{e}\right]\,\left[U^{\dagger}(\overline{\bm{z}}^{\prime})\,U(\bm{z}^{\prime})\right]^{fa}\,U^{de}(\bm{\overline{z}})\\
&\simeq\,\mathcal{Q}\left(\bm{z}^{\prime},\,\overline{\bm{z}}^{\prime},\,\overline{\bm{z}},\,\bm{y}\right)\,\mathcal{S}\left(\overline{\bm{z}}^{\prime},\,\bm{z}^{\prime}\right)\,\mathcal{S}\left(\overline{\bm{x}},\,\bm{z}\right),
\end{split}\end{equation}

 \begin{equation}\begin{split}\label{wils4}
&S_{qqgg}^{(1)}\left(\overline{\bm{w}},\,\bm{x},\,\bm{z},\,\bm{z}^{\prime}\right)\,\equiv\,\frac{1}{C_{F}^{2}\,N_{c}}\,\mathrm{tr}\left[V^{\dagger}(\overline{\bm{w}})\,t^{c}\,t^{d}V(\bm{x})\,t^{b}\,t^{a}\right]\,U^{db}(\bm{z})\,U^{ca}(\bm{z}^{\prime})\\
&=\,\frac{1}{4C_{F}^{2}}\,\left(N_{c}^{2}\,\mathcal{S}(\overline{\bm{w}},\,\bm{z}^{\prime})\,\mathcal{S}(\bm{z}^{\prime},\,\bm{z})\,\mathcal{S}(\bm{z},\,\bm{x})\,-\,\mathcal{S}(\overline{\bm{w}},\,\bm{z})\,\mathcal{S}(\bm{z},\,\bm{x})\,-\,\mathcal{S}(\overline{\bm{w}},\,\bm{z}^{\prime})\,\mathcal{S}(\bm{z}^{\prime},\,\bm{x})\,+\,\frac{1}{N_{c}^{2}}\,\mathcal{S}(\overline{\bm{w}},\,\bm{x})\right)\\
&\simeq\,\mathcal{S}(\overline{\bm{w}},\,\bm{z}^{\prime})\,\mathcal{S}(\bm{z}^{\prime},\,\bm{z})\,\mathcal{S}(\bm{z},\,\bm{x}).
\end{split}\end{equation}

\smallskip
$\bullet$ {\bf Definitions which appear in \ref{qggtrijbb}}
 
 \begin{equation}\begin{split}\label{wils5}
&S_{qggqgg}^{(2)}(\overline{\bm{x}},\,\overline{\bm{z}},\,\overline{\bm{z}}^{\prime},\,\bm{x},\,\bm{z},\,\bm{z}^{\prime})\,\equiv\,\frac{1}{C_{F}\,N_{c}^{2}}\,f^{rmn}\,f^{abc}\,\left[U^{\dagger}(\overline{\bm{z}})\,U(\bm{z})\right]^{nc}\,\left[U^{\dagger}(\overline{\bm{z}}^{\prime})\,U(\bm{z}^{\prime})\right]^{mb}\,\mathrm{tr}\left(V^{\dagger}(\overline{\bm{x}})\,V(\bm{x})\,t^{a}\,t^{r}\right)\\
&\simeq\,\frac{1}{2}\left(\mathcal{Q}\left(\overline{\bm{z}},\,\bm{z},\,\bm{z}^{\prime},\,\overline{\bm{z}}^{\prime}\right)\,\mathcal{Q}\left(\overline{\bm{x}},\,\bm{x},\,\bm{z},\,\overline{\bm{z}}\right)\,\mathcal{S}\left(\overline{\bm{z}}^{\prime},\,\bm{z}^{\prime}\right)+\mathcal{Q}\left(\bm{z},\,\overline{\bm{z}},\,\overline{\bm{z}}^{\prime},\,\bm{z}^{\prime}\right)\,\mathcal{Q}\left(\overline{\bm{x}},\,\bm{x},\,\bm{z}^{\prime},\,\overline{\bm{z}}^{\prime}\right)\,\mathcal{S}\left(\overline{\bm{z}},\,\bm{z}\right)\right),
  \end{split}\end{equation}

 \begin{equation}\begin{split}
&S_{qggqg}^{(2)}(\overline{\bm{x}},\,\overline{\bm{z}},\,\overline{\bm{z}}^{\prime},\,\bm{x},\,\bm{y})\,\equiv\,\frac{1}{C_{F}\,N_{c}^{2}}\,f^{rmn}\,f^{bde}\,U^{en}(\overline{\bm{z}})\,U^{dm}(\overline{\bm{z}}^{\prime})\,U^{ba}(\bm{y})\,\mathrm{tr}\left(V^{\dagger}(\overline{\bm{x}})\,V(\bm{x})\,t^{a}\,t^{r}\right)\\
&\simeq\,\frac{1}{2}\left(\mathcal{Q}\left(\bm{y},\,\overline{\bm{z}},\,\overline{\bm{x}},\,\bm{x}\right)\,\mathcal{S}\left(\overline{\bm{z}},\,\overline{\bm{z}}^{\prime}\right)\,\mathcal{S}\left(\overline{\bm{z}}^{\prime},\,\bm{y}\right)+\mathcal{Q}\left(\bm{y},\,\overline{\bm{z}}^{\prime},\,\overline{\bm{x}},\,\bm{x}\right)\,\mathcal{S}\left(\overline{\bm{z}},\,\bm{y}\right)\,\mathcal{S}\left(\overline{\bm{z}}^{\prime},\,\overline{\bm{z}}\right)\right),
  \end{split}\end{equation}

 \begin{equation}\begin{split}
&S_{qgqgg}^{(2)}(\overline{\bm{x}},\,\overline{\bm{y}},\,\bm{x},\,\bm{z},\,\bm{z}^{\prime})\,\equiv\,\frac{1}{C_{F}\,N_{c}^{2}}\,f^{nde}\,f^{abc}\,U^{nr}(\overline{\bm{y}})\,U^{ec}(\bm{z})\,U^{db}(\bm{z}^{\prime})\,\mathrm{tr}\left(V^{\dagger}(\overline{\bm{x}})\,V(\bm{x})\,t^{a}\,t^{r}\right)\\
&\simeq\,\frac{1}{2}\left(\mathcal{Q}\left(\overline{\bm{x}},\,\bm{x},\,\bm{z}^{\prime},\,\overline{\bm{y}}\right)\,\mathcal{S}\left(\overline{\bm{y}},\,\bm{z}\right)\,\mathcal{S}\left(\bm{z},\,\bm{z}^{\prime}\right)\,+\,\mathcal{Q}\left(\overline{\bm{x}},\,\bm{x},\,\bm{z},\,\overline{\bm{y}}\right)\,\mathcal{S}\left(\overline{\bm{y}},\,\bm{z}^{\prime}\right)\,\mathcal{S}\left(\bm{z}^{\prime},\,\bm{z}\right)\right),
  \end{split}\end{equation}

 \begin{equation}\begin{split}\label{wils6}
&S_{qqgg}^{(2)}(\overline{\bm{w}},\,\bm{x},\,\bm{z},\,\bm{z}^{\prime})\,\equiv\,\frac{1}{C_{F}\,N_{c}^{2}}\,f^{abc}\,f^{rde}\,U^{ec}(\bm{z})\,U^{db}(\bm{z}^{\prime})\,\mathrm{tr}\left(V^{\dagger}(\overline{\bm{w}})\,t^{r}\,V(\bm{x})\,t^{a}\right)\\
&=\,\frac{1}{4N_{c}\,C_{F}}\,\left(N_{c}^{2}\,\mathcal{S}(\overline{\bm{w}},\,\bm{z}^{\prime})\,\mathcal{S}(\bm{z}^{\prime},\,\bm{z})\,\mathcal{S}(\bm{z},\,\bm{x})\,+\,N_{c}^{2}\,\mathcal{S}(\overline{\bm{w}},\,\bm{z})\,\mathcal{S}(\bm{z},\,\bm{z}^{\prime})\,\mathcal{S}(\bm{z}^{\prime},\,\bm{x})\,-\,\mathcal{H}\left(\overline{\bm{w}},\,\bm{z}^{\prime},\,\bm{z},\,\bm{x},\,\bm{z}^{\prime},\,\bm{z}\right)\right.\\
&\left.\,-\,\mathcal{H}\left(\overline{\bm{w}},\,\bm{z},\,\bm{z}^{\prime},\,\bm{x},\,\bm{z},\,\bm{z}^{\prime}\right)\right)\,\simeq\,\frac{1}{2}\left(\mathcal{S}(\overline{\bm{w}},\,\bm{z}^{\prime})\,\mathcal{S}(\bm{z}^{\prime},\,\bm{z})\,\mathcal{S}(\bm{z},\,\bm{x})\,+\,\mathcal{S}(\overline{\bm{w}},\,\bm{z})\,\mathcal{S}(\bm{z},\,\bm{z}^{\prime})\,\mathcal{S}(\bm{z}^{\prime},\,\bm{x})\right).
  \end{split}\end{equation}

$\bullet$ {\bf Definitions which appear in \ref{qggtrijab}}
 \begin{equation}\begin{split}\label{wils7}
&S_{qggqgg}^{(3)}\left(\overline{\bm{x}},\,\overline{\bm{z}},\,\overline{\bm{z}}^{\prime},\,\bm{x},\,\bm{z},\,\bm{z}^{\prime}\right)\,\equiv\,\frac{2i}{C_{F}\,N_{c}^{2}}f^{abc}\,\left[U^{\dagger}(\overline{\bm{z}})\,U(\bm{z})\right]^{mb}\,\left[U^{\dagger}(\overline{\bm{z}}^{\prime})\,U(\bm{z}^{\prime})\right]^{rc}\,\mathrm{tr}\left(V^{\dagger}(\overline{\bm{x}})\,V(\bm{x})\,t^{a}\,t^{r}\,t^{m}\right)\\
&\simeq\,\mathcal{Q}\left(\overline{\bm{z}},\,\bm{z},\,\bm{z}^{\prime},\,\overline{\bm{z}}^{\prime}\right)\,\mathcal{Q}\left(\overline{\bm{x}},\,\bm{x},\,\bm{z},\,\overline{\bm{z}}\right)\,\mathcal{S}\left(\overline{\bm{z}}^{\prime},\,\bm{z}^{\prime}\right),
  \end{split}\end{equation}

 \begin{equation}\begin{split}
&S_{qggqg}^{(3)}\left(\overline{\bm{x}},\,\overline{\bm{z}},\,\overline{\bm{z}}^{\prime},\,\bm{x},\,\bm{y}\right)\,\equiv\,\frac{2i}{C_{F}\,N_{c}^{2}}f^{bde}\,U^{dm}(\overline{\bm{z}})\,U^{er}(\overline{\bm{z}}^{\prime})\,U^{ba}(\bm{y})\,\mathrm{tr}\left(V^{\dagger}(\overline{\bm{x}})\,V(\bm{x})\,t^{a}\,t^{r}\,t^{m}\right)\\
&\simeq\,\mathcal{Q}\left(\overline{\bm{x}},\,\bm{x},\,\bm{y},\,\overline{\bm{z}}\right)\,\mathcal{S}\left(\overline{\bm{z}},\,\overline{\bm{z}}^{\prime}\right)\,\mathcal{S}\left(\overline{\bm{z}}^{\prime},\,\bm{y}\right),
  \end{split}\end{equation}

 \begin{equation}\begin{split}
&S_{qgqgg}^{(3)}\left(\overline{\bm{y}},\,\overline{\bm{z}}^{\prime},\,\bm{x},\,\bm{z},\,\bm{z}^{\prime}\right)\,\equiv\,\frac{2i}{C_{F}\,N_{c}^{2}}f^{abc}\,\left[U^{\dagger}(\overline{\bm{z}}^{\prime})\,U(\bm{z}^{\prime})\right]^{rc}\,U^{db}(\bm{z})\,\mathrm{tr}\left(V^{\dagger}(\overline{\bm{y}})\,t^{d}V(\bm{x})\,t^{a}\,t^{r}\right)\\
&\simeq\,\mathcal{Q}\left(\overline{\bm{y}},\,\bm{z},\,\bm{z}^{\prime},\,\overline{\bm{z}}^{\prime}\right)\,\mathcal{S}\left(\bm{z},\,\bm{x}\right)\,\mathcal{S}\left(\overline{\bm{z}}^{\prime},\,\bm{z}^{\prime}\right),
  \end{split}\end{equation}

 \begin{equation}\begin{split}\label{wils8}
&S_{qqgg}^{(3)}\left(\overline{\bm{w}},\,\bm{x},\,\bm{z},\,\bm{z}^{\prime}\right)\,\equiv\,\frac{2i}{C_{F}\,N_{c}^{2}}\,f^{abc}\,U^{ec}(\bm{z}^{\prime})\,U^{db}(\bm{z})\,\mathrm{tr}\left(V^{\dagger}(\overline{\bm{w}})\,t^{e}\,t^{d}\,V(\bm{x})\,t^{a}\right)\,=\,\frac{1}{2C_{F}\,N_{c}}\\
&\times\left(N_{c}^{2}\mathcal{S}\left(\overline{\bm{w}},\,\bm{z}^{\prime}\right)\mathcal{S}\left(\bm{z}^{\prime},\,\bm{z}\right)\mathcal{S}\left(\bm{z},\,\bm{x}\right)\,-\,\mathcal{H}\left(\overline{\bm{w}},\,\bm{z}^{\prime},\,\bm{z},\,\bm{x},\,\bm{z}^{\prime},\,\bm{z}\right)\right)\,\simeq\,\mathcal{S}\left(\overline{\bm{w}},\,\bm{z}^{\prime}\right)\mathcal{S}\left(\bm{z}^{\prime},\,\bm{z}\right)\mathcal{S}\left(\bm{z},\,\bm{x}\right).
  \end{split}\end{equation}
  
 \begin{equation}\begin{split}
&S_{qggq}\left(\overline{\bm{x}},\,\overline{\bm{z}},\,\overline{\bm{z}}^{\prime},\,\bm{w}\right)\,\equiv\,\frac{2i}{C_{F}\,N_{c}^{2}}\,f^{ade}\,U^{dm}(\overline{\bm{z}})\,U^{er}(\overline{\bm{z}}^{\prime})\,\mathrm{tr}\left(V^{\dagger}(\overline{\bm{x}})\,t^{a}\,V(\bm{w})\,t^{r}\,t^{m}\right)\\
&=\,\frac{1}{2C_{F}\,N_{c}}\,\left(N_{c}^{2}\,\mathcal{S}(\overline{\bm{x}},\,\overline{\bm{z}})\,\mathcal{S}(\overline{\bm{z}},\,\overline{\bm{z}}^{\prime})\,\mathcal{S}(\overline{\bm{z}}^{\prime},\,\bm{w})\,-\,\mathcal{H}\left(\overline{\bm{x}},\,\overline{\bm{z}}^{\prime},\,\overline{\bm{z}},\,\bm{w},\,\overline{\bm{z}}^{\prime},\,\overline{\bm{z}}\right)\right)\,\simeq\,\mathcal{S}(\overline{\bm{x}},\,\overline{\bm{z}})\,\mathcal{S}(\overline{\bm{z}},\,\overline{\bm{z}}^{\prime})\,\mathcal{S}(\overline{\bm{z}}^{\prime},\,\bm{w}),
  \end{split}\end{equation}


\providecommand{\href}[2]{#2}\begingroup\raggedright\endgroup

\end{document}